%% file: PIPXXXII.tex
\documentclass[twocolumn,traditabstract,longauth]{aa}

\input{polar_definitions}

\input{Planck.tex}

\usepackage{natbib}
\usepackage{float}
\usepackage{color}
\usepackage{hyperref}
\usepackage{url}
\usepackage{graphicx,amsmath}
\usepackage{fixltx2e}
\usepackage{graphicx,epsfig}
\usepackage{txfonts}
\usepackage{amsmath}
\newcommand{\hi}{\ion{H}{i}} 
 
\newcommand{\healpix}{{\tt HEALPix}}
\newcommand{\planck}{\textit{Planck}}

\def\ben{\begin{enumerate}}
\def\een{\end{enumerate}}
\def\bi{\begin{itemize}}
\def\ei{\end{itemize}}
\def\be{\begin{equation}}
\def\ee{\end{equation}}
\def\bea{\begin{eqnarray}}
\def\eea{\end{eqnarray}}

\begin{document}

\input{PIP_94_Boulanger_authors_and_institutes.tex}
   \title{\textit{{\Planck}} intermediate results. XXXII. The
     relative orientation between the magnetic field and 
     structures traced by interstellar dust}


  \abstract{
The role of the magnetic field in the formation of the filamentary
structures observed in the interstellar medium (ISM) is a debated
topic owing to the paucity of relevant observations needed to test existing models.
The \Planck\ all-sky maps of linearly polarized emission from dust at
$353\,$GHz provide the required combination of imaging and statistics to study the correlation between the structures
of the Galactic magnetic field and of interstellar matter over the whole sky, both in the diffuse ISM and in molecular clouds. 
The data reveal that structures, or ridges, in the intensity map have counterparts in the Stokes $\StokesQ$ and/or $\StokesU$ maps. 
We focus our study on structures at intermediate and high Galactic latitudes, which cover two orders of magnitude in column density, from $10^{20}$ to $10^{22}$
cm$^{-2}$. 
We measure the magnetic field orientation on the plane of the sky from the polarization data, 
and present an algorithm to estimate the orientation of the ridges from the dust intensity map. We use analytical models to account  for projection effects.
Comparing polarization angles on and off the structures, we estimate the mean ratio between the strengths of 
the turbulent and mean components of the magnetic field to be between
0.6 and 1.0, with a preferred value of 0.8. We find that the ridges
are usually aligned with the magnetic field measured on the structures. This statistical trend becomes more striking
for increasing polarization fraction and decreasing column density. There is no alignment for the highest column density ridges. 
We interpret the increase in alignment with polarization fraction  as a consequence of projection effects. 
We present maps to show that the decrease in alignment for high column density is not due to a loss of correlation between the distribution of matter and the geometry of the magnetic field. In molecular complexes, we also observe structures perpendicular to the magnetic field, 
which, statistically, cannot be accounted for by projection effects. 
This first statistical study of the relative orientation between the matter structures and the magnetic field in the ISM points out that, at the angular scales probed by 
\Planck, the field geometry projected on the plane of the sky is correlated with the distribution of matter. 
In the diffuse ISM, the structures of matter are usually aligned with the magnetic field, while perpendicular structures appear in molecular clouds. We discuss our results in the context of 
models and MHD simulations, which attempt to describe the respective roles of turbulence, magnetic field, and self-gravity in 
the formation of structures in the magnetized ISM. 
}

\titlerunning{Coherent structures in polarization 
maps of the diffuse ISM}

\authorrunning{Planck Collaboration}
\maketitle

\section{ Introduction }
\label{sec:intro}

The filamentary appearance of the interstellar medium (ISM) has been revealed over the last decades by observations of
dust emission, stellar reddening, and gas line emission, mainly CO and {\ion{H}{i}} \citep[see][for a recent ISM review]{Hennebelle12}. 
Most recently, {\it Herschel} maps of dust emission at far-infrared wavelengths  have identified gravitationally bound  filaments as the 
loci where stars form \citep{Andre10}.  
Filaments are ubiquitous in interstellar space and are essential to star formation, but our understanding of how they form is still fragmentary.  

Filaments are striking features in numerical  
simulations of the diffuse ISM and molecular clouds \citep[e.g.][]{Heitsch05,Nakamura08,Gong11,Hennebelle13}. 
They are present in both hydrodynamic and magneto-hydrodynamic (MHD) simulations, 
but they are more conspicuous in the latter.  These studies relate the filamentary appearance of the ISM to 
compression and shear driven by turbulence, and the anisotropic infall of gravitationally unstable structures. 
\citet{Soler13} find that, statistically, the orientation changes from parallel to perpendicular for gravitationally bound structures 
in simulations where the magnetic field is dynamically important. 

The role of the magnetic field in the ISM dynamics depends on the field strength with respect to gravitational and turbulent energies. 
In the diffuse ISM, the magnetic energy is observed to be comparable with the turbulent kinetic energy of the gas \citep{Heiles05,Basu2013} and to dominate its self-gravity \citep{Crutcher2010}, while stars form where and when gravity prevails. On what spatial and time scales does this transition in the ratio between magnetic and gravitational energies occur? This question has been addressed by theorists in several ways.  Ambipolar diffusion \citep{Ciolek93}, including turbulence \citep{Zweibel02} or magnetic reconnection \citep{Lazarian99}, can decouple matter from the magnetic field.  Furthermore, gas motions along field lines contribute to condensing the matter without increasing the magnetic flux. This has been suggested for the formation of molecular clouds  \citep{Blitz80,Hartmann01,Hennebelle08,Heitsch09,Inoue09}
and of gravitationally bound filaments within gas sheets \citep{Nagai98}. Because of magnetic tension \citep{Hennebelle00,Heyer12}, the gas is expected to flow 
mainly along field lines where turbulence is sub-Alfv{\'e}nic.

The challenge faced by observers is to gather the data necessary to characterize the interplay between gravity, turbulence, and magnetic fields from the diffuse ISM to star-forming molecular clouds.
A wealth of data is already available to quantify the gas self-gravity and turbulence \citep{Hennebelle12}, 
but we have comparatively little information on the magnetic field strength and its structure 
within interstellar clouds.  The dearth of data on the magnetic field follows from the difficulty of performing 
the relevant observations. Measurements of the magnetic field components along the line of sight and on the plane of the sky using the Zeeman effect and 
dust and synchrotron linear polarizations, respectively, are notoriously difficult \citep{Crutcher12,Haverkorn2015}. 

Synchrotron emission and Faraday rotation have been used to estimate the strength of the magnetic field and the ratio between its random and regular components in the Milky Way and external galaxies \citep{Haverkorn2004,Beck2007,Schnitzeler2007,Houde2013}. A spatial correlation between the magnetic field structure and that of interstellar matter has been observed at kpc-scales in external galaxies from synchrotron radio polarization \citep{Beck2005,Patrikeev2006,Fletcher2011}. This correlation has been observed to depend on the gas density and star formation rate \citep{Chyzy2008}. 
However, the interplay between the structure of the field and that of matter on smaller scales in the solar neighbourhood is still highly debated.

A number of studies, using the polarization of background starlight caused by dichroic absorption, have targeted filaments in dark clouds \citep[e.g.][]{Goodman90,Goodman95,Pereyra04,Alves08,Chapman11,Cashman14}, and
in the diffuse ISM at lower column densities \citep{McClure06,Clark13}. Studying the relative orientation between the main axis of elongated molecular clouds and the orientation of the magnetic field inferred from starlight polarimetry, \citet{Li2013} present evidence for a bimodal distribution of relative orientations being either parallel or perpendicular. 
Most of these studies rely on polarization observations for discrete 
lines of sight selected by the availability of background stars, and often the magnetic field orientation is not measured at the position of the matter structures but on nearby 
lines of sight.

The \Planck\footnote{\Planck\ (\url{http://www.esa.int/Planck}) is a project of the European Space Agency (ESA) with instruments provided by two scientific consortia funded by ESA member states (in particular the lead countries France and Italy), with contributions from NASA (USA) and telescope reflectors provided by a collaboration between ESA and a scientific consortium led and funded by Denmark.} satellite has recently completed the first all-sky map of dust polarization in
emission.  This is an immense step forward in brightness sensitivity and statistics from 
earlier polarization observations at sub-mm wavelengths \citep[e.g.][]{Benoit2004,Ward09,Koch10,Poidevin14,Matthews14}.
While only ground-based observations provide the angular resolution required to measure the polarization of pre-stellar cores \citep{Matthews09,Tang2009} and to image dust polarization in distant molecular clouds \citep{Li03,Tassis09}, the {\planck} data are unique in their ability to map 
the dust polarization of filamentary structures in the solar neighbourhood. 
For the first time, we have the  data needed to characterize statistically the structure
of the Galactic magnetic field and its coupling to interstellar matter and turbulence at physical scales relevant to the formation of interstellar filaments.

The data are revealing a new view of the sky that we have started to explore.  
A first description of the \planck\ polarization maps at $353\,$GHz is presented in \citet{planck2014-XIX} and \citet{planck2014-XX}.
These first two papers describe the statistics of the polarization angle $\psi$ (perpendicular to the magnetic field orientation projected on the plane of the sky) and polarization fraction $p$. \citet{planck2014-XX}
show that the statistics of the data on $\psi$ and $p$  compare well with those measured on a MHD simulation of 
a molecular cloud formed at the interface of two colliding flows. A major finding of this paper is that the statistics of $\psi$ and $p$ depend on the direction of the mean magnetic field. 
Here, we pursue our analysis of the \planck\  dust polarization sky, 
focusing on the polarization properties of the localized filamentary structures in the solar neighbourhood, alternatively called ridges, identified in the Stokes $\StokesI $ map.
We use the \planck\ data to determine and compare the orientation of the filamentary structures and that of the magnetic field projected on the plane of the sky. 

The paper is organized as follows. The \planck\ data we use are introduced in Sect.~\ref{sec:planckdata}. 
Section~\ref{sec:coherent_structures} 
presents sky images that emphasize the correlation between structures in polarization and corresponding  
features in intensity. The selection and characteristics of 
the regions where we compare the orientations of the magnetic field and that of the structures of matter are described in Sect.~\ref{sec:filaments}. Section~\ref{sec:magfield} presents the magnetic field properties of the selected structures.
Sections~\ref{sec:alignment} and \ref{sec:align_nh} focus on quantifying the relative
orientation of the magnetic field and the ridges in the diffuse ISM and molecular clouds. We discuss our results 
in the context of our present understanding of the formation of structures in the magnetized ISM
in Sect.~\ref{sec:discussion}. The main results  are summarized in Sect.~\ref{sec:summ}. The paper has two appendices. 
Appendix~\ref{appendix:Hessian} details how we  
measure the local orientation of the structures in the dust emission map and 
quantify uncertainties.
In Appendix~\ref{appendix:gaussian}, we present the model that we use 
to quantify projection effects and interpret the statistics of the 
angle between the magnetic field and the brightness ridges on the sky.


\section{Data Sets}
\label{sec:planckdata}

\Planck\  observed the sky polarization in seven frequency bands
from 30 to 353\,GHz \citep{planck2013-p01}.  In this paper, we only use the data from the
High Frequency Instrument \citep[HFI,][]{Lamarre:2010} at the highest frequency, 353\,GHz,
where the dust emission is the brightest. This is the best-suited \Planck\ map for studying the structure of the dust polarization 
sky \citep{planck2014-XIX,planck2014-XX}. 

We use the Stokes $\StokesQ$ and $\StokesU$  maps and the associated noise maps made with the five independent consecutive sky surveys of the \Planck\ cryogenic mission, which correspond to 
the DR3 (delta-DX9) internal data release. We refer to previous \planck\ publications for the data processing,
map-making, photometric calibration, and photometric uncertainties
\citep{planck2013-p02,planck2013-p03,planck2013-p02b,planck2013-p03f}.
The $\StokesQ$ and $\StokesU$ maps are corrected for spectral leakage as described in \citet{planck2014-XIX}. As in this first \Planck\ polarization paper, 
we use the IAU convention for the polarization angle, measured from the local direction
to the north Galactic pole with positive values towards the east. 

For the dust total intensity at $353\,$GHz we use the model map, $D_{353}$, and the associated noise map, derived from a fit with a modified blackbody
of the \planck\ data at $\nu \ge 353\,$GHz, and {\it IRAS} at $\lambda = 100\,\mu$m \citep{planck2013-p06b}. This map has a lower noise than the corresponding 
$353\,$GHz Stokes $\StokesI$  \Planck\ map.  Furthermore, $D_{353}$ is the dust specific intensity corrected for zodiacal emission, cosmic microwave background  anisotropies, and the cosmic infrared background  monopole. 

The  $\StokesQ$ and $\StokesU$ maps are initially at 4\parcm8 resolution, and  $D_{353}$ at $5\arcmin$. The three maps 
are in \healpix\ format\footnote{\citet{Gorski05}, \tt{http://healpix.sf.net}}
with a pixelization $N_{\rm side} = 2048$. To increase the signal-to-noise ratio of extended emission, we smooth the three maps to
$15'$ resolution using the Gaussian approximation to the \Planck\ beam. We 
reduce the  \healpix\  resolution to $N_{\rm side} = 512$ (7\parcm1 pixels) after smoothing. 
To finely sample the beam, we  also use the smoothed $D_{353}$ map with $N_{\rm side} = 1024$ in Appendix~\ref{appendix:Hessian}.
For the polarization maps,
we apply the ``\ensuremath{\tt ismoothing}" routine of \healpix\ that decomposes the $\StokesQ$ and $\StokesU$ maps  into $E$ and $B$ maps, applies the Gaussian smoothing in harmonic space, 
and transforms the smoothed $E$ and $B$  back into $\StokesQ$ and $\StokesU$ maps at $N_{\rm side}=512$ resolution.
Most of our analysis is based on the $\StokesQ$, $\StokesU$, and $D_{353}$ maps, but   
we also use the maps of the de-biased polarization fraction $p$ and angle $\psi$ produced by \citet{planck2014-XIX}. The contribution of the CMB polarization to the
$\StokesQ$ and $\StokesU$ maps at $353\, $GHz is negligible  for this study.

\section{Structures in the polarization maps}
\label{sec:coherent_structures}

We introduce dust polarization (Sect.~\ref{sec:framework})
and  present all-sky images highlighting localized structures in
the dust Stokes $\StokesQ$ and $\StokesU$ maps correlated with corresponding  
features in intensity (Sect.~\ref{sec:visu}).

\begin{figure*}[h!]
\centerline{\includegraphics[width=18cm]{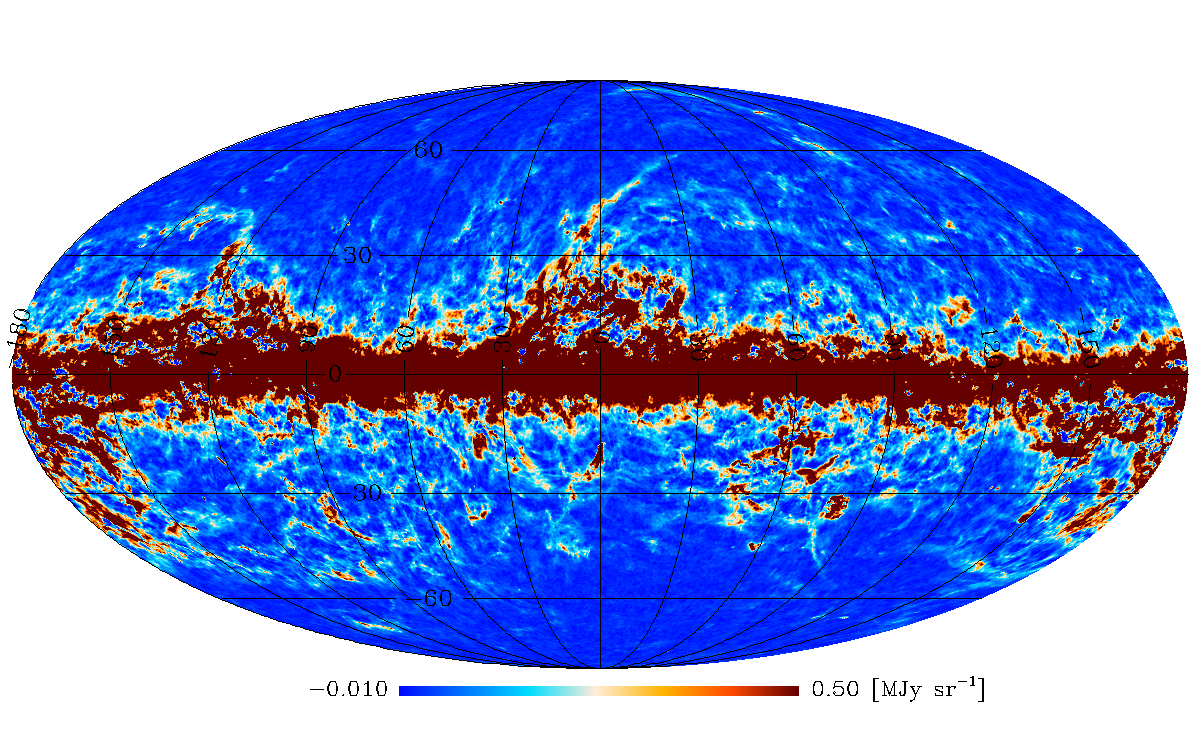}}
\centerline{\includegraphics[width=18cm]{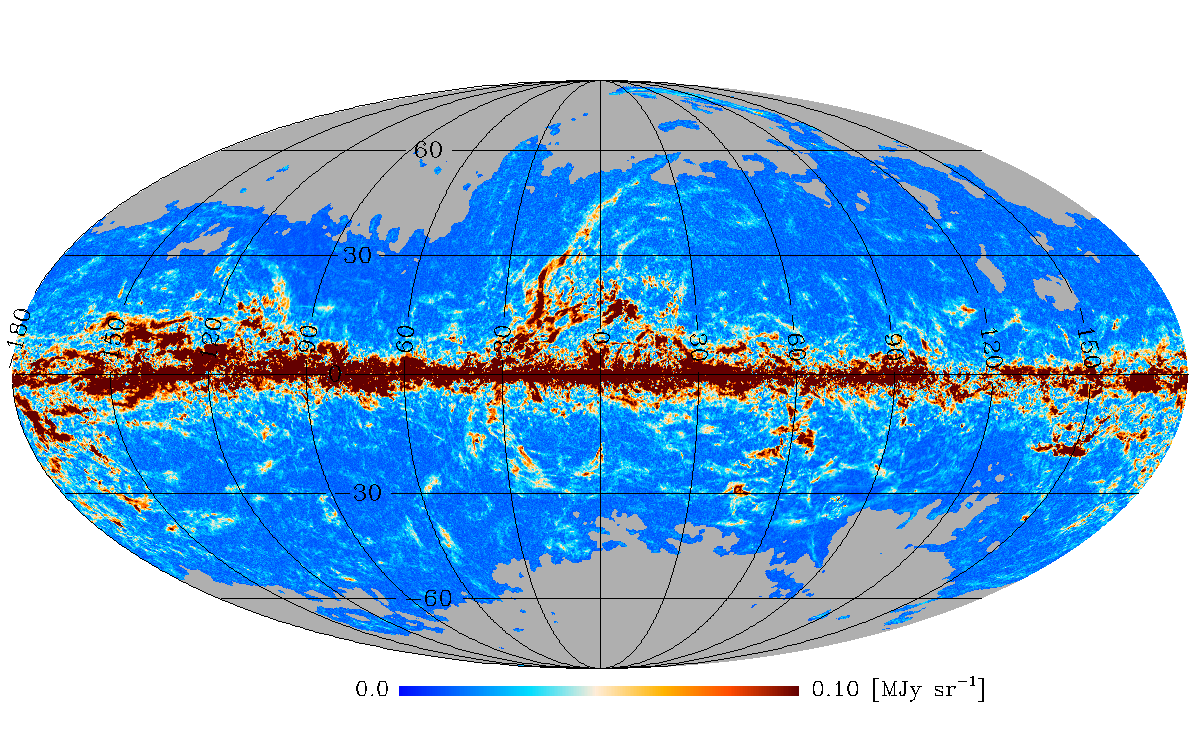}}
\caption{  {\it Top:} all-sky Mollweide display of the dust
  emission intensity at 353~GHz after background subtraction,  $D_{\rm
    353}^{\rm Dif}$ (see Sect.~\ref{sec:visu}).
 {\it Bottom:} corresponding difference map for the polarized
 emission,  $P_{\rm 353}^{\rm Dif}$ in Eq.~(\ref{eq:pcon}). The regions of low polarization
 signal at high Galactic latitude are masked. These
images include a grid of Galactic coordinates in degrees.
 }
\label{fig:Pmap}
\end{figure*}
\begin{figure*}[h!]
\centerline{\includegraphics[width=18cm]{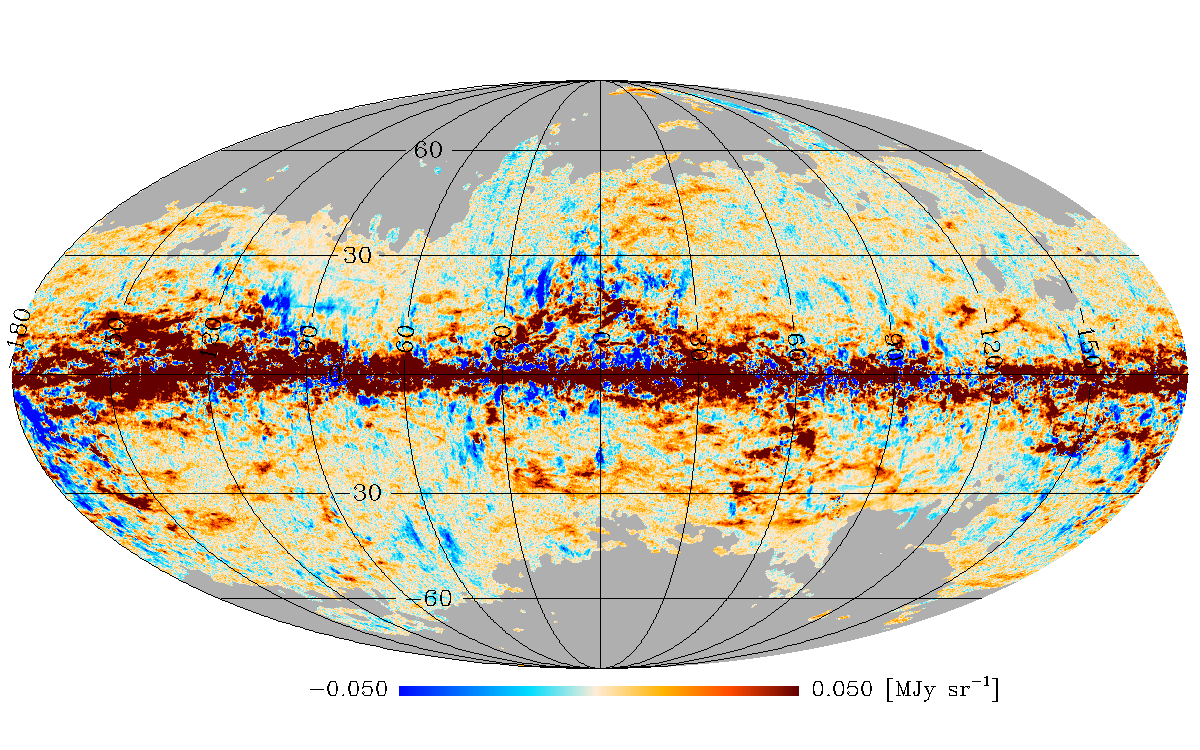}}
\centerline{\includegraphics[width=18cm]{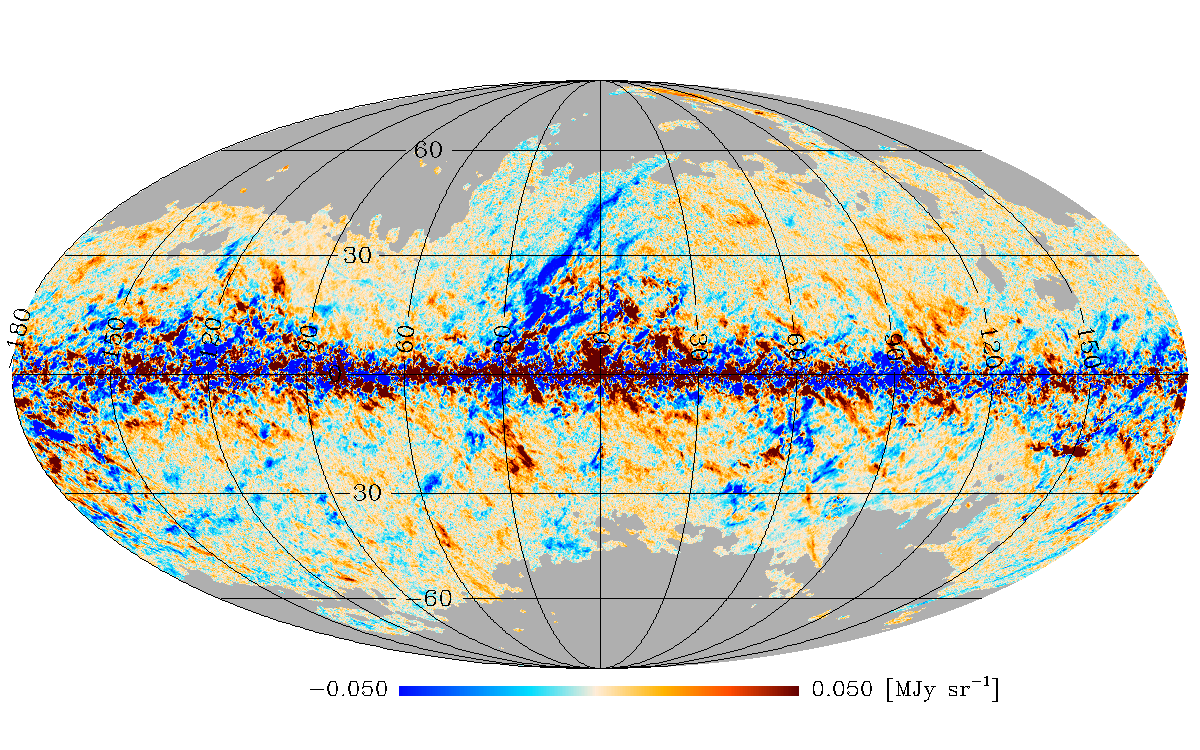}}
\caption{ 
All-sky Mollweide display of the  Stokes maps 
$Q_{353}^{\rm Dif}$  ({\it top}) and $U_{353}^{\rm Dif}$  ({\it bottom}) at  $353\,$GHz after background 
subtraction. The regions of low polarization
 signal at high Galactic latitude are masked. These
images include a grid of Galactic coordinates in degrees. The map
shown in the bottom panel represents the Stokes $\StokesU$ parameter in
\healpix\ convention, which corresponds to $-\StokesU$ in IAU convention, as
described in Eq.~(\ref{eq:Stokes_parameters}).}  
\label{fig:QUmaps}
\end{figure*}

\subsection{Dust polarization over the sky}
\label{sec:framework}

The integral equations of the Stokes parameters for linear dust
polarization are defined in \citet{planck2014-XX} (their Eqs.~5 to 7). 
For constant magnetic field orientation and polarization fraction along the line of sight, the
$\StokesQ$ and $\StokesU$ parameters can be related to the total intensity, $I$ , through 
\begin{align}
\label{eq:Stokes_parameters}
& Q = p_{0}\,\cos^2{\gamma}\cos{(2\,\psi)}\, I,\nonumber \\
& U = p_{0}\,\cos^2{\gamma}\sin{(2\,\psi)}\, I, 
\end{align}
where $\gamma$ is the angle between the
magnetic field and the plane of the sky and $\psi$  the
polarization angle. The intrinsic polarization fraction of dust emission, $p_{0}$,  
is given by

\begin{equation}
\label{eq:intrinsic_pol}
p_{0}=\frac{C_{\rm pol}}{C_{\rm avg}}\, R,
\end{equation}
where $C_{\rm pol}$ and $C_{\rm avg}$ are the polarization
and the average cross-sections of
dust, as defined in Appendix~B of \citet{planck2014-XX}, and $R$ is the Rayleigh reduction factor, which
characterizes the degree of dust grain alignment with the local
magnetic field \citep{Lee85,Hildebrand88}. The observed polarization
fraction is 
\begin{equation}
\label{eq:observed_pol}
p= p_{0} \cos^2{\gamma}.
\end{equation}

From Eq.~(\ref{eq:Stokes_parameters}), the localized structures in maps of dust polarization can come either
from local variations of the magnetic field orientation, the intrinsic polarization
fraction $p_{0}$, or the total emission $\StokesI$ map. The structure of the
polarization sky also depends on the depolarization associated with the magnetic field structure in the volume sampled by
the beam \citep{Lee85,planck2014-XX}, which is ignored in Eq.~(\ref{eq:Stokes_parameters}).
The filamentary structures revealed by the maps of the local dispersion of the polarization angle, presented in 
\citet{planck2014-XIX} and \citet{planck2014-XX}, are structures associated with changes in 
the orientation of the magnetic field because they do not correlate
with structures in the dust $\StokesI$ map. As pointed
  out by \citet{planck2014-XIX}, these structures morphologically
  resemble those found in maps of normalized gradients of polarized
synchrotron emission at radio frequencies \citep{Burkhart2012,Iacobelli2014}.

On the other hand, this paper presents a complementary analysis of the \Planck\
polarization sky, focusing on structures that have a counterpart in
the $\StokesI$  map.

\subsection{Visualization of the structures}
\label{sec:visu}

Henceforth, we use Eq.~(\ref{eq:Stokes_parameters}) with the $D_{353}$ map for the total intensity $I$, 
and the $\StokesQ$ and  $\StokesU$ maps at $353\,$GHz, which we write as $Q_{353}$ and $U_{353}$.
We consider localized structures, 
which appear in the $D_{353}$ map as 
contrasted ridges with respect to the local and more diffuse emission,
hereafter referred to as the background. 
Previous works at radio frequencies already faced the problem of separating
  the signal of localized structures in the Galaxy from the background
emission. \citet{Sofue1979} introduced the unsharp-mask
method to investigate the structure of the North Polar Spur from radio
continuum observations at $1420$ MHz obtained with the $100$-m telescope. 
To present the contrasted structures we follow a similar, but simpler, approach.

We produce a low
resolution background map, $D_{\rm 353}^{\rm BG}$, from $D_{\rm 353}$.
For each sky pixel, we compute a histogram of  $D_{\rm 353}$ within a
circular aperture of radius $2\pdeg5$ (this is not a critical value of
the data analysis, repeating the background estimate with an aperture
of $5^\circ$ does not significantly change our results). The background value at this
position is estimated from the mean of the $20\,\%$ lowest values.  We
also show that our choice of the $20\,\%$ fraction is not a critical aspect of the data
analysis in Sects.~\ref{sec:magfield} and \ref{subsec:alignment}. The top panel in Fig.~\ref{fig:Pmap} shows the difference
\begin{equation}
\label{eq:Dstr}  
D_{\rm 353}^{\rm Dif} = D_{\rm 353}- D_{\rm 353}^{\rm BG},
\end{equation}
which highlights localized features in the sky from low to high
Galactic latitudes. 

We also make the background maps, $Q_{\rm 353}^{\rm BG}$ and $U_{353}^{\rm BG}$, 
computing the mean values of the $Q_{353}$ and $U_{353}$ maps over the same pixels used to
compute $D_{\rm 353}^{\rm BG}$, as well as the difference maps
for the Stokes parameters
\begin{align}
\label{eq:QUstr}  
& Q_{\rm 353}^{\rm Dif} = Q_{353} - Q_{\rm 353}^{\rm BG}\nonumber \\
& U_{\rm 353}^{\rm Dif} = U_{353} - U_{\rm 353}^{\rm BG}.
\end{align}
The $Q_{\rm 353}^{\rm Dif}$ and $U_{\rm 353}^{\rm Dif}$  maps are presented in
Fig.~\ref{fig:QUmaps}. The  results of the background subtraction on the polarization data are illustrated
by the map defined by
\begin{equation}
\label{eq:pcon}
 P_{\rm 353}^{\rm Dif} = \sqrt{(Q_{\rm 353}^{\rm Dif})^2 +(U_{\rm
     353}^{\rm Dif})^2},
\end{equation}
shown in the bottom panel of Fig.~\ref{fig:Pmap}.
We present the maps after applying the same all-sky mask
defined in \citet{planck2014-XIX}. They show pixels where the systematic uncertainties are
small, and where the dust signal dominates the total emission
\citep[see Sect.~2.4 of][]{planck2014-XIX}. We note that
 $ P_{\rm 353}^{\rm Dif} $ is only used for
visualization purposes, and we stress that it is not used for data analysis.
In the data analysis we make use of the polarization
fraction map described by \citet{planck2014-XIX} (see
Sect.~\ref{sec:planckdata}), which is corrected for the positive bias
due to noise. 

The polarization maps show localized structures that are spatially
coincident with comparable features in $D_{\rm 353}^{\rm Dif}$.
However, there is not a one-to-one correspondence between the polarization and intensity maps. 
The localized features in $D_{\rm 353}^{\rm Dif}$ appear with different contrast and sign in the polarization maps. 
These differences trace changes in the polarization fraction and angle, 
which are observed to vary.

\section{Ridges in the dust emission map}
\label{sec:filaments}

In this section we describe how we identify and select the ridges in the $D_{353}$ map, where
we will later compare the orientation of the magnetic field and that of the
matter structures (Sects.~\ref{subsec:ridges} and \ref{subsec:mask}).
The selected structures are characterized in Sect.~\ref{subsec:column}.

\subsection{Detection of the ridges}
\label{subsec:ridges}

Deciding where to compare the orientations of the
magnetic field and structures of matter is an important step of our data analysis. 
We need an algorithm that selects pixels on
localized structures, providing the orientation
at each position on the sky. Thus, unlike what was done in
analysing {\it Herschel} maps of molecular clouds in the Gould Belt
\citep[][]{Arzoumanian2011},  we do not seek to identify filaments as
coherent structures, and we do not need to restrict the analysis to
the crest of the filaments. 

Anisotropic wavelet techniques, like those applied by
  \citet{Patrikeev2006} to investigate the spiral arms of M51, can be
  used to measure the relative orientation between the magnetic field
  and the matter structure, although they are not optimal for tracing
  complicated and intricate patterns.

These distinct requirements led us to make use of a different algorithm than that applied in these earlier studies. 
To identify the structures, we use a Hessian analysis of $D_{353}$. The
Hessian matrix was also used to analyse {\it Herschel} images of the L1641 cloud in Orion
\citep{Polychroni13}, and is related to analyses of the {\it
  cosmic-web} in cosmological large-scale structures \citep{Pogosyan2009}.
This algorithm detects elongated ridges using a local determination of the curvature of the dust
emission intensity. We compute the Hessian matrix of the
  unfiltered $D_{353}$ map (the estimate of the local curvature is
  independent of the background subtraction). For each pixel of this map, we estimate the first and second derivatives with respect to the local
Galactic coordinates ($l$, $b$) in order to build the corresponding Hessian
matrix, 
   \begin{equation}
     \label{eq:hessI} 
        H(x,y)\, \equiv \, \left ( \begin{array}{cc} H_{xx} & H_{xy }\\
            H_{yx} & H_{yy} \end{array} \right ),
    \end{equation}   
where {\it x} and {\it y} refer to the Galactic coordinates ({\it l}, {\it b}) as $x=b$ and $y=l\,\cos{b}$, so that
the {\it x}-axis is pointing towards the north Galactic pole. The second-order
partial derivatives are $H_{xx}=\partial^2 D_{353} / \partial x^2$,
$H_{xy}=\partial^2 D_{353} / \partial x \partial y$,
$H_{yx}=\partial^2 D_{353} / \partial y \partial x$,
$H_{yy}=\partial^2 D_{353} / \partial y^2$. 
The Hessian matrix would be nearly the same if we used $D^{\rm Dif}_{353}$
instead of $D_{353}$. Indeed, the difference between the two maps,
$D^{\rm BG}_{353}$, does not have significant structure at the scales over which
the derivatives are computed.  
 
By solving the characteristic equation of the Hessian matrix, we
find the two eigenvalues,
\begin{equation}
\label{eq:lambda}
\lambda_{\pm}=\frac{(H_{xx}+H_{yy}) \pm \sqrt{(H_{xx}-H_{yy})^2+4H_{xy}H_{yx}}}{2}.
\end{equation}
The two eigenvalues define the local
curvature of the intensity. The map of the minimum eigenvalue,
$\lambda_-$, shown in the upper panel of 
Fig.~\ref{fig:lambdaf}, highlights filamentary structures in
$D_{353}$. The Hessian matrix encodes the information about the local orientation of the ridges. The angle between the north direction and the eigenvector
corresponding to $\lambda_-$ is perpendicular to the orientation angle $\theta$ of the
crest of the ridge with respect to the north Galactic pole. This angle $\theta$ can be derived as
\begin{equation}
\label{eq:dirfil}
\theta=\frac{1}{2}{\rm tan}^{-1}{\frac{H_{xy}+H_{yx}}{H_{xx}-H_{yy}}}.
\end{equation}
The computation of the orientation angle $\theta$ and its uncertainty, 
over the whole sky,  is detailed in Appendix~\ref{appendix:Hessian}. 
The Appendix also presents an independent algorithm, based on a method used by
 \citet{Hennebelle13}  to analyse results from numerical simulations, 
 where the orientation of structures is computed from the inertia matrix of the dust $D_{353}$ map.
The two independent estimates of the orientation angle  are in good agreement.

\subsection{Selection procedure}
\label{subsec:mask}

The $\lambda_-$ curvature map highlights a complex bundle of filamentary
structures, where the most significant ridges in $D_{353}$ intersect underlying features
owing to noise and background emission. 
To select interstellar matter structures, we build a
mask based on three local criteria: intensity contrast with respect to the background map $D_{\rm 353}^{\rm BG}$, curvature 
and signal-to-noise of the polarization fraction. 
The Magellanic Clouds and the Galactic
plane within $\pm 5^{\circ}$ in latitude are masked to focus on
structures located in the solar neighbourhood.
Hereafter, the masked pixels
are the ones that we do not consider in the analysis. 
We also mask single-pixel regions produced by the selection criteria. 
Our final sample of ridges amounts to $4\,\%$ of the sky.

The details of the masking procedure
are discussed in Appendix~\ref{appendix:Hessian}. Here, we
give a description of the main points. 
The first criterion defines a structure as a contrasted ridge in $D_{353}$ with respect to the
local background. We introduce a threshold,  $\zeta$, on the brightness contrast:
${D^{\rm Dif}_{353}}/{D^{\rm BG}_{353}} > \zeta$.
We set $\zeta=1$ and checked that changing the value of $\zeta$
to 0.5 or 2 does not change our statistical results.

The second criterion eliminates the contribution of  
background emission to the curvature. 
We use a toy model of the sky to define a  threshold, $C_{\rm T}$, 
which depends on the brightness of the background (see Eq.~(\ref{eq:threshold}) in Appendix~\ref{appendix:Hessian}),
on the negative curvature: $\lambda_- < -C_{\rm T}$. This criterion has its main impact at high Galactic latitudes.

The third selection criterion ensures a sufficient
  accuracy in the polarization angle. The uncertainty in the polarization angle directly
depends on the uncertainty in polarized intensity $P$ \citep[see Eq.~(B.5)
in][]{planck2014-XIX}; however, $p/\sigma_{p}\approx
P/\sigma_{P}$ (where $p$ and $\sigma_{p}$ are the polarization
fraction and the corresponding error described in
\citet{planck2014-XIX}), when $D_{353}/\sigma_{D_{353}} \gg
  P/\sigma_{P}$, which is true for all the pixels that meet our first
  two criteria. Thus, we select pixels with
$p/\sigma_{p} > 3$, so that the uncertainties in the polarization
angle are smaller than $10^{\circ}$, with a median value of
$3^{\circ}$.

\begin{figure*}[]
\centerline{\includegraphics[width=18cm]{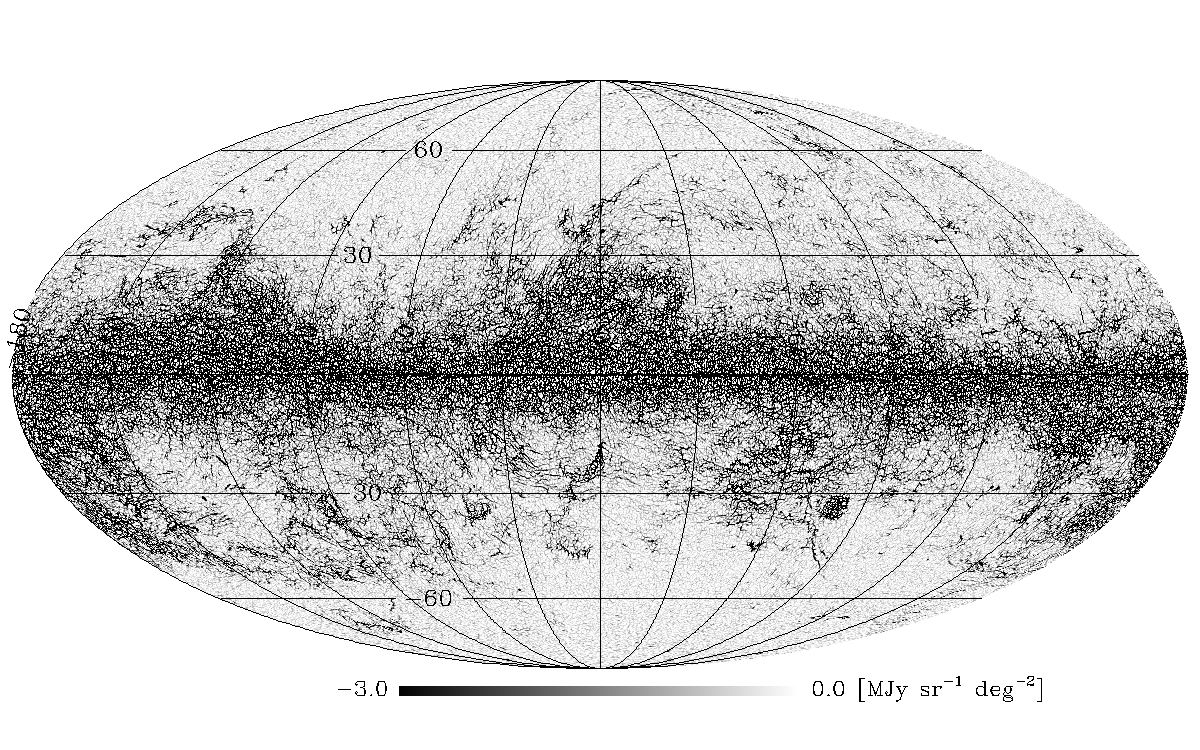}}
\centerline{\includegraphics[width=18cm]{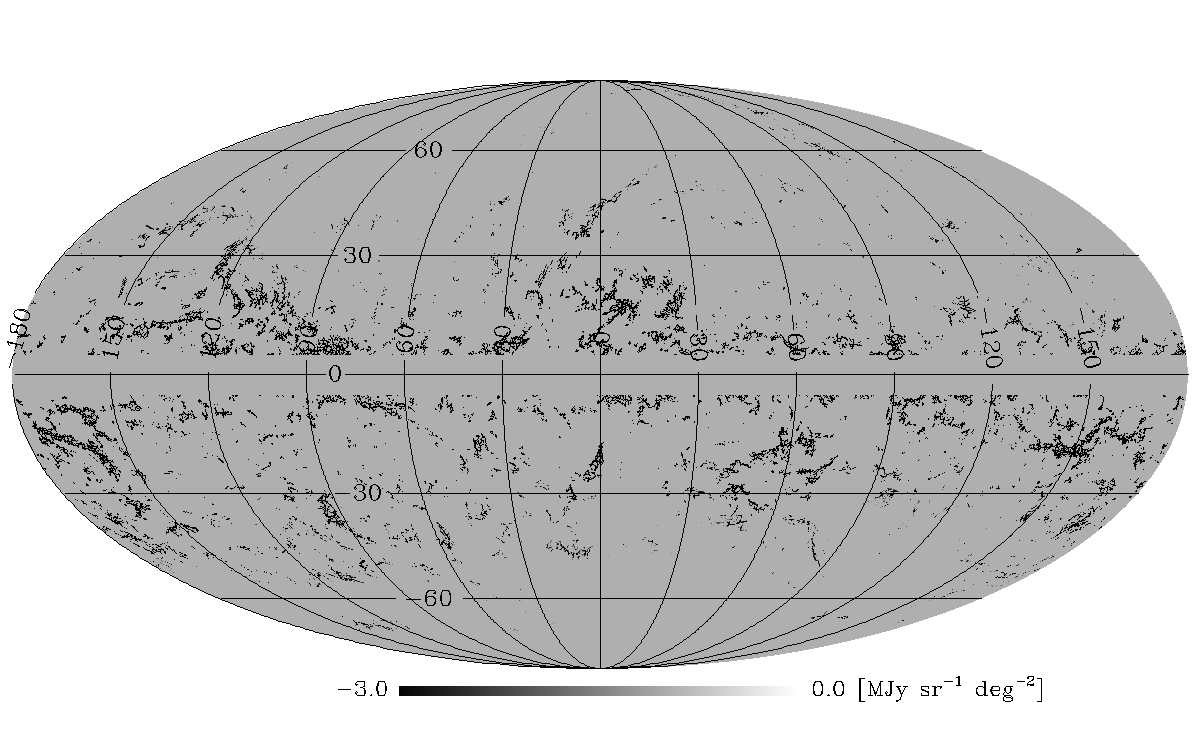}}
\caption{ {\it Top:}  
all-sky map of the negative curvature, $\lambda_{-}$, of $D_{353}$.  {\it Bottom:}
same as in the top panel where only the selected pixels are shown (see
Sect.~\ref{subsec:mask}).} 
\label{fig:lambdaf}
\end{figure*}

\begin{figure*}[h!]
  \centering
  \begin{tabular}{r l}
      \includegraphics[width=8.8cm,trim=30 0 15 0,clip=true]{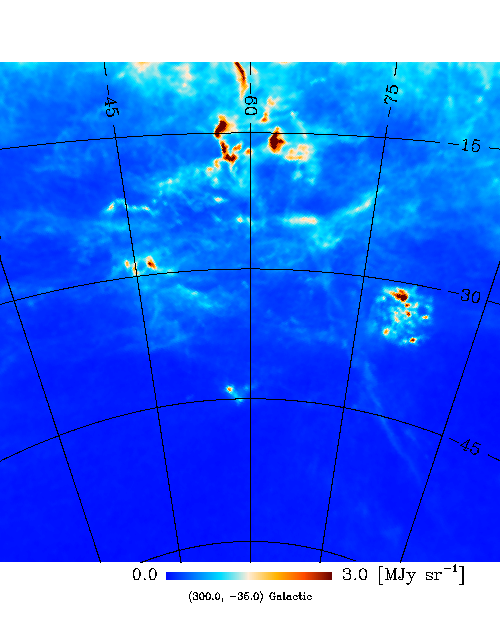} & \includegraphics[width=8.8cm,trim=30 0 15 0,clip=true]{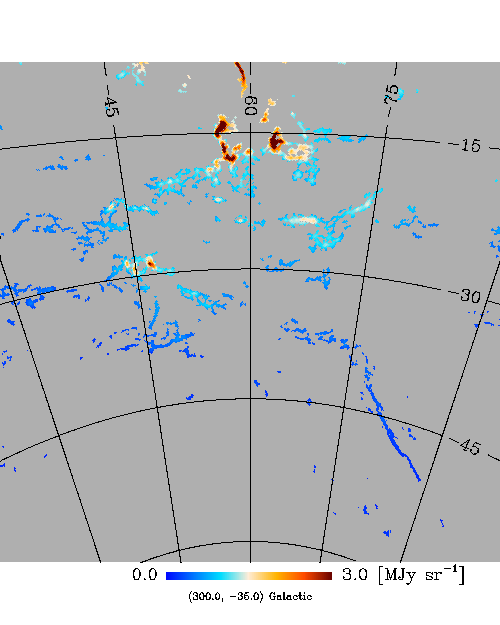}
    \end{tabular}
       \caption{{\it Left:} expanded view  of the Chamaeleon
  complex in $D_{353}$. The centre of the map is at
  $(l,b)=(300^{\circ},-35^{\circ})$. {\it Right:}
  same as for the left panel with the masked pixels in grey. 
  The Magellanic Clouds are masked. }
      \label{fig:mask_cut}
\end{figure*}

Two figures illustrate the selection procedure. The bottom panel in Fig.~\ref{fig:lambdaf}
presents the all-sky curvature map, where only the selected pixels are
shown. As can be seen, our procedure does not bias the selection of the structures towards
specific regions in the sky, but covers a wide range of Galactic latitudes. 
Figure~\ref{fig:mask_cut} illustrates an expanded view of the Chamaeleon
complex, highlighting the selected ridges in $D_{353}$.

\begin{figure*}[!h]
\centering
  \begin{tabular}{r l}
     \includegraphics[width=8.8cm,trim=50 0 22 0,clip=true]{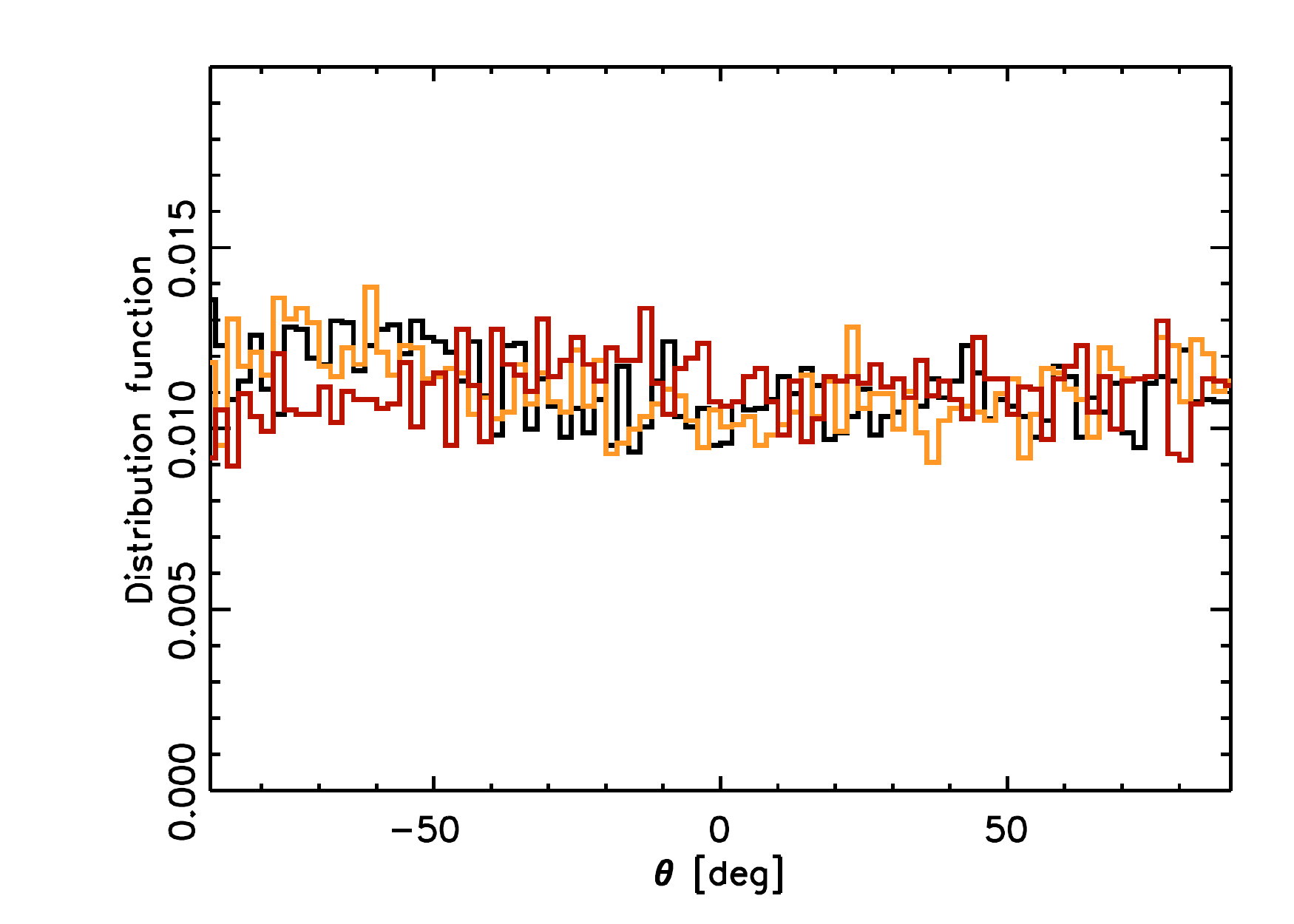}
&\includegraphics[width=8.8cm,trim=50 0 22 0,clip=true]{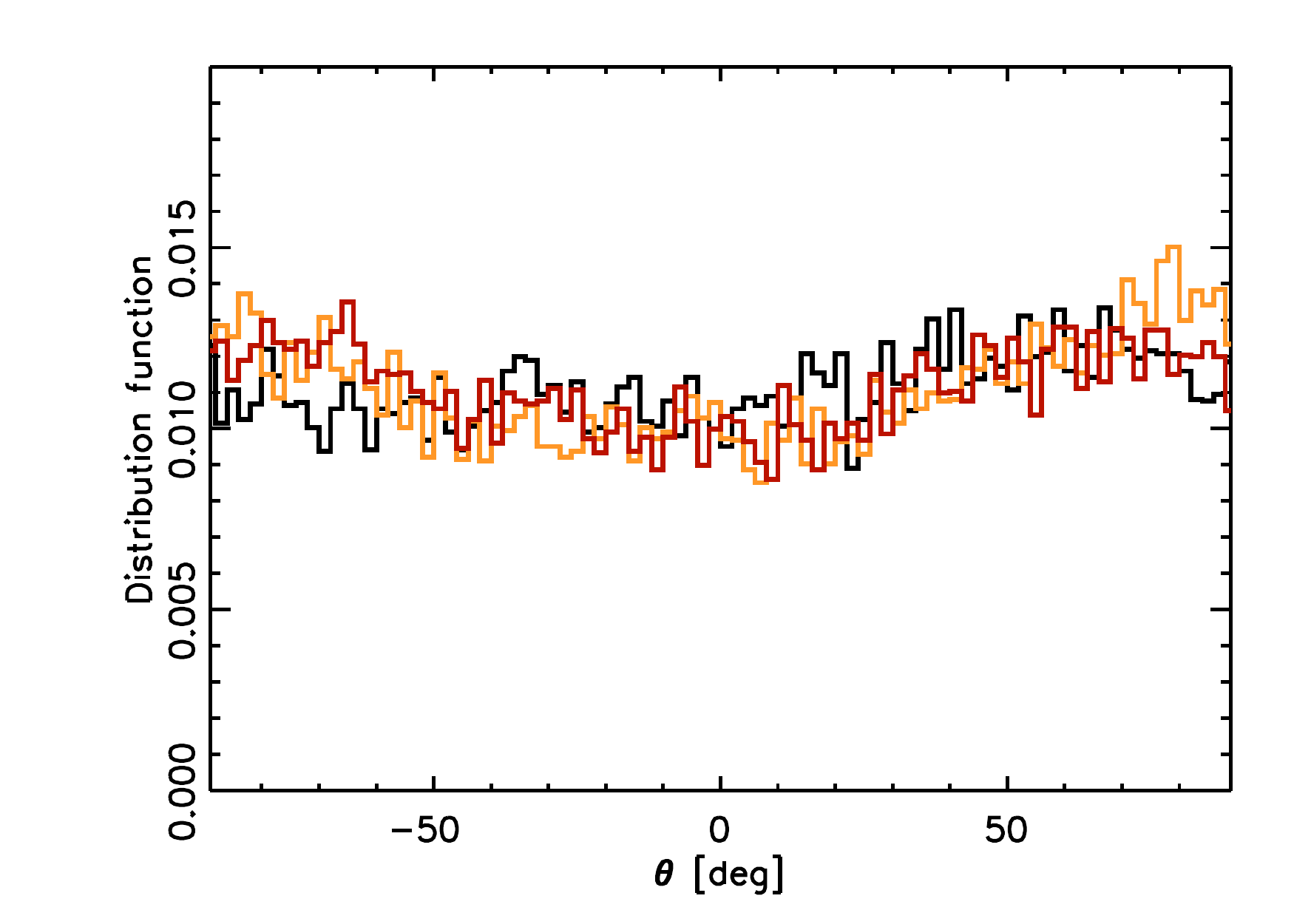}\\
    \end{tabular}
\caption{{\textit{Left:}} distribution functions of the orientation angle $\theta$ from Eq.~(\ref{eq:dirfil}) 
for the northern Galactic hemisphere  and for 3 mean values of the Galactic latitude:
  $b= \, 7^{\circ}$ (black curve), $15^{\circ}$
(orange curve) and $33^{\circ}$ (red curve). 
{\textit{Right:}} same as for the left panel but for the southern hemisphere. The mean values
of $b$ are $-8^{\circ}$ (black curve), $-20^{\circ}$
(orange curve) and $-39^{\circ}$ (red curve).}
\label{fig:thetahem}
\end{figure*}

Figure~\ref{fig:thetahem} shows the distribution function (DF) of the orientation angle $\theta$
from Eq.~(\ref{eq:dirfil}) for the selected structures as a function of Galactic latitude. 
The normalization of the DF is done by dividing the number of ridges in each  bin of $\theta$
by the total number of selected ridges within each latitude bin. 
This total number is the same for all latitude bins.  
The DF does not present any preferred orientation
of the structures in the northern hemisphere, but there is 
a slight dip at  $\theta = 0^\circ $ in the southern hemisphere, most noticeable for
the highest latitudes. 

\subsection{Characterization of ridges}
\label{subsec:column}

\begin{figure}[h!]
\centerline{\includegraphics[width=8.8cm,trim=50 0 22 0,clip=true]{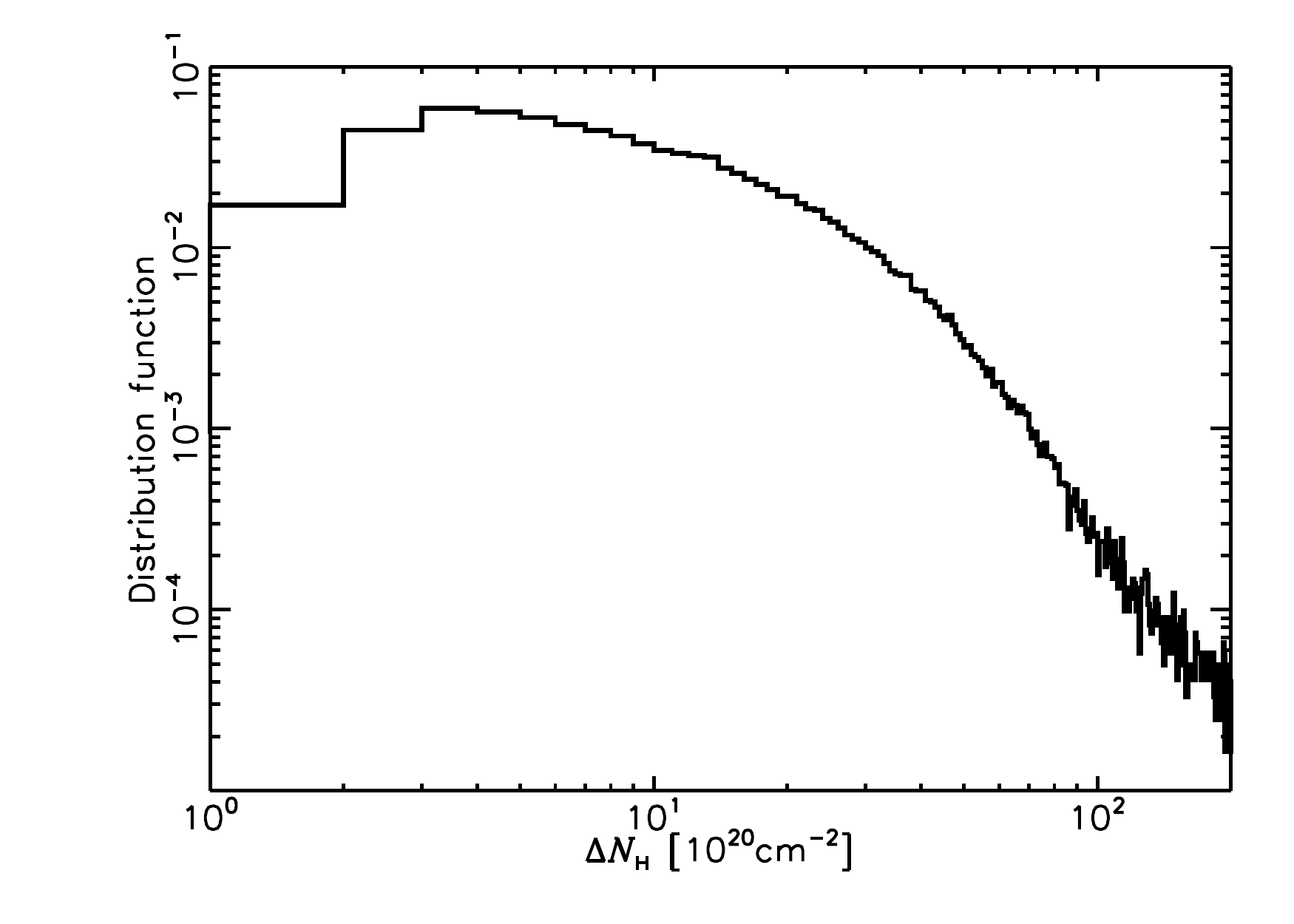}}
\caption{Distribution function of the excess column density,
  $\Delta N_{\rm H}$ in Eq.~(\ref{eq:DnH}), computed for the selected pixels.}
\label{fig:nH}
\end{figure}

For each selected pixel, we compute the excess column density
defined as 
\begin{equation}
\label{eq:DnH}
\Delta N_{\rm H}=N_{\rm H} \, \frac{D^{\rm Dif}_{353}}{D_{353}} = 8.7\times 10^{25} \, \tau_{353} \, \frac{D^{\rm Dif}_{353}}{D_{353}} \, {\rm cm}^{-2},
\end{equation}
where the opacity at $353\,$GHz, $ \tau_{353}$, is taken from \citet{planck2013-p06b}.
The conversion factor to the hydrogen column density, 
$N_{\rm H}$, is the value measured  from the comparison with \hi\ data
at high Galactic latitudes
\citep{planck2013-XVII,planck2013-p06b}. Going from high to
intermediate Galactic latitudes we ignore the decrease in the ratio
between $N_{\rm H}$ and $\tau_{353}$ by a factor
of $\sim2$ for increasing column densities, reported in \citet{planck2013-p06b}.
Figure~\ref{fig:nH} presents the DF of $\Delta N_{\rm H}$. The distribution 
covers two orders of magnitude, from $10^{20}$ to $10^{22}\, {\rm
  cm}^{-2}$, with a median value of $1.2\times 10^{21} {\rm cm}^{-2}$.

\begin{figure}[h!]
\centerline{\includegraphics[width=8.8cm,trim=50 0 22 0,clip=true]{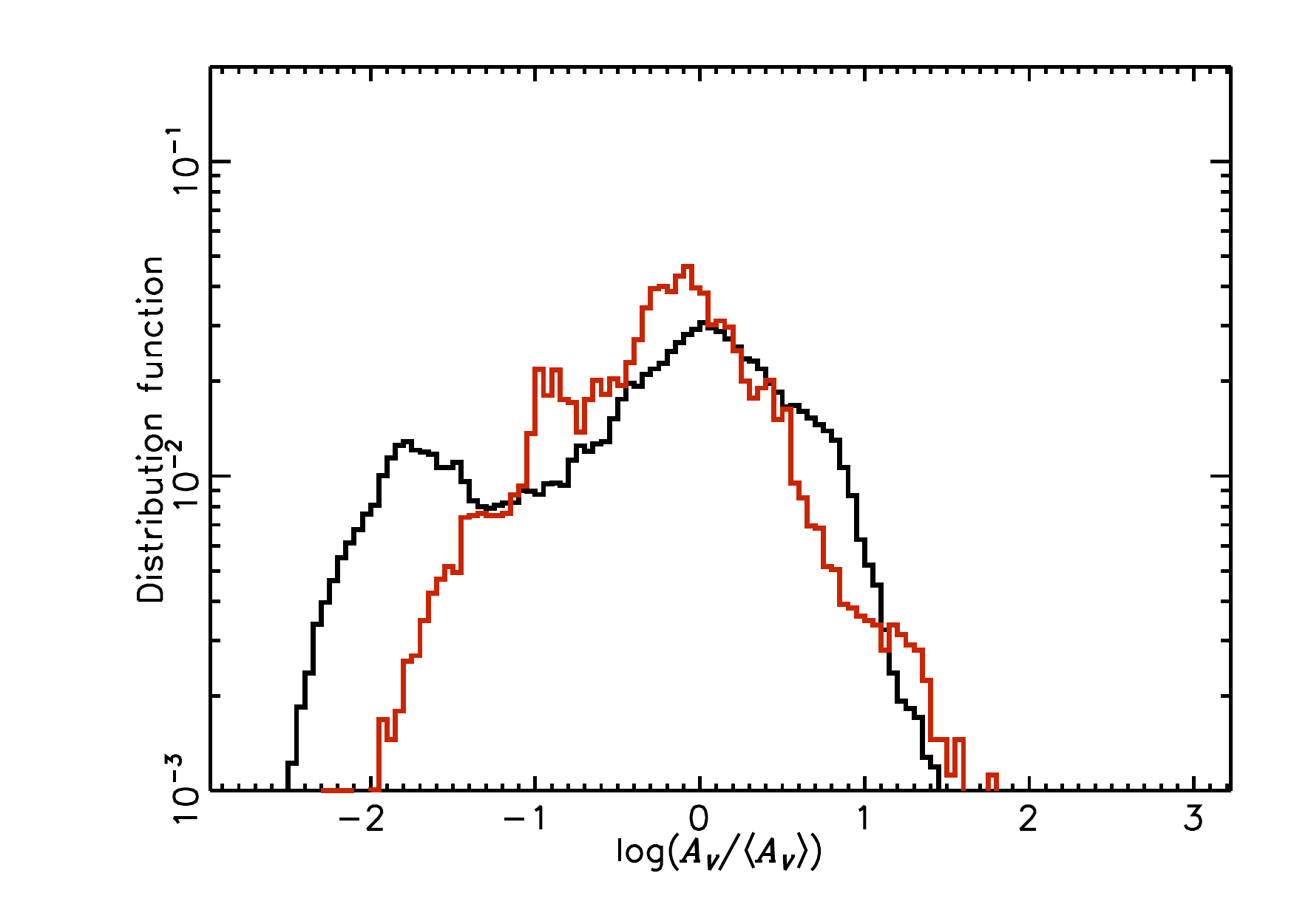}}
\caption{Distribution functions of the extinction within the
  Chamaeleon complex. Here, $\langle A_{V} \rangle$ is the mean value of
  extinction over the whole Chamaeleon field. The figure shows the
 comparison between the DF of the total field  (black line) and that
 relative to the selected pixels (red line). In both cases the
 Magellanic clouds are discarded.}
\label{fig:av}
\end{figure}

In Fig.~\ref{fig:av} we compare the DFs of the extinction $A_{V}$ derived  from $\tau_{353}$ for the ridges and all the pixels in the Chamaeleon complex.
We use the relation $A_{V}=R_{V}\, E(B-V) = 1.49\times10^4 \,
R_{V}\,  \tau_{353}$ from \citet{planck2013-p06b} and $R_{V}=3.1$ \citep{Jones2011,Schlafly2011,Mortsell2013}.
The figure shows that the selected pixels 
cover most of the range of $A_{V} $ measured over the entire
Chamaeleon complex, except the lowest values.  

Here, we explain how we estimate the mean gas density of the ridges using the
curvature map. On the crest of a ridge, the first derivatives of the sky
brightness with respect to Galactic coordinates are zero. The second
derivative in the direction perpendicular to the ridge is
$\lambda_{-}$. Along this direction, the local variation of the brightness over an angular
distance $\epsilon$ may be approximated by a second-order Taylor
expansion as
\begin{equation}
\label{eq:taylor}  
\delta D_{353}  \approx 0.5 | \lambda_{-} |  \epsilon ^2.
\end{equation}
The Hessian algorithm tends to select ridges with a
  thickness close to the $20\arcmin$ angular distance over which the
  derivatives of $D_{353}$ are computed (Appendix~\ref{appendix:Hessian}). Since this angular distance is
  barely larger than the angular resolution of the map ($15\arcmin$),
  the width of some of the structures that we analyse is not resolved in the
  smoothed $D_{353}$ map. For all the selected structures we compute $\delta D_{353}$ for $\epsilon=20\arcmin$.
 The physical thickness in parsecs corresponding to $\epsilon$ is $\delta \approx \epsilon\,d$, where $d$ is the distance to the
  ridges. We estimate the distance from the scale height of the \hi\ emission of the cold
neutral medium (CNM) in the solar neighbourhood, $h$, as $d={h}/|\sin{b}\,|$. 
For $h=100\,$pc \citep[see Fig.~14 in][]{Kalberla07}, we find a mean
distance over the latitudes of the selected pixels of
$\bar{d}= 430 \, {\rm pc}$, and a mean thickness of $\bar{\delta}=2.5\, {\rm pc}$.

To estimate the mean gas density we first convert the brightness variation $\delta
D_{353}$ into column density variation 
\begin{equation}
\label{eq:delta_NH}  
\delta N_{\rm H}= \frac{\delta
D_{353}}{D_{353}}N_{\rm H}. 
\end{equation}

Provided that the extent of the ridges along the line of sight is, on
average, comparable to their thickness in the sky ($\delta$),
the mean density may be expressed as
\begin{equation}
\label{eq:n_est}
  \langle n_{\rm H} \rangle =\left \langle \frac{\delta N_{\rm H}}{\delta}  \right \rangle,
\end{equation}
where the mean, $\langle\, ...\, \rangle\,$, is computed over the
selected pixels.
We find $\langle n_{\rm H} \rangle = 300 \, {\rm cm}^{-3}$, 
a value within the range of CNM gas densities.
This conclusion is true even if the extent of the ridges along the 
line of sight is larger than their thickness in the sky. We stress
that the value of $\langle n_{\rm H} \rangle$ only provides a
rough estimate of the mean volume density, which we use to show that we are selecting CNM structures. 

We have visually compared our map of ridges with the southern sky of the
GASS survey \citep{Kalberla2010} and we notice that many of the selected
structures are also seen in \hi\, but their column densities, and our density estimate, are high
enough that a significant fraction of the gas must be molecular
\citep{planck2011-7.0,planck2011-7.12,Wolfire10}, even if many of the selected ridges do not have a counterpart
in CO maps \citep{dame2001,planck2013-p03a}.

\section{The dispersion of magnetic field orientations}
\label{sec:magfield}

The orientation of the magnetic field in interstellar clouds has often been inferred
from the polarization of starlight occurring in the environment around
the clouds, e.g. \citet{Li2013} for molecular clouds and \citet{Clark13} for the diffuse ISM. 
The {\Planck} maps allow us to compare the polarization angles
on the filamentary ridges with those measured on the nearby
background. We present the DF of the difference of polarization angles in
Sect.~\ref{subsec:disp_polangles}, and its modelling in Sect.~\ref{subsec:disp_models}
to estimate the ratio between the random and mean components of
the magnetic field.


\subsection{Difference between local and background polarization angles}
\label{subsec:disp_polangles}

We compute the polarization angles, for the ridges and the background, making use of $Q^{\rm Dif}_{353}$,
$U^{\rm Dif}_{353}$ and $Q^{\rm BG}_{353}$, $U^{\rm BG}_{353}$,
respectively. Inverting Eq.~(\ref{eq:Stokes_parameters}), we obtain 
\begin{align}
\label{eq:polangle}
& \psi^{\rm Dif}_{353} = \frac{1}{2}{\rm tan}^{-1}(U^{\rm Dif}_{353},Q^{\rm Dif}_{353}),\nonumber \\
& \psi^{\rm BG}_{353} = \frac{1}{2}{\rm tan}^{-1}(U^{\rm BG}_{353},Q^{\rm BG}_{353}).
\end{align}
The difference between the two polarization angles in
Eq.~(\ref{eq:polangle}), accounting for the $180^{\circ}$ degeneracy that
characterizes both, can be expressed as
\begin{equation}
 \label{eq:diffang_st_bg} 
  \delta\psi  = \frac{1}{2}{\rm tan}^{-1}\left ( {\frac{\sin{2\alpha}\cos{2\beta}-\cos{2\alpha}\sin{2\beta}}{\cos{2\alpha}\cos{2\beta}+\sin{2\alpha}\sin{2\beta}}} \right ), 
\end{equation} 
where $\alpha = \psi^{\rm Dif}_{353}$, $\beta =
\psi^{\rm BG}_{353}$. The values of $\delta\psi$ are computed from $-180^{\circ}$ to $180^{\circ}$ 
matching both the sine and the cosine values.

The DF of $\delta \psi$ is presented in Fig.~\ref{fig:diff_st_bg}.
On the plane of the sky, the magnetic field orientation
is perpendicular to the polarization angle. Thus, the DF of $\delta \psi$
characterizes the difference between the magnetic field orientations determined at two
different scales: that of the ridges, at $15\,\arcmin$ ($2\,{\rm pc}$
at the mean distance of $430\,$pc), and that of
the local background, at $5^\circ$ ($40\,{\rm pc}$).  
The DF of $\delta \psi$ has a mean value of $0^\circ$ and a standard deviation of 
$40^{\circ}$, much larger than what we expect from data noise (Sect.~\ref{subsec:mask}). 
Thus, we conclude that the magnetic field on
the ridges is statistically aligned with the background field, but with a significant scatter. 

We check that the DF of $\delta \psi$ does not depend on the method we used to compute the 
background maps. 
For our selection of pixels, the polarization angle,
$\psi^{\rm Dif}_{353}$, is close to $\psi_{353}$, the polarization angle at $353\,$GHz
without background subtraction, computed  with $Q_{353}$ and $U_{353}$ in Eq.~(\ref{eq:polangle}). 
The DF of the difference between these two polarization angles, $\delta
\psi_{\rm str}$, computed with Eq.~(\ref{eq:diffang_st_bg}) where
$\alpha=\psi^{\rm Dif}_{353}$ and $\beta=\psi_{353}$, is shown
in Fig.~\ref{fig:Deltapsi1}. This distribution has a standard deviation of
$15^{\circ}$, which is smaller than that measured for 
$\delta \psi$.  This result follows from the fact that, on the selected ridges, the local structures have statistically a dominant contribution to the
values of the Stokes $Q_{353}$ and $U_{353}$ parameters.
We also compute the same DF varying the percentage of pixels used to estimate the background maps 
(Sect.~\ref{sec:visu}) from $20\,\%$ to 10\,\% and $40\,\%$. The three DFs of $\delta
\psi_{\rm str}$ are compared in Fig.~\ref{fig:Deltapsi1}. They are identical and much narrower than the DF of 
$\delta \psi$ in Fig.~\ref{fig:diff_st_bg}.

\subsection{Modelling of the distribution function}
\label{subsec:disp_models}

We present a model, detailed in Appendix~\ref{appendix:gaussian}, which 
takes into account the projection onto the plane of the sky and 
relates the width of the DF of $\delta \psi$  to the ratio between the random and mean components of
the magnetic field. 

\begin{figure}[h!]
\centerline{\includegraphics[width=8.8cm,trim=50 0 22 0,clip=true]{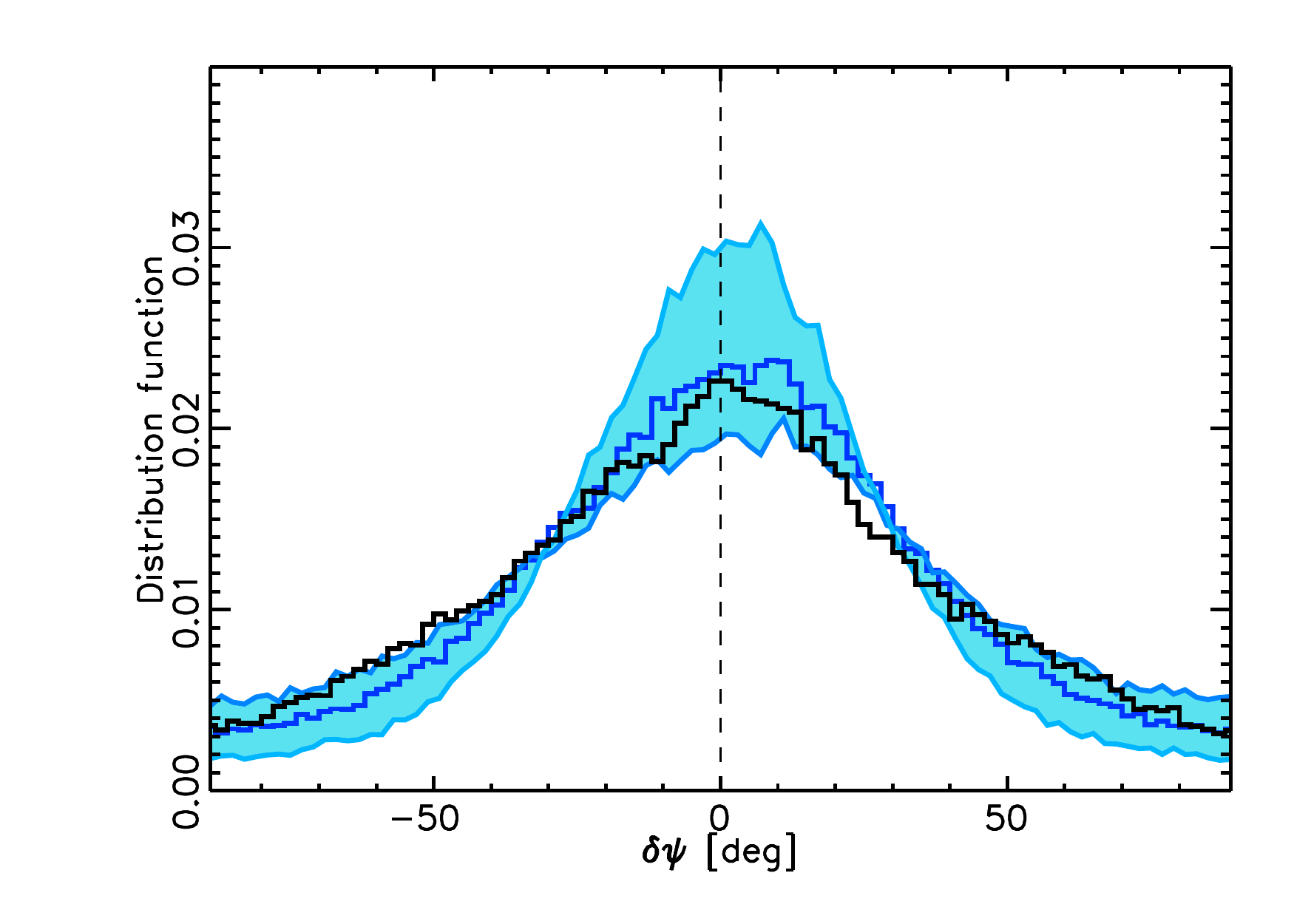}}
\caption{Distribution function of $\delta\psi $, the difference between the polarization angle with
  background subtraction $\psi^{\rm Dif}_{353}$ and that of the background
  $\psi^{\rm BG}_{353}$, in black. 
 The data are compared with Gaussian models computed for  
$f_{\rm M} \in [0.6, 1.0] $  in the light-blue band. The blue solid line represents
the Gaussian model for $f_{\rm M} = 0.8$. The model parameter $f_{\rm M}$ 
measures the ratio between the strengths of the turbulent and mean components of the magnetic field.}
\label{fig:diff_st_bg}
\end{figure}


The model is built from 3D vectors
$\vec{V}_{\rm M}$ (hereafter, the subscript M refers to
  the model) with a Gaussian distribution of orientations about a mean reference vector,
$\vec{V}_{\rm M0}$. Each component of $\vec{V}_{\rm M}$  is an independent realization of a Gaussian field on the sphere, with an angular power spectrum
equal to a power law of index $\alpha_{\rm M}=-1.5$, to which we add the corresponding
component of $\vec{V}_{\rm M0}$.  By construction, the mean of $\vec{V}_{\rm M}$ is $\vec{V}_{\rm
  M0}$. The spectral index of the power
spectrum allows us to introduce fluctuations about
the mean direction correlated across the sky. 
This stochastic description of the field follows the early models proposed by \citet{Jokipii69}.
The degree of alignment between $\vec{V}_{\rm M}$  and $\vec{V}_{\rm M0}$  is parametrized by $f_{\rm M} $,  the 
standard deviation of the modulus of the random component of $\vec{V}_{\rm M}$ normalized
by $|\vec{V}_{\rm M0}|$. 
The  DF of the angles between $\vec{V}_{\rm M}$  and $\vec{V}_{\rm M0}$ 
in 3D, per unit solid angle, is close to Gaussian with a standard deviation,
$\sigma_{\rm M}$,  which increases with $f_{\rm  M} $. The models quantify 
statistically the  projection of the 3D direction of the magnetic field onto the 2D celestial sphere. 
They do not include any averaging due to the superposition of uncorrelated structures along the line of sight, as done by
\citet{Jones89} and \citet{Myers91} for the modelling of polarization data towards the Galactic plane and molecular clouds.

For each model, we compute  the 
projections of $\vec{V}_{\rm M}$  and $\vec{V}_{\rm
  M0}$ on the sky and the angle maps $\psi_{\rm M}$ and $\psi_{\rm M0}$ 
 with respect to the local direction of the north
Galactic pole. We use Eq.~(\ref{eq:diffang_st_bg}) with 
$\alpha=\psi_{\rm M}$ and $\beta=\psi_{\rm M0}$ to compute
the difference, $\delta \psi_{\rm M}$, between these two angle maps.
In Fig.~\ref{fig:diff_st_bg}, we show the DFs of $\delta \psi_{\rm M} $ for $f_{\rm M}
\in [0.6, 1.0]$ as well as the model with $f_{\rm M} = 0.8$ that best matches the data.
This value agrees with that inferred from the modelling of near-IR stellar polarization in the Galactic plane and molecular clouds by
\citet{Jones89,Jones1992}, and from synchrotron observations \citep{Beck2007,Houde2013,Haverkorn2015}.  
The comparison between the models and the data provides an estimate of the  
ratio between the amplitudes of the random (turbulent) and mean
components of the magnetic field. The analogy with the data is such that $\psi^{\rm Dif}_{353}$
corresponds to the turbulent component of the field at $20\arcmin$ scale, and
$\psi^{\rm BG}_{353}$ to the mean component  at a few degrees scale, on the sky.

This method is similar to the one proposed by
  \citet{Hildebrand09} to measure the local difference of polarization
  angles in molecular clouds in order to separate the mean and turbulent 
components of the magnetic field. However, there
  are two main differences with what is described by this earlier work. First, we do not compute, nor
  fit, the dependence of the variance of the angle difference on the
  angular distance. Second, by measuring the dispersion of
  polarization angles over the whole sky, and by comparing the data
  with the Gaussian models, we obtain a 3D estimate of the ratio
  between the turbulent and mean components of the field, corrected
  for the projection of the magnetic field on the plane of the sky. 

Equipartition between kinetic energy from turbulence and magnetic energy
is found in the diffuse ISM from Zeeman \hi\ observations, which
implies that turbulence in the CNM is trans-Alfv\'enic \citep[][]{Myers95,Heiles05}. 

Thus, our result fits with the Chandrasekhar and Fermi description of turbulence in the diffuse and
magnetized ISM in terms of Alfv\'en
waves\citep{Chandra1953,Ostriker01,Hildebrand09}. This
framework assumes energy equipartition between the kinetic
energy of the gas and the energy of the random component of the magnetic field. Our
analysis suggests that this non-trivial assumption applies in the diffuse ISM. 


\begin{figure}[h!]
\centerline{\includegraphics[width=8.8cm,trim=50 0 22 0,clip=true]{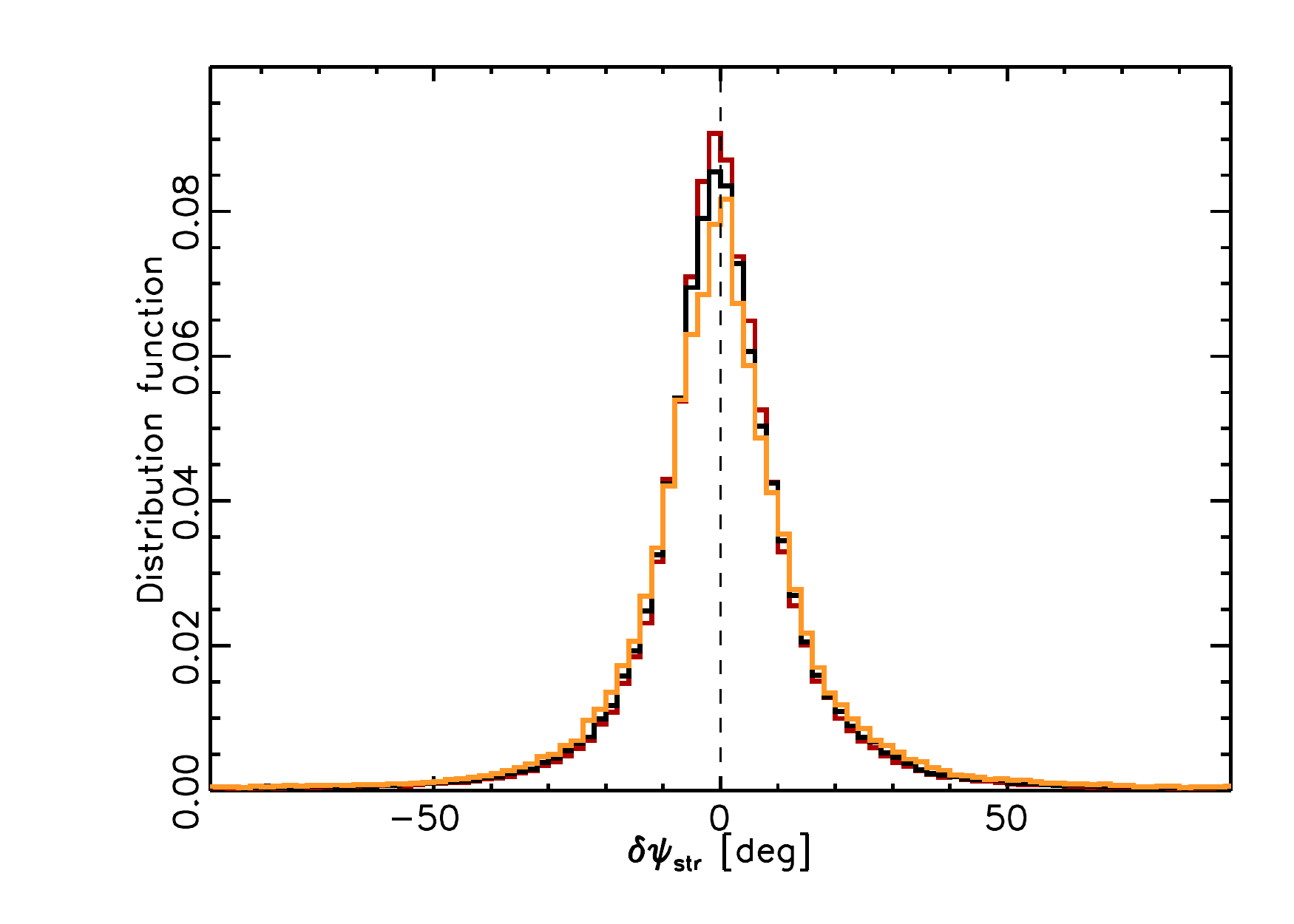}}
\caption{Distribution function of $\delta\psi_{\rm str}$, the difference between the polarization
 angles with and without background subtraction, $\psi_{353}^{\rm Dif}
 $ and $\psi_{353}$ respectively, computed over the selected pixels. The three curves show the DFs when we compute the
  local background among the 10 (red), 20 (black), and $40\,\%$ (orange)
  lowest values of $D_{353}$ (see Sect.~\ref{sec:visu}).}
\label{fig:Deltapsi1}
\end{figure}

\section{Alignment of the magnetic field
  and the matter structures in the diffuse ISM}
\label{sec:alignment}

We quantify the relative orientation between the magnetic field and ridges in the dust emission map.  
We present and discuss the statistical results from our data analysis 
in light of the Gaussian model in Appendix~\ref{appendix:gaussian}, which takes into account projection effects.
The global statistics presented in this section refer to the diffuse ISM because only a small fraction of the selected structures are within molecular clouds.

\subsection{The alignment between the magnetic field and the matter structures}
\label{subsec:alignment}

\begin{figure}[h!]
\centerline{\includegraphics[width=8.8cm,trim=40 0 20 0,clip=true]{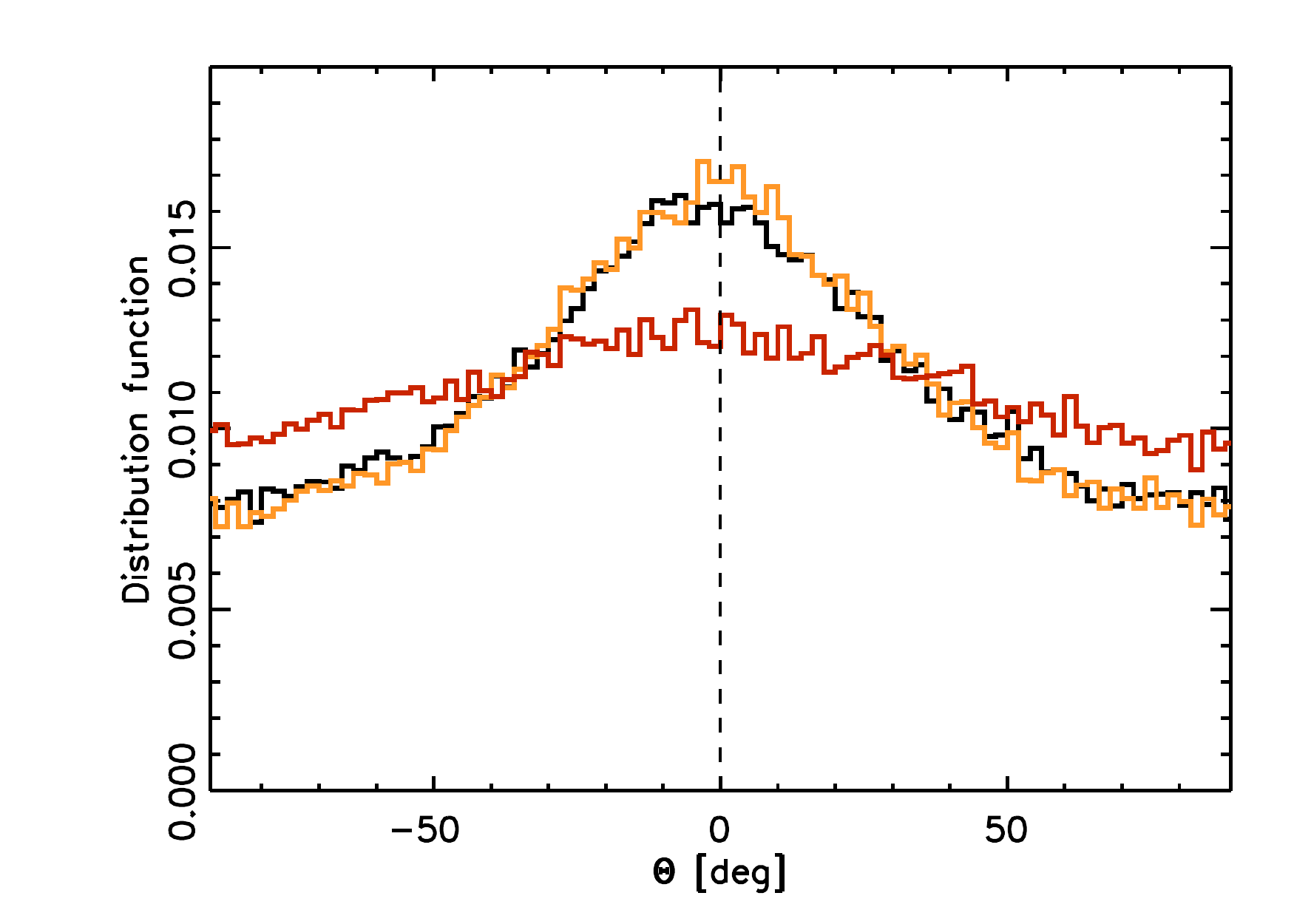}}
\caption{Distribution function of $\Theta$, the difference between the inferred orientation angle of the magnetic field 
and that of the ridges, for the selected pixels. The orange line and the black line represent the {\it Planck} data 
with ($\psi^{\rm Dif}_{353}$) and without ($\psi_{353}$) background subtraction, respectively.
The red line refers to the polarization angle of the subtracted
background ($\psi^{\rm BG}_{353}$). The structures of matter appear as
statistically aligned with the orientation of the magnetic field
projected on the plane of the sky.}
\label{fig:diff_allsky_fil_bg}
\end{figure}

In order to calculate the relative orientation between the magnetic field
and the ridges, we make a pixel-by-pixel comparison of the polarization angle $\psi $ 
and the orientation angle $\theta$ of the ridges. We compute the difference, $\Theta$,  between the  orientation of the ridge 
and that of the magnetic field inferred from the polarization angle, using
Eq.~(\ref{eq:diffang_st_bg}) with $\alpha= \psi -90^\circ$ and $\beta =
\theta$.
In Fig.~\ref{fig:diff_allsky_fil_bg}, we show the DFs of
$\Theta$ for the selected pixels, computed with $\psi = \psi^{\rm
  Dif}_{353}$, $\psi_{353}$, and $\psi^{\rm BG}_{353}$.

A preferred alignment is observed for the two DFs computed with $\psi^{\rm Dif}_{353}$ and 
$\psi_{353}$, while that computed with $\psi^{\rm BG}_{353}$ is much broader.
The comparison of the DFs in Fig.~\ref{fig:diff_allsky_fil_bg} leads to two
main conclusions. First, the similarity between the
DFs computed with and without background subtraction tells us 
that the background subtraction is not a critical aspect of our data
analysis. This
follows from the fact that, for the selected pixels, the polarized signal is dominated by the contribution of
the ridges. 
Second, the fact that the DF obtained when comparing  $\theta$ with $\psi^{\rm BG}_{353}$ is almost flat 
indicates that the matter structures are preferentially aligned with the local magnetic field, rather than with the background field.

\subsection{Correlation between alignment and polarization fraction}

In spite of the predominant alignment of the interstellar matter structures with
the magnetic field, all DFs in
Fig.~\ref{fig:diff_allsky_fil_bg} show a
broad dispersion, with a significant probability up to $90^\circ$  from the central peak.
The widths of the DFs are much larger than those  computed for
the uncertainty in the polarization angle, and in the direction of the ridges in Appendix~\ref{appendix:Hessian}.

The DFs of $\Theta $ combine the intrinsic scatter in the relative orientations between 
the matter structures and the magnetic field in 3D with the projection onto the plane of the sky. 
Thus, we expect the shape of the DF to depend on the orientation of the magnetic field with respect to the line of sight. 
Where the magnetic field orientation is close to the line of sight, the polarization 
angle on the plane of the sky does not strongly constrain the orientation of the field and, 
thereby, its relative orientation with the ridges in the dust map.

\begin{figure}[h!]
\centerline{\includegraphics[width=8.8cm,trim=40 0 20 0,clip=true]{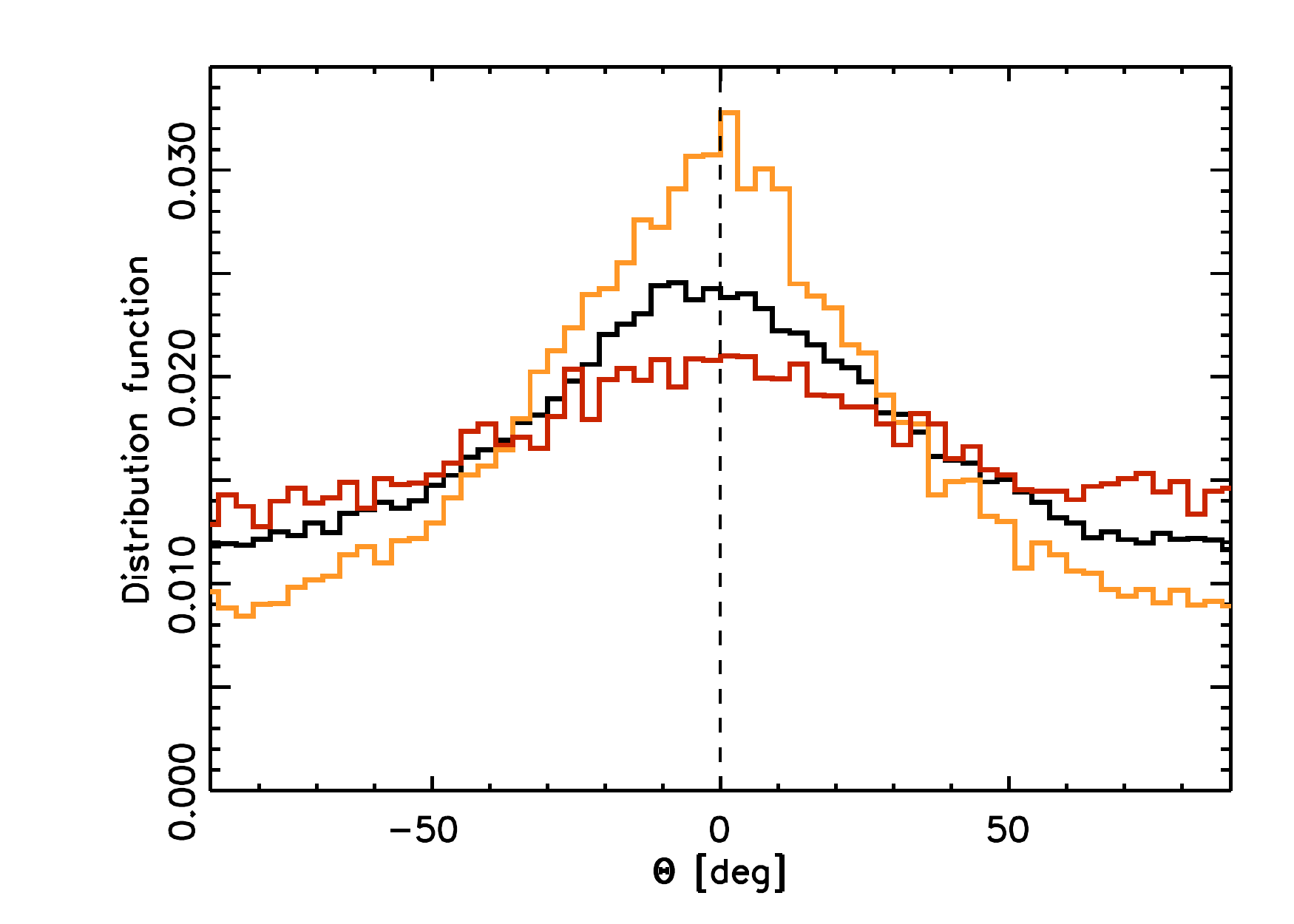}}
\caption{Distribution function of $\Theta $ computed with $\psi^{\rm Dif}_{353}$ as a function of
 the polarization fraction $p$ of the selected structures. The black line
  represents all the selected pixels. The orange line refers to the
  pixels with the $30\,\%$ highest values of $p$ and the red line to the
  $30\,\%$ lowest. The distribution of relative orientations is the sharpest
  for the highest values of $p$.}
\label{fig:diff_allsky_p}
\end{figure}

The analysis of the polarization maps, built from the 3D MHD simulation in
\citet{planck2014-XX}, shows that the polarization fraction $p$ in Eq.~(\ref{eq:observed_pol}) traces
$\cos^2{\gamma}$ averaged over
the line of sight (see their Fig.~21). Although $p$ also depends
on changes of the magnetic field orientation along the line of sight
\citep{planck2014-XIX,planck2014-XX}, depolarization along the line of sight
and within the beam does not preclude the use of $p$ to
statistically test the impact of projection effects on the DF of $\Theta$.
We do find that the relative orientation between the matter structures
and the magnetic field depends on  $p$. In Fig.~\ref{fig:diff_allsky_p},
we compare the DFs computed with $\psi^{\rm Dif}_{353}$
for all the selected ridges, for those with the $30\,\%$ highest values of
$p$ and for the lowest $30\,\%$. 
The higher the polarization fraction, the sharper the peak at $0^{\circ}$ of the DF of $\Theta $.

\begin{figure}[h!]
\centerline{\includegraphics[width=8.8cm,trim=40 0 20 0,clip=true]{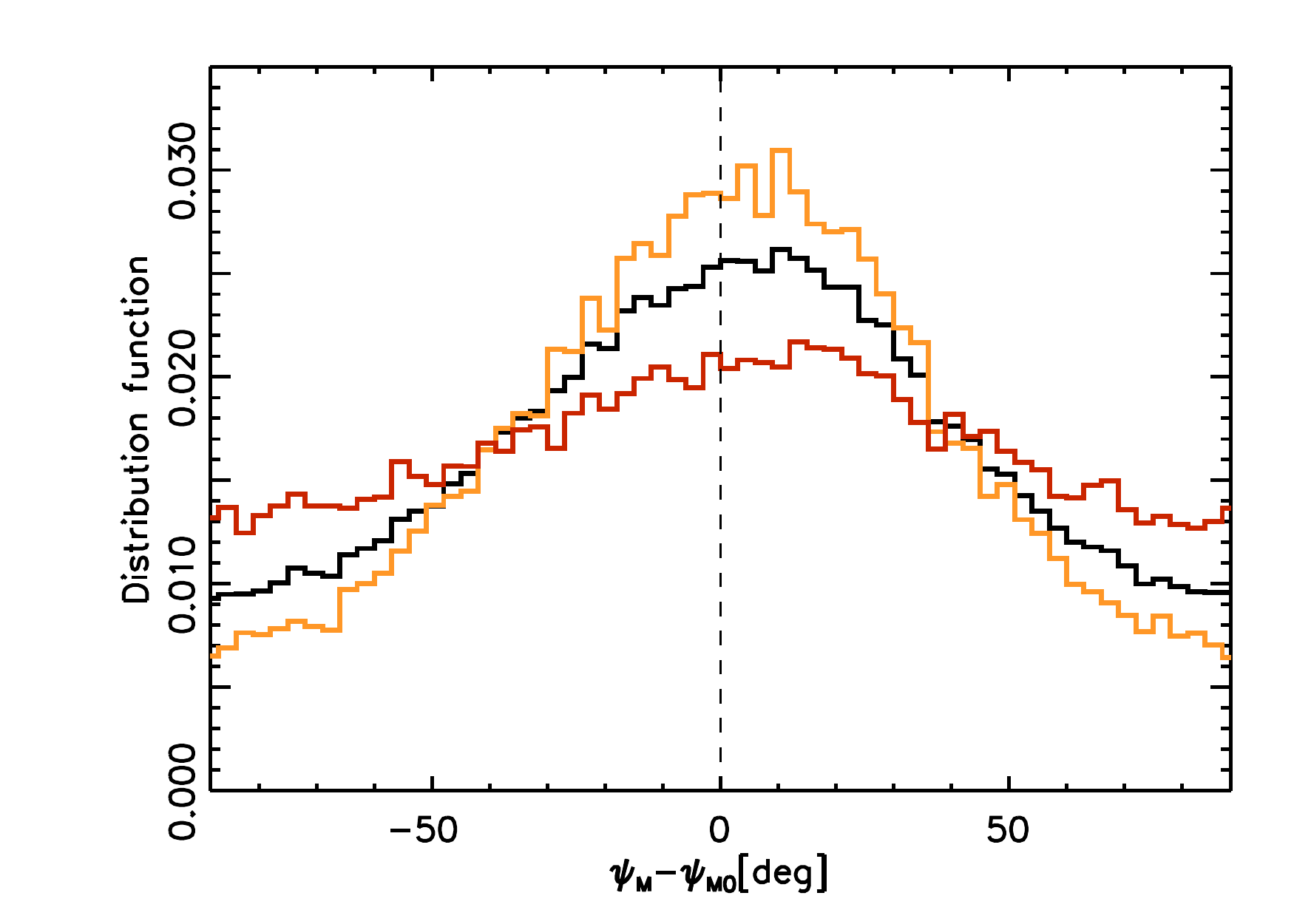}}
\caption{Distribution functions of relative orientations between $\vec{V}_{\rm M}$ and $\vec{V}_{\rm M0}$ for the Gaussian model described in the text and
  Appendix~\ref{appendix:gaussian} for $\sigma_{\rm M} = 33^{\circ}$. 
For the model, the polarization fraction is the projection factor $\cos^2{\gamma_{\rm M}}$.
We use for the models the same mask as for the data.
The black line represents all the selected pixels. The orange (red) line refers to the
 structures with the $30\,\%$ highest (lowest) values of
 $\cos^2{\gamma_{\rm M}}$. This figure shows that projection effects may
 reproduce the same trends found for the data.}
\label{fig:diff_allsky_ps}
\end{figure}

To quantify the projection effects on the relative orientation, we make use of the model
presented in Sect.~\ref{sec:magfield} and 
Appendix~\ref{appendix:gaussian}.
Here, we make the assumption that the
magnetic field can be decomposed into a component aligned with the orientation of the
ridges and a random component with zero mean.
 In the model, the orientation of the ridges is fixed to the constant vector $\vec{V}_{\rm M0}$.
The degree of alignment between the field and the ridges 
is parametrized by $\sigma_{\rm M}$ the standard deviation of the DF per unit solid angle 
of the angle between  $\vec{V}_{\rm M}$ and $\vec{V}_{\rm M0}$.  
As an example, Fig.~\ref{fig:diff_allsky_ps} shows the DFs of the angle differences between the projection onto the plane of the sky
of $\vec{V}_{\rm M}$ ($\psi_{\rm M}$) and $\vec{V}_{\rm M0}$ ($\psi_{\rm M0}$) for
$\sigma_{\rm M} = 33^{\circ}$.
The plot compares the relative orientations for all
the selected pixels and for those with the
$30\,\%$ highest and lowest values of $\cos^2{\gamma_{\rm M}}$, hereafter the projection factor, 
where $\gamma_{\rm M}$ is the angle between  $\vec{V}_{\rm M}$ and the plane of the sky. 
These three DFs compare well with those of $\Theta $ in 
Fig.~\ref{fig:diff_allsky_p}. The Gaussian model, which takes into account projection effects, 
reproduces the main characteristics of the DFs computed on the data, including the
dependence on the polarization fraction. 
The data and the model are further compared in the next section. 

\subsection{The $\xi$ parameter: the degree of alignment}

To quantify the variation of the DF of $\Theta $, $H(\Theta)$,
with the polarization fraction, we introduce a normalized version of the parameter 
used by \citet{Soler13} to study the relative orientation between magnetic fields and
density structures in MHD simulations.
We compute an estimator of the probability of having $\Theta $ near
$0^\circ$ as
\begin{equation}
\label{Ainn}
 A_{\rm in} = \int_{-20^{\circ}}^{20^{\circ}} H(\Theta) \mathrm{d}\Theta, 
\end{equation}
where the subscript `in' stands for inner range, and near $\pm 90^\circ$ as 
\begin{equation}
\label{Aout}
A_{\rm out} = \int_{-90^{\circ}}^{-70^{\circ}} H(\Theta) \mathrm{d}\Theta +
\int_{70^{\circ}}^{90^{\circ}} H(\Theta) \mathrm{d}\Theta,
\end{equation}
where the subscript `out' stands for outer range.
We define the degree of alignment $\xi$ as  
\begin{equation}
\label{eq:xi}
\xi=\frac{A_{\rm in}-A_{\rm out}}{A_{\rm in}+A_{\rm out}}.
\end{equation}
The $\xi$ parameter spans values between $-1$ and $1$, depending on whether the DF
peaks in the outer or inner range of angles, respectively.

\begin{figure}[h!]
\centerline{\includegraphics[width=8.8cm,trim=50 0 15 0,clip=true]{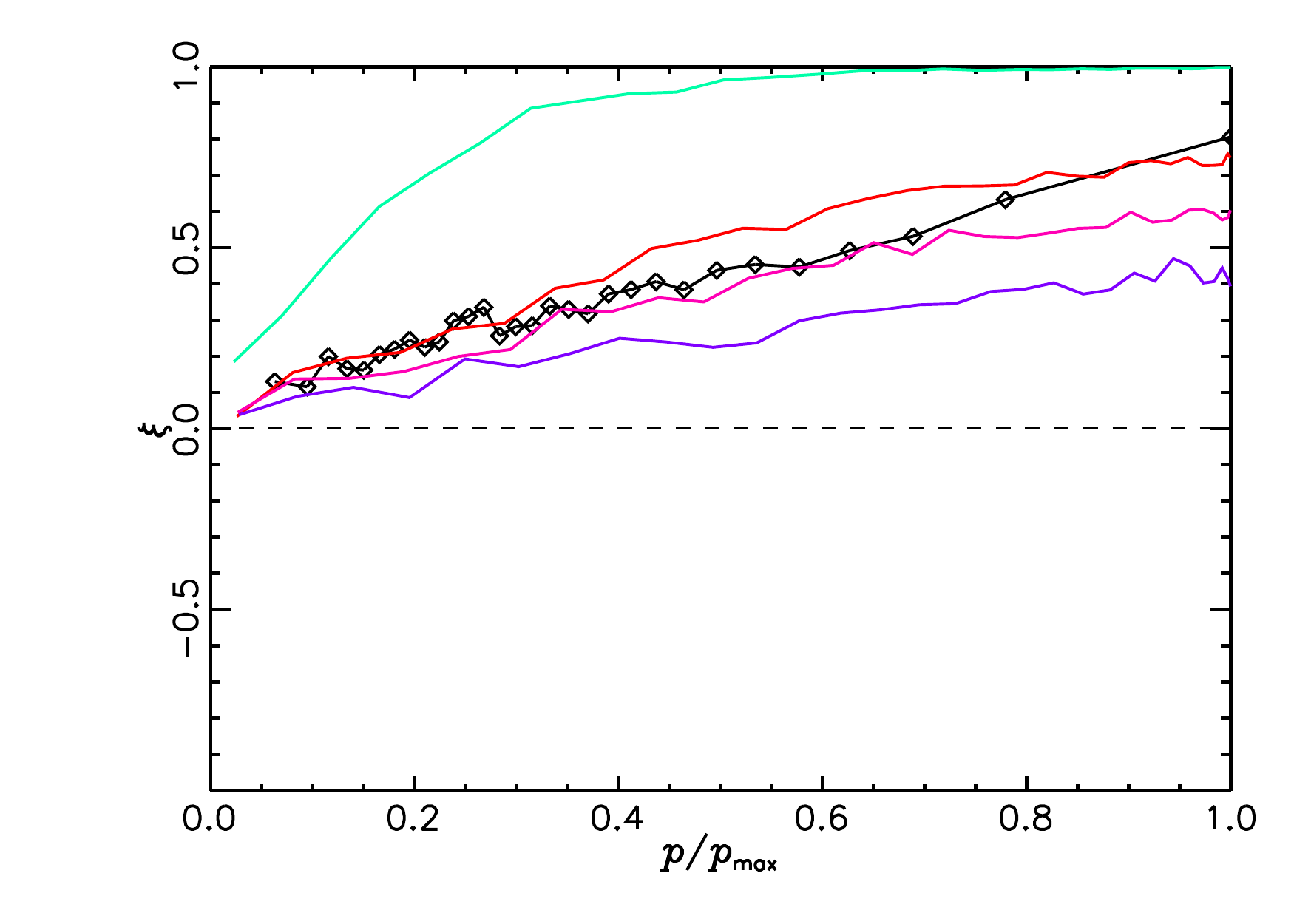}}
\caption{Correlation between the degree of alignment, $\xi$, and the
  polarization fraction, $p$, for the selected pixels, both for data
  (black squares) and for the Gaussian
  models. The models are characterized by the following values of $\sigma_{\rm M}$:
  $15^{\circ}$ in green,  $29^{\circ}$ in red,  $33^{\circ}$ in
  magenta and  $38^{\circ}$ in purple.
  The data values of $p$ are normalized to the maximum value, $p_{\rm
    max}$, within the sample. For the models, $p =
  \cos^2{\gamma_{\rm M}}$. This figure shows that projection effects, probed by the
  Gaussian models, are likely to be the main cause  of the  correlation between $\xi$
and $p$.}
\label{fig:eps_p}
\end{figure}

In Fig.~\ref{fig:eps_p}, we study the dependence of $\xi$ on $p$ by binning the
latter and keeping a constant number of pixels in each
bin. For each bin of $p$,  we  compute $H(\Theta)$ and $\xi$ and we find
that, on the sky, $\xi$ increases with $p$. 
We check that this dependence is not affected by 
noise. Using the error map from \citet{planck2014-XIX}, we compute a Gaussian realization of the noise in the polarization
angle that we add to the data to increase the noise
level by a factor of $\sqrt{2}$. We find that the dependence of $\xi$ on
$p$, obtained for the noisier angle map, is the same as that in Fig.~\ref{fig:eps_p}. 

In Fig.~\ref{fig:eps_p}, the data are compared with model results that show how $\xi$ varies with 
$\cos^2{\gamma_{\rm M}}$ for increasing
values of $\sigma_{\rm M}$. 
We find that $\xi$ correlates
with the projection factor. The model that best matches
the data has $\sigma_{\rm M}=33^{\circ}$. This value of $\sigma_{\rm  M}$ corresponds to a preferred alignment, where the angle between matter
structures and the magnetic field in 3D is within $45^{\circ}$  for
about $80\,\%$ of the selected ridges.

\begin{figure}[h!]
\centerline{\includegraphics[width=8.8cm,trim=40 0 20 0,clip=true]{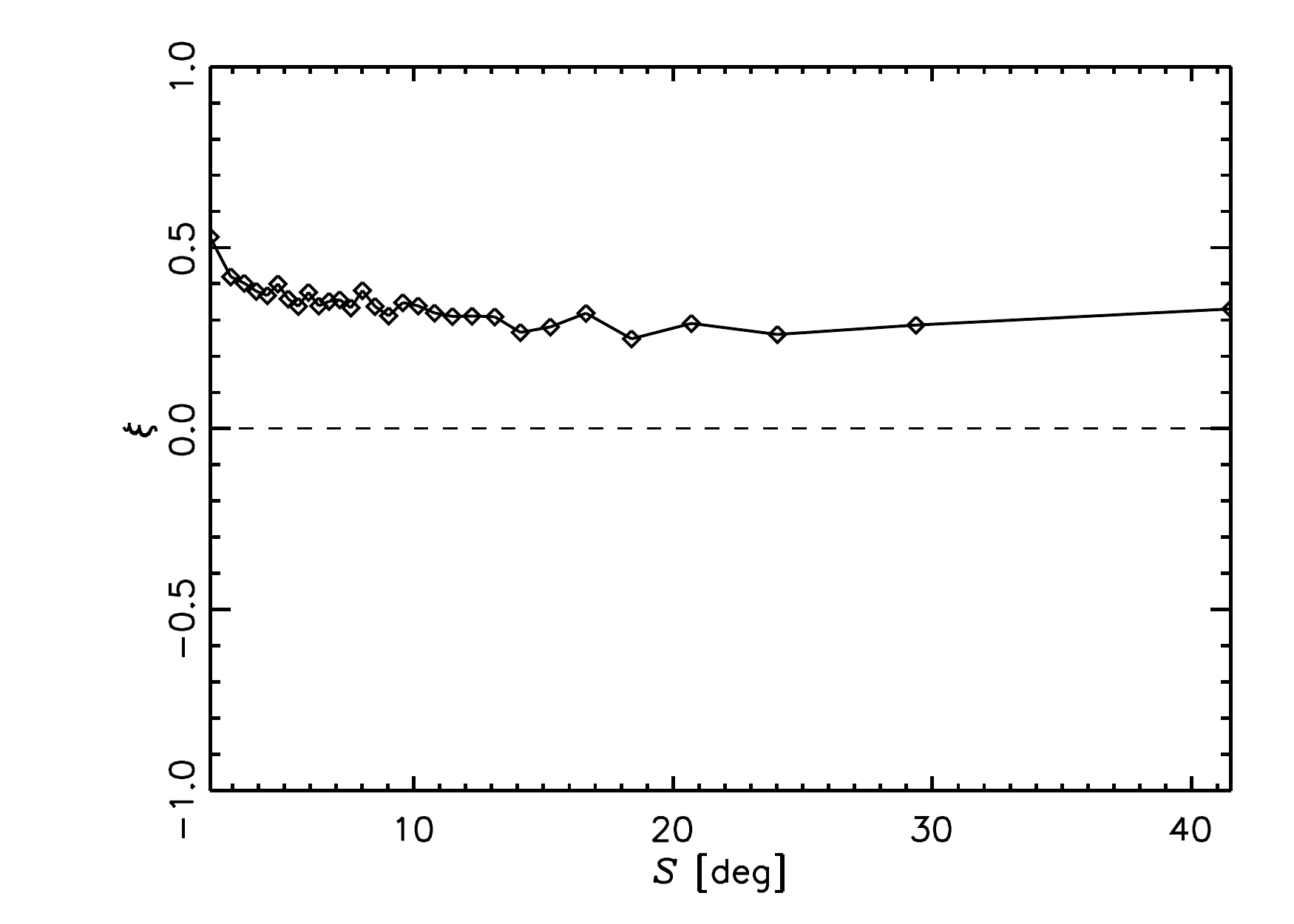}}
\caption{Dependence of the degree of alignment, $\xi$, on $\mathcal{S}$, the local dispersion of the polarization angle, an empirical tracer 
of depolarization along the line of sight and within the beam. This plot shows that 
$\xi$  does not significantly depend on $\mathcal{S}$.}
\label{fig:eps_S}
\end{figure}

For the data, unlike for the Gaussian model where $p=\cos^2{\gamma_{\rm M}}$, 
the  polarization fraction depends on the line-of-sight depolarization. 
Depolarization results from the dispersion of the magnetic field
orientation along the line of sight, and within the beam \citep{Fiege00}. 
\citet{planck2014-XIX} used maps of the dispersion
of the polarization angle, $\mathcal{S}$, to quantify local variations
of the magnetic field orientation. They defined $\mathcal{S}$ as 
\begin{equation}
\label{eq:deltapsi}
 \mathcal{S}(\vec{x},\delta)=\sqrt{\frac{1}{N}\sum^{N}_{i=1}[\psi(\vec{x})-\psi(\vec{x}+\vec{\delta_i})]^2},
\end{equation}
where $|\vec{\delta_i}| = \delta$.
\citet{planck2014-XIX}  report a general trend where the regions in the sky with high (low) values of
$\mathcal{S}$ have low (high) polarization fraction; the fractional
variation of $\mathcal{S}$ is equal to that of $p$.
$\mathcal{S}$ is an indicator of the depolarization
along the line of sight associated with the tangling of the field within the beam. This interpretation is supported by  \citet{planck2014-XX},
who point out that $\mathcal{S}$, computed from their 3D MHD
simulation, does not vary with the mean projection factor
  for $\mathcal{S}>5^{\circ}$ (see their Fig.~24).  
In Fig.~\ref{fig:eps_S}, we plot $\xi$ as a function of $\mathcal{S}$ computed on the data using the 
same binning procedure applied for $p$. We find a slight decrease in $\xi$ versus $\mathcal{S}$ much smaller than 
the increase in $\xi$ versus $p$.  This result supports our interpretation of 
the dependence of $\xi$ on $p$, which  results primarily from the orientation of the field with respect to the plane of the sky. 

\begin{figure}[h!]
\centerline{\includegraphics[width=8.8cm,trim=40 0 20 0,clip=true]{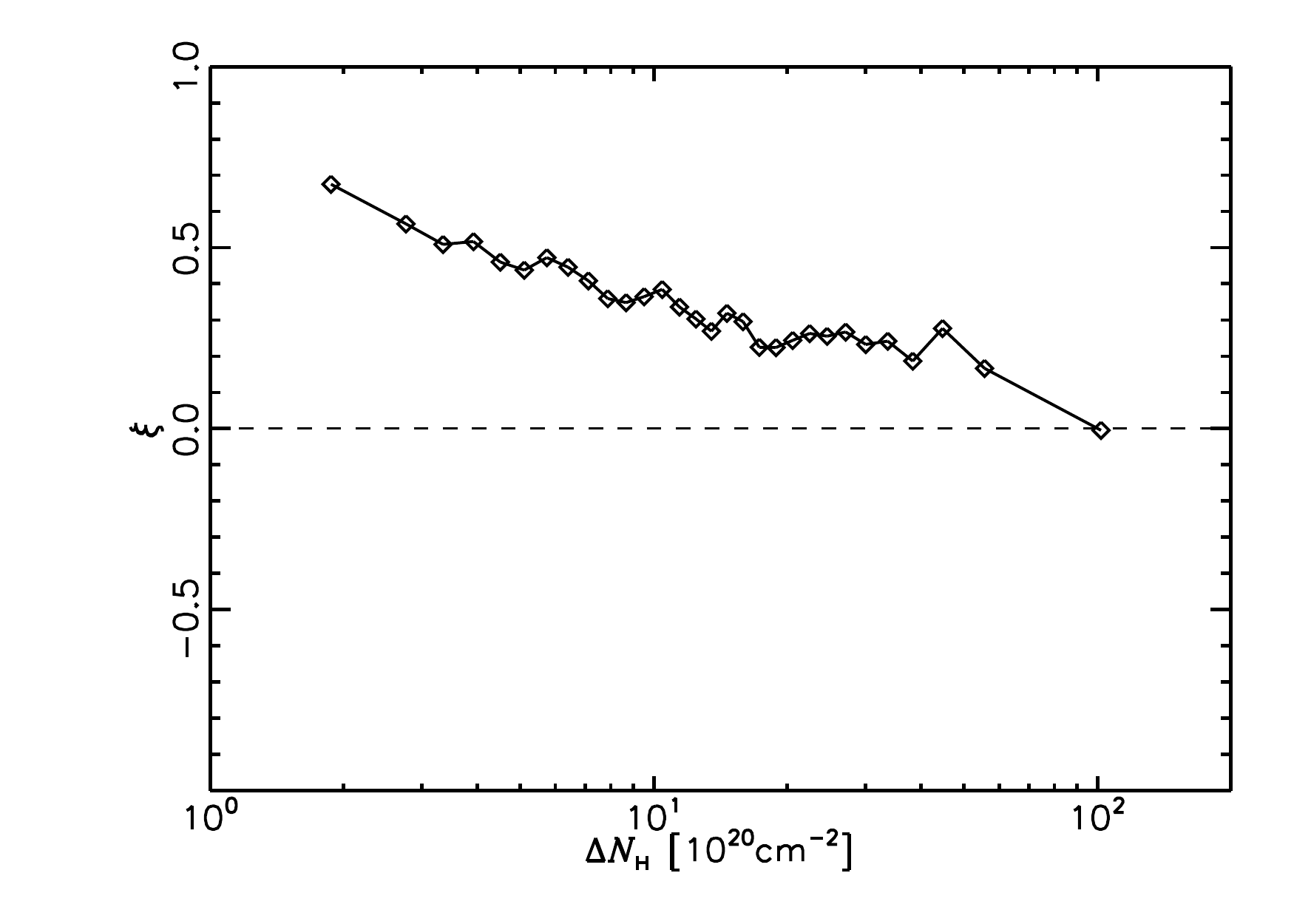}}
\caption{Variation of the degree of alignment, $\xi$, as a function of the excess column
  density, $\Delta N_{\rm H}$, for the selected pixels. The degree of
alignment decreases for increasing values of the column density.}
\label{fig:eps_nh}
\end{figure}


\section{Relative orientation between the magnetic field and the matter structures in molecular clouds}
\label{sec:align_nh}

We extend our statistical analysis to molecular clouds 
characterizing how the degree of alignment between the matter structures and the magnetic field varies with  column density. 
In Sect.~\ref{subsec:NH_dependence}, we show that the degree of alignment decreases for increasing column density.
Maps of the relative orientation are presented for the Chamaeleon and Taurus molecular clouds in Sect.~\ref{subsec:bimodal}.

\subsection{$\xi$ versus $\Delta N_{\rm H}$ over the whole sky}
\label{subsec:NH_dependence}

\begin{figure}[h]
\centerline{\includegraphics[width=8.8cm,trim=10 0 10 0,clip=true]{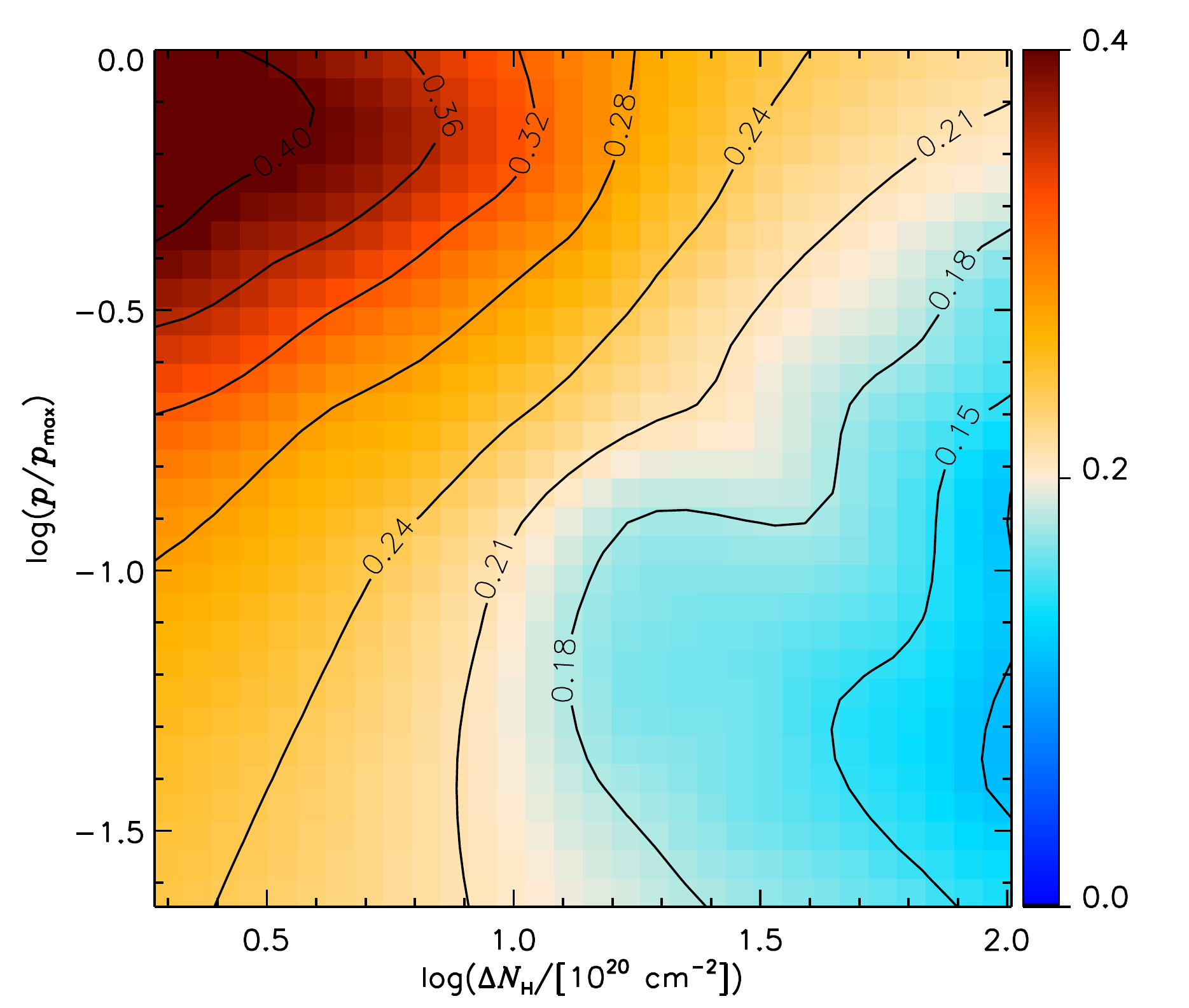}}
\caption{Map and contours of the degree of alignment, $\xi$, as a
  function of $p$ and $\Delta N_{\rm H}$. Only the selected pixels are
  taken into account for computing $\xi$. This figure shows that $\xi$ 
  depends on both quantities, $p$ and $\Delta N_{\rm H}$.}
\label{fig:eps_nh_p}
\end{figure}

\begin{figure*}[h!]
  \centering
  \begin{tabular}{r l}
      \includegraphics[width=8.5cm,trim=5 0 5 0,clip=true]{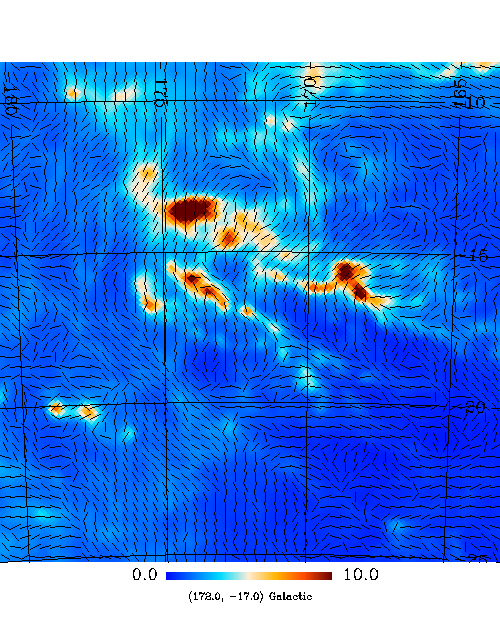}
      & \includegraphics[width=8.5cm,trim=5 0 5 0,clip=true]{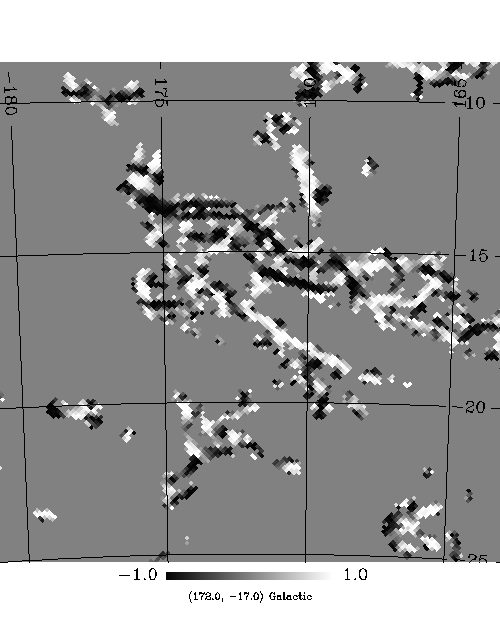}\\
      \includegraphics[width=8.5cm,trim=5 0 5 0,clip=true]{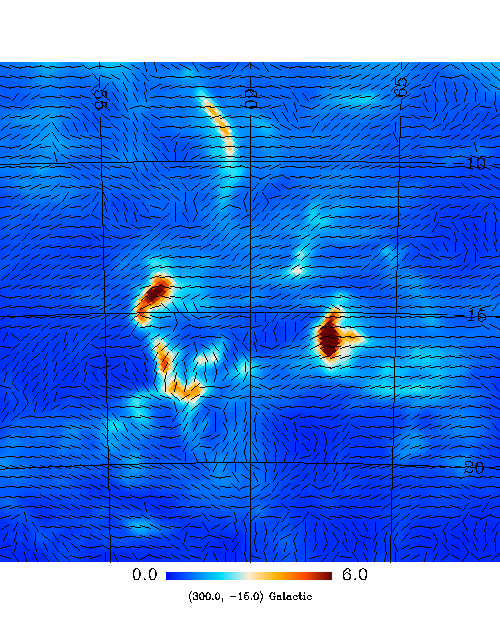}
      & \includegraphics[width=8.5cm,trim=5 0 5 0,clip=true]{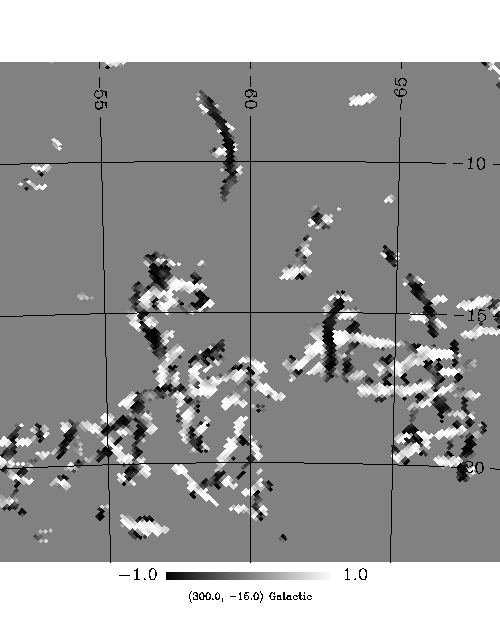}\\
    \end{tabular}
       \caption{{\it Left}: maps of the
         visual extinction $A_V$ derived from the sub-mm dust opacity from \citet{planck2013-p06b} at the $15\arcmin$ resolution of our analysis for the Taurus molecular
         cloud ({\it top}),  and the Chamaeleon molecular complex
         including the  Musca filament ({\it bottom}). The vectors
         tracing the magnetic field orientation, computed from $Q^{\rm Dif}_{353}$
         and $U^{\rm Dif}_{353}$, are plotted with a fixed length. {\it Right}: corresponding maps of the relative orientation between 
         matter structures and magnetic field quantified by
         $\cos{2 \Theta }$ for the selected structures (Sect.~\ref{sec:filaments}). This figure shows coherent
         structures where the cosine is  either positive
         or negative, 
         corresponding to a magnetic field 
         aligned with, or perpendicular to the structures.}
      \label{fig:tau_mus}
\end{figure*}

We quantify the dependence of the degree of alignment $\xi$ on
the excess column density of the selected ridges $\Delta N_{\rm H}$ (see
Sect.~\ref{sec:filaments}). We bin $\Delta N_{\rm H}$, applying the
same binning procedure as for $p$. 
We find that $\xi$ is anti-correlated with $\Delta N_{\rm H}$, as shown
in Fig.~\ref{fig:eps_nh}. 

The {\Planck} polarization data show an overall anti-correlation
between polarization fraction and column density \citep{planck2014-XIX,planck2014-XX}.
We test that the decrease in $\xi$ with $\Delta N_{\rm H}$ does not result from 
this variation of $p$ with column density.
In Fig.~\ref{fig:eps_nh_p}, we present a map
that characterizes the variations of $\xi$ both
as a function of $p$ and $\Delta N_{\rm H}$. We bin
the selected pixels first in $\Delta N_{\rm H}$ and then in $p$, 
ensuring that we have the same number of elements for each bin of both variables. Given a 
2D bin,  we compute the corresponding
$H(\Theta)$ and $\xi$. We then interpolate the $\xi$ map 
over a regular grid of values for $p$ and $\Delta N_{\rm H}$.
The map of $\xi$  confirms the decrease in
alignment from low to high excess column densities and from high
to low polarization fractions. The degree of alignment
clearly depends on both $p$ and $\Delta N_{\rm H}$. 

\subsection{A glimpse into molecular clouds}
\label{subsec:bimodal}

\begin{figure*}[h!]
  \centering
  \begin{tabular}{r l}
      \includegraphics[width=8.5cm,trim=5 0 5 0,clip=true]{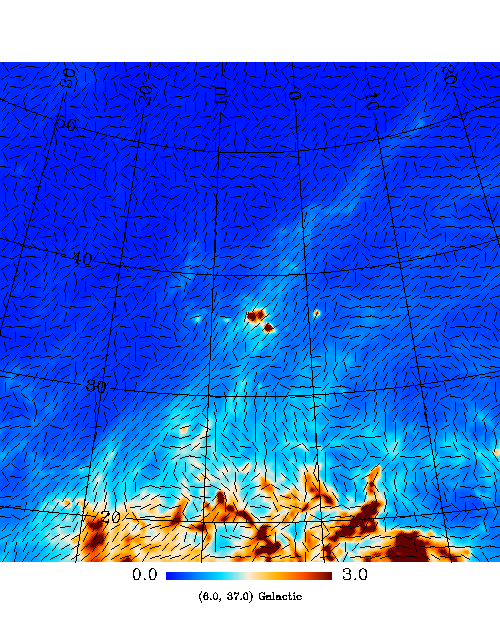}
      & \includegraphics[width=8.5cm,trim=5 0 5 0,clip=true]{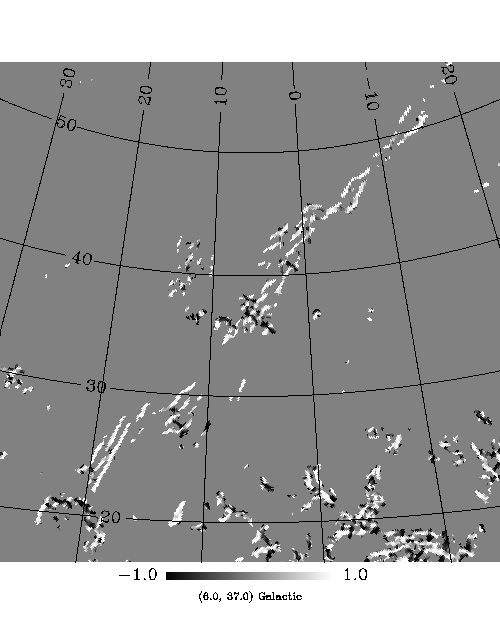}\\
      \includegraphics[width=8.5cm,trim=5 0 5 0,clip=true]{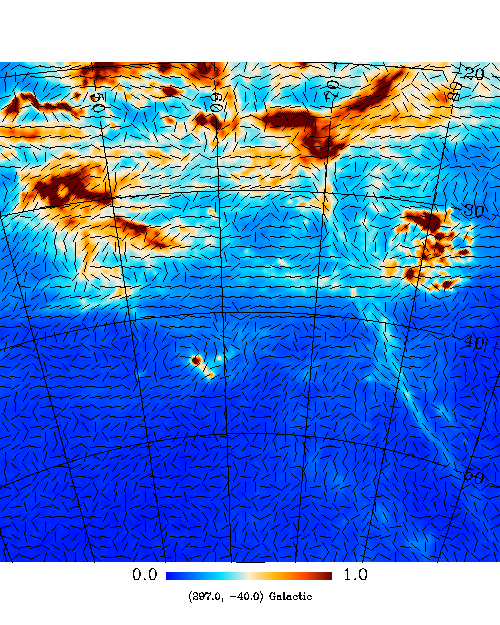}
      & \includegraphics[width=8.5cm,trim=5 0 5 0,clip=true]{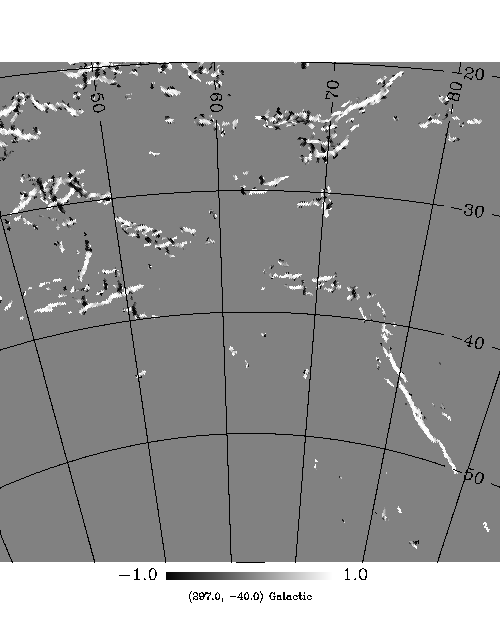}\\
    \end{tabular}
       \caption{ Same as in Fig.~\ref{fig:tau_mus} but for two fields at intermediate
         Galactic latitudes sampling the diffuse ISM. The central pixel for
         the top panels corresponds to
         ($l,b$)=($6^{\circ},37^{\circ}$). The central pixel for
         the bottom panels corresponds to
         ($l,b$)=($295^{\circ},-40^{\circ}$).  The Magellanic Clouds
         in the bottom extinction map (left) are masked. Most of
       the structures in the relative orientation maps appear as
       parallel to the magnetic field.}
      \label{fig:diffuse}
\end{figure*}

The anti-correlation found between $\xi$ and $\Delta N_{\rm H}$ suggests that at high column densities, within molecular
clouds, the degree of alignment between
the magnetic fields and matter structures decreases.  To discuss this result we present maps of 
the Taurus and the Chamaeleon molecular clouds, as representative examples of molecular complexes in the solar neighbourhood.
For comparison, Fig.~\ref{fig:diffuse} illustrates two fields at intermediate Galactic latitudes, which sample the diffuse ISM. 

The four panels in Figure~\ref{fig:tau_mus} show the extinction maps
derived from the dust sub-mm opacity in \citet{planck2013-p06b}
next to the corresponding maps of alignment quantified by 
$\cos{2 \Theta }$. The cosine function spans values between $-1$
and $1$, identifying structures that are perpendicular and parallel
to the magnetic field, respectively. 
We stress that the cosine representation,
chosen for visualization, stretches the
contrast of the $\Theta $ distribution towards the extrema.
On the extinction maps, we plot the vectors tracing the magnetic
field orientation inferred from $Q^{\rm Dif}_{353}$ and $U^{\rm Dif}_{353}$. 

The maps of relative orientation in the Taurus and Chamaeleon clouds (Fig.~\ref{fig:tau_mus})
reveal some coherent structures where the magnetic field tends to be perpendicular to the interstellar ridges, in particular
to those with the highest extinction, while in the diffuse ISM (Fig.~\ref{fig:diffuse}), 
in agreement with the DFs of $\Theta $ presented in Sect.~\ref{subsec:alignment},  there is 
a predominance of structures parallel to the magnetic field. Hence, 
the flattening of $H(\Theta)$ as a function of $\Delta N_{\rm H}$ might
be associated with the presence of matter structures that are
perpendicular to, rather than aligned with, the magnetic field, and not
related to a loss of correlation between the field and
the structure of matter. 

In Appendix~\ref{appendix:gaussian}, we show maps of $\cos{2 \Theta }$ in Taurus and Chamaeleon computed 
with a Gaussian model where we set $\vec{V}_{\rm M0}$ 
to zero in order to quantify the effect of projection onto the plane of  the sky when the orientations of the magnetic field and
the matter structures are uncorrelated. The spectral index used in the Gaussian realizations, $\alpha_{\rm M}=-1.5$, introduces a correlation in the 
orientation of $\vec{V}_{\rm M}$ over the sky, which is  independent of the structure of matter.
The maps of $\cos{2 \Theta }$ computed with this model, shown
in Fig.~\ref{fig:relativemodel}, present
small black or white structures that appear perpendicular or
parallel with respect to the magnetic field orientation. However,
prominent elongated features, with the magnetic field oriented
preferentially orthogonal to the matter structures, seen in 
the sky images in Fig.~\ref{fig:tau_mus}, such as the Musca filament, 
are absent in the model images in Fig.~\ref{fig:relativemodel}. This difference suggests that 
in molecular clouds there is a significant 
number of matter structures, which tend to be perpendicular to the
magnetic field, explaining the decrease in $\xi$ with $\Delta N_{\rm H}$.  

Dust polarization in molecular clouds allows us to trace the field
 orientation in high column-density
 structures, as detailed in \citet{doris2014},
 where three examples are analysed. Thus, we consider it unlikely that the decrease in $\xi$ with $\Delta N_{\rm H}$ is due to a loss
of polarization from either enhanced turbulence or reduced dust grain
alignment efficiency for increasing column density \citep{Falceta08,Whittet2008,Jones2015}. Furthermore, this alternative interpretation is not supported by the
fact that $\xi$ decreases for increasing $\Delta N_{\rm H}$,
independently of its dependence on $p$ (see Fig.~\ref{fig:eps_nh_p}).

The \Planck\ images should not be
interpreted, however, as evidence for two distinct orientations with
respect to that of the magnetic field, with
depletion at intermediate angles. 
Such a bimodality was suggested  by previous studies 
based on extinction data for dark clouds \citep{Li2013} and MHD
simulations \citep{Soler13}, but questioned by other
observational studies \citep{Goodman90,Houde04}.
Our all-sky analysis does not show a significant turn-over in the
statistics of relative orientation between the matter structures and the magnetic field,
from the diffuse ISM to molecular clouds. 

\begin{figure}[h!]
\centerline{\includegraphics[width=8.8cm,trim=40 0 20 0,clip=true]{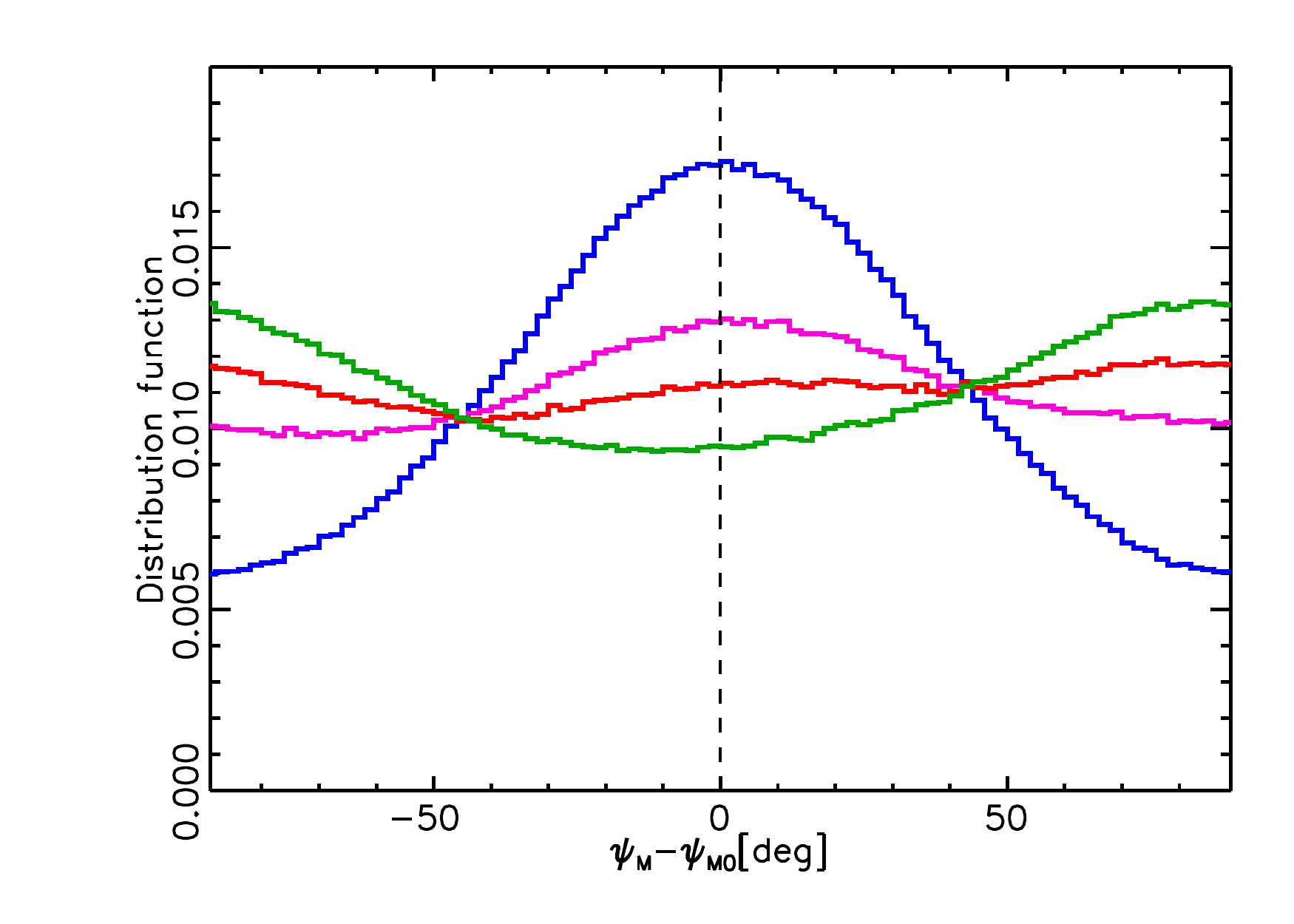}}
\caption{Distribution functions of relative orientations for the Gaussian model with $\sigma_{\rm M}=33^{\circ}$. 
The distributions show the effects of projection onto the plane of the sky of
the 3D relative orientations between $\vec{V}_{\rm M}$ and
$\vec{V}_{\rm M0}$. The curves refer to four different
configurations depending on $\eta$, the fraction of 3D perpendicular
orientations between the two vectors. For the blue curve,
$\eta=0$. For the pink, red, and green curves, $\eta$ is 0.5, 0.7 and
0.9, respectively.} 
\label{fig:perpstr}
\end{figure}

Because of projection effects, it is difficult to identify a bimodal distribution between
magnetic fields and matter structures in a statistical way. 
To quantify this statement, we introduce a bimodal
configuration of relative orientations in 3D between $\vec{V}_{\rm M}$
and $\vec{V}_{\rm M0}$ in the Gaussian models (Appendix~\ref{appendix:gaussian})
using a new model parameter $\eta$, which represents the fraction of sky pixels where $\vec{V}_{\rm M}$
is distributed about a second reference direction  perpendicular to $\vec{V}_{\rm M0}$. 
Essentially, if $\eta$
is the probability of having the mean of $\vec{V}_{\rm M}$ perpendicular to
$\vec{V}_{\rm M0}$, then $1-\eta$ represents the probability
of having the mean of $\vec{V}_{\rm M}$ parallel to $\vec{V}_{\rm M0}$. So far, 
throughout the paper, the models have been used with $\eta=0$. 
The dispersion of the distribution of angles, $\sigma_{\rm M}$, is the same for both directions.  
In Fig.~\ref{fig:perpstr}, we show the impact of
the value of $\eta$ on the DF of $\delta \psi_{\rm M}$, for 
$\sigma_{\rm M}=33^{\circ}$. Up to $\eta=0.7$, the DF is nearly flat and does not indicate any turn-over in the relative
orientations. 

\section{Discussion}
\label{sec:discussion}

This paper presents the first analysis across the whole sky comparing column density
structures in the ISM with the orientation of the Galactic magnetic field. 
Previous studies focused on dark clouds \citep[e.g.][]{Goodman95,Pereyra04,Alves08,Chapman11} 
where the magnetic field was found mainly, but not systematically, perpendicular to the long axis of the clouds.
Our sample of  ridges, built from the \Planck\ dust map,
is dominated by structures in the CNM of the diffuse ISM. 
For these structures we find a preferred alignment with the magnetic
field projected on the plane of the sky. This trend disappears for the highest column
densities in molecular clouds,where the data show coherent structures, which tend to be perpendicular to the
magnetic field orientation.

In the next paragraphs, we discuss these observational results in
light of models and MHD simulations, which attempt to describe the respective roles of the magnetic field, turbulence and 
the gas self-gravity in  the formation of structures in the magnetized ISM. 

The alignment between the magnetic field and matter structures in the diffuse ISM reported in Sect.~\ref{sec:alignment} could be a signature of the formation
of CNM filaments through turbulence. 
Their formation could be initiated by a local compression that would trigger the condensation of cold gas out
of the warm neutral phase \citep{Audit05,Inoue09,Heitsch09,Saury14}. The shear of
the turbulent flow  would then stretch the gas condensations
into structures, such as sheets and filaments, which would appear 
elongated in column density maps. These structures will tend to be aligned with the magnetic field if the 
gas velocity is dynamically aligned with the field \citep{Brandenburg13}.
Furthermore, where the velocity shear stretches 
matter into filaments, the field is stretched in the same direction, creating alignment because the field is frozen into 
matter.\footnote{The formation of filaments by shear is illustrated in Fig.~3 of  \citet{Hennebelle13}.}
This interpretation was proposed by \citet{Hennebelle13}  who found  a strong correlation between
the orientation of the density structures and that of the maximum shear of the
velocity field in his MHD simulations. 
Although beyond the scope of
this paper, from an observational point of
view, the correlation between density structures and shear can be
tested by looking for line-of-sight velocity gradients along 
filaments in \hi\,  and CO surveys.  Where the filaments  
are inclined with respect to the plane of the sky, we can measure both the 
polarization angle and the radial component of the gas velocity.

For supersonic turbulence, gas sheets and filaments can also be formed by gas compression in shocks. For sub-Alfv\'enic
turbulence (strong magnetic field with respect to turbulence), 
compression preferentially occurs  where gas flows along the magnetic field lines, creating, thereby, structures 
perpendicular to the field. This cannot be the dominant process because we observe preferred alignment between matter structures and the magnetic field.
For super-Alfv\'enic turbulence (weak magnetic field), gas compression  also occurs for all directions of the shock velocity
enhancing the component of the magnetic field perpendicular to the shock velocity, 
because of magnetic flux conservation and freezing into matter.\footnote{See Fig.~4 in \citet{Hartmann01} for an illustration.}
In this case, shocks tend to form structures aligned with the field.

For the highest column density ridges in molecular clouds, the
interpretation must also involve self-gravity, which
is known to amplify anisotropic structure \citep{Lin65}.   A gravitationally unstable cloud
first collapses along the shortest dimension, forming a sheet, which
subsequently breaks into elongated filamentary structures. The
presence of a large-scale magnetic field can influence this effect,
as the collapse can preferentially be along the mean field direction. 
Here, it is necessary to distinguish 
between sub-critical and super-critical structures \citep{Mouschovias76}.
When a gravitationally bound structure forms, gravitational and turbulent energies are comparable and  turbulence is sub-Alfv\'enic (super-Alfv\'enic) for 
sub-critical (super-critical) structures. 
For sub-critical structures,  the magnetic field is dynamically important and gravity pulls matter  preferentially along  field lines. 
As a consequence, in sub-critical clouds 
gravitationally bound sheets and filaments are expected to be perpendicular to the
magnetic field. The presence of striations orthogonal to the high
column density filaments in {\it Herschel} maps of nearby molecular
clouds, such as Taurus, supports this scenario
\citep{Palmeirim13}.  

A bimodal distribution of orientations of structures with respect to the magnetic
field is  observed in numerical simulations. In their MHD
simulations of molecular clouds, \citet{Soler13}
find a change in the relative orientation between 
matter structures and the magnetic field, from parallel to perpendicular, for
gravitationally bound structures. This change is most significant for their simulation with the highest magnetization.
We will need to combine our polarization data with velocity and column
density measurements in order to test whether the perpendicularity 
between the magnetic field and the matter structures is a sign of filaments formed
by self-gravity in magnetically dominated interstellar clouds.

\section {Summary and perspectives}
\label{sec:summ}

The {\Planck} $353\,$GHz all-sky polarization maps provide unprecedented information,
on the structure of the Galactic magnetic field and its correlation
with interstellar matter.
 
Most of the structures (i.e. ridges) observed in the dust intensity map outside the Galactic plane
are also seen as coherent structures in the dust Stokes $\StokesQ$ and $\StokesU$ maps.
We performed a statistical analysis of the  {\Planck} data on these  structures of matter, measuring
their orientation from a Hessian analysis of the dust intensity map, 
and that of the magnetic field from the dust polarization. 
Our data analysis  characterizes the variation of the magnetic field orientation between 
the structures and their local background, and the relative orientation between the structures  
and the magnetic field, with unprecedented statistics. 
Our sample of structures covers roughly $4\,\%$ of the
sky and spans two orders of magnitude in column density from $10^{20}$ to $10^{22}$
cm$^{-2}$.   In the following, we summarize the main results 
of our analysis.

We separated the polarized emission of the structures from that of the surrounding Galactic background within a few degrees on the sky. 
Comparing polarization angles, we estimate the ratio between the typical strengths of the turbulent and mean components of the field to be
between 0.6 and 1.0, with a preferred value of 0.8. This result is in agreement with 
Zeeman \hi\ observations indicating an approximate equipartition between
turbulent and magnetic energies in the diffuse ISM.

We find that the interstellar matter structures are preferentially aligned with the
magnetic field inferred from the polarization angle measured at the
position of the structures.
We introduced a parameter $\xi$ that quantifies the degree of alignment 
between the orientation of the magnetic field and matter structures.
We find that $\xi$  increases with the polarization fraction $p$. We
interpret this correlation in light of Gaussian models, which take into account
the projection onto the plane of the sky of the 3D configuration between the magnetic field and
the structures. Where the polarization fraction is low, the field tends
to be close to the line of sight. In this configuration, the orientation
of the magnetic field is not well constrained by the observations because its projection onto the
plane of the sky is a minor component of the field. This geometric interpretation
of the correlation between $\xi$ and $p$
is supported by the weakness of the dependence of $\xi$  on the local
dispersion of the polarization angle.
The Gaussian models best match the data for a standard deviation
between the orientation of the magnetic field and that of the structures of matter of $33^\circ$ in 3D.

We find that  $\xi$ decreases for increasing column
density, and that there is no alignment for the highest column
density.
This result does not reflect an absence of correlation between the structures of matter and the magnetic field, at high $N_{\rm H}$. 
Structures that tend to be perpendicular to the magnetic field appear
in molecular clouds, where they contribute to the
statistics of relative orientations. We show maps
of the Taurus and the Chamaeleon molecular
complexes that support this interpretation.

We have presented the first analysis on the relative
orientation between the filamentary density structures of the ISM and the
Galactic magnetic field across the whole sky.
The main outcome of this study is that, at the angular scales probed
by {\Planck}, the field geometry, 
projected on the plane of the sky, is correlated with the distribution of matter in the solar neighbourhood. 
In the diffuse ISM, the structures of matter are preferentially aligned with the magnetic field, while perpendicular structures appear in molecular clouds. 

Our results support a scenario of formation of
structures in the ISM where turbulence organizes matter parallel to the magnetic field in the diffuse
medium, and the gas self-gravity produces perpendicular structures in the densest and
magnetically dominated regions.
This tentative interpretation on the role of turbulence in structuring interstellar matter may be tested by 
comparing the relative orientation of structures and the magnetic field with line-of-sight velocity gradients.
It will also be interesting to apply our statistical analysis to MHD simulations of the
diffuse ISM. Statistical analysis also needs to be complemented by  
detailed studies of specific structures in the diffuse ISM and in molecular clouds.



\begin{acknowledgements}
The development of \planck\ has been supported by: ESA; CNES and
CNRS/INSU-IN2P3-INP (France); ASI, CNR, and INAF (Italy); NASA and DoE
(USA); STFC and UKSA (UK); CSIC, MICINN, JA, and RES (Spain); Tekes,
AoF, and CSC (Finland); DLR and MPG (Germany); CSA (Canada); DTU Space
(Denmark); SER/SSO (Switzerland); RCN (Norway); SFI (Ireland);
FCT/MCTES (Portugal); and PRACE (EU).  A description of the Planck
Collaboration and a list of its members, including the technical or
scientific activities in which they have been involved, can be found
at
\url{http://www.sciops.esa.int/index.php?project=planck&page=Planck_Collaboration}.
The research leading to these results has received funding from the European Research 
Council under the European Union's Seventh Framework Programme (FP7/2007-2013) / ERC grant agreement n$^\circ$ 267934.
\end{acknowledgements}

\bibliographystyle{aa}

\input{PIPXXXII.bbl}
\appendix

\section{Hessian analysis of the sky}
\label{appendix:Hessian}

In this appendix, we detail the implementation of the Hessian analysis on
the \Planck\ map of specific dust intensity at $353\,$GHz  $D_{353}$. 
We explain how we compute the derivatives of the sky brightness 
in Sect.~\ref{subsec:method}. The analysis is applied to a  toy model of the dust emission to quantify the impact of the data noise 
and to define the threshold on the curvature we use to identify and select ridges (Sect.~\ref{subsec:gaussian}).
In Sect.~\ref{subsec:inertia}, we compare the orientations derived 
from the Hessian analysis with those obtained from an alternative method, the inertia matrix 
used by \citet{Hennebelle13} for MHD simulations.

\subsection{Implementation of the Hessian analysis}
\label{subsec:method}

A general description of the Hessian analysis is presented in Sect.~\ref{sec:filaments}.
Here, we detail how we compute the Hessian matrix for the dust map $D_{353}$ 
on its \healpix\ grid. We compute the first  derivatives of the
$D_{353}$ map by performing a bilinear interpolation. We locally fit the dust emission with a linear combination of 
$x=b$ and $y=l \, \cos(b)$ plus a constant, where $l$ and $b$ are the
Galactic longitude and latitude. In order to well sample the
computation of the derivatives, bilinear interpolation is
performed on a region within a disc of $10\arcmin$ radius about each
pixel of the \healpix\ map. We select 
roughly 30 pixels at $N_{\rm side}=1024$, around each pixel at $N_{\rm side}=512$.  
The fit, performed with no weights, is equivalent to 
a first-order Taylor expansion of $D_{353}$. We repeat this procedure on the maps of the first derivatives to obtain maps of the four elements of the
Hessian matrix in Eq.~(\ref{eq:hessI}). The orientation angle $\theta $
of the brightness structures is computed using
Eq.~(\ref{eq:dirfil}).

\begin{figure}[h!]
\centerline{\includegraphics[width=8.8cm,trim=50 0 20 0,clip=true]{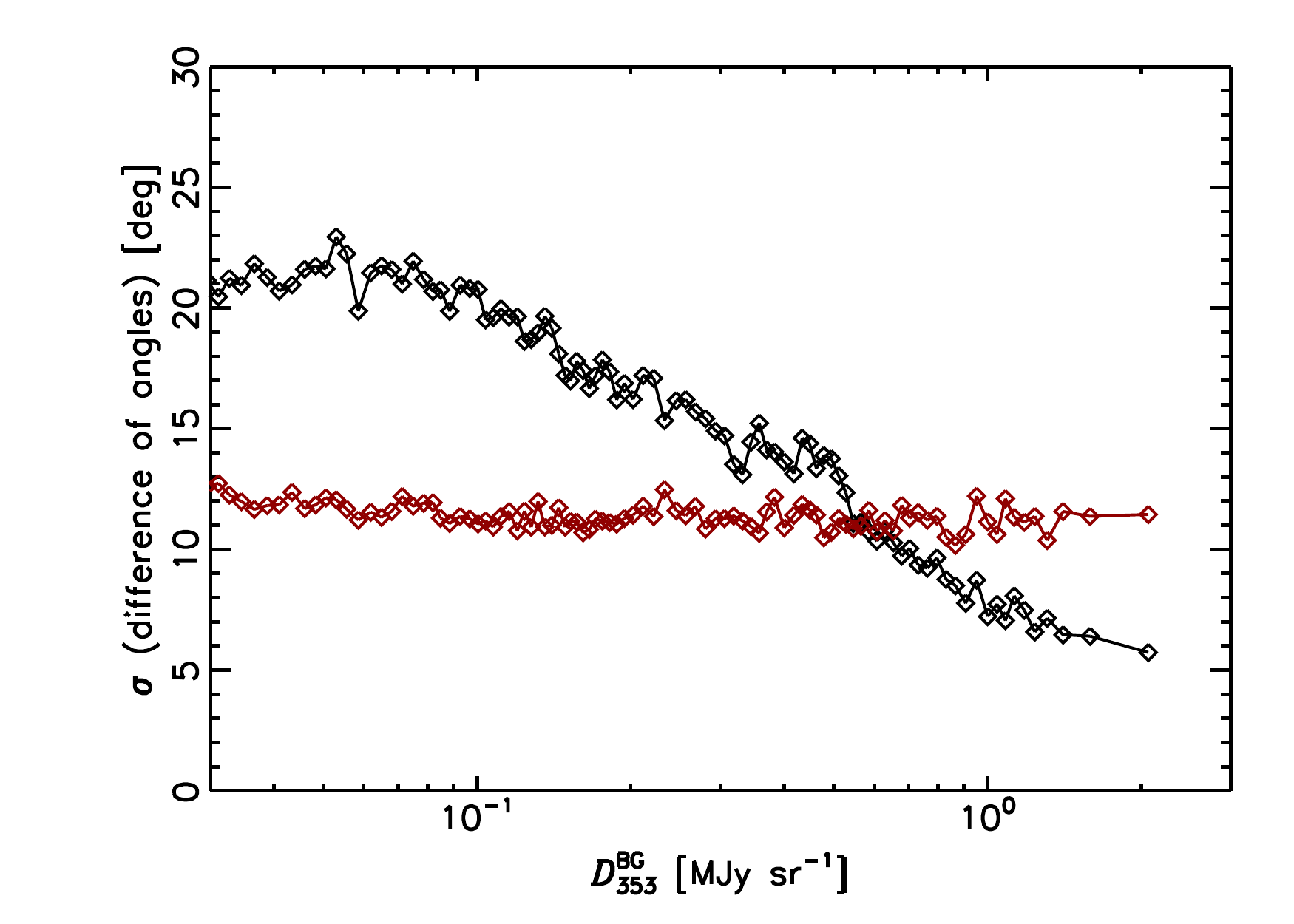}}
 \caption{Two independent estimates of the uncertainties in the determination 
 of angles from the Hessian analysis, computed within bins of  values
 of the  $D^{\rm BG}_{353}$ map.
 The red squares represent the noise obtained by propagation of the errors on the
 coefficients of the Hessian matrix. The noise computed  with the toy model (Sect.~\ref{subsec:gaussian}) 
 is plotted with black squares.   }
 \label{fig:sig_Hes}
\end{figure}

The bilinear interpolation provides error maps for the four coefficients
of the matrix, computed under the assumption
that the linear model is exact, i.e. assuming that the $\chi^2$ of the fit per degree of freedom is 1. 
The errors on the elements of the Hessian matrix are propagated to
$\theta $. Since Eq.~(\ref{eq:dirfil}) is non-linear, we compute the error map on
 $\theta $ from  Gaussian realizations of the errors on the coefficients of the matrix.
The uncertainty in $\theta $, computed within bins of values of the  $D_{353}$ map, is $12.6^\circ$, independent of the intensity at $353\,$GHz (Fig.~\ref{fig:sig_Hes}).

\begin{figure}[h!]
 \centerline{\includegraphics[width=8.8cm,trim=50 0 22 0,clip=true]{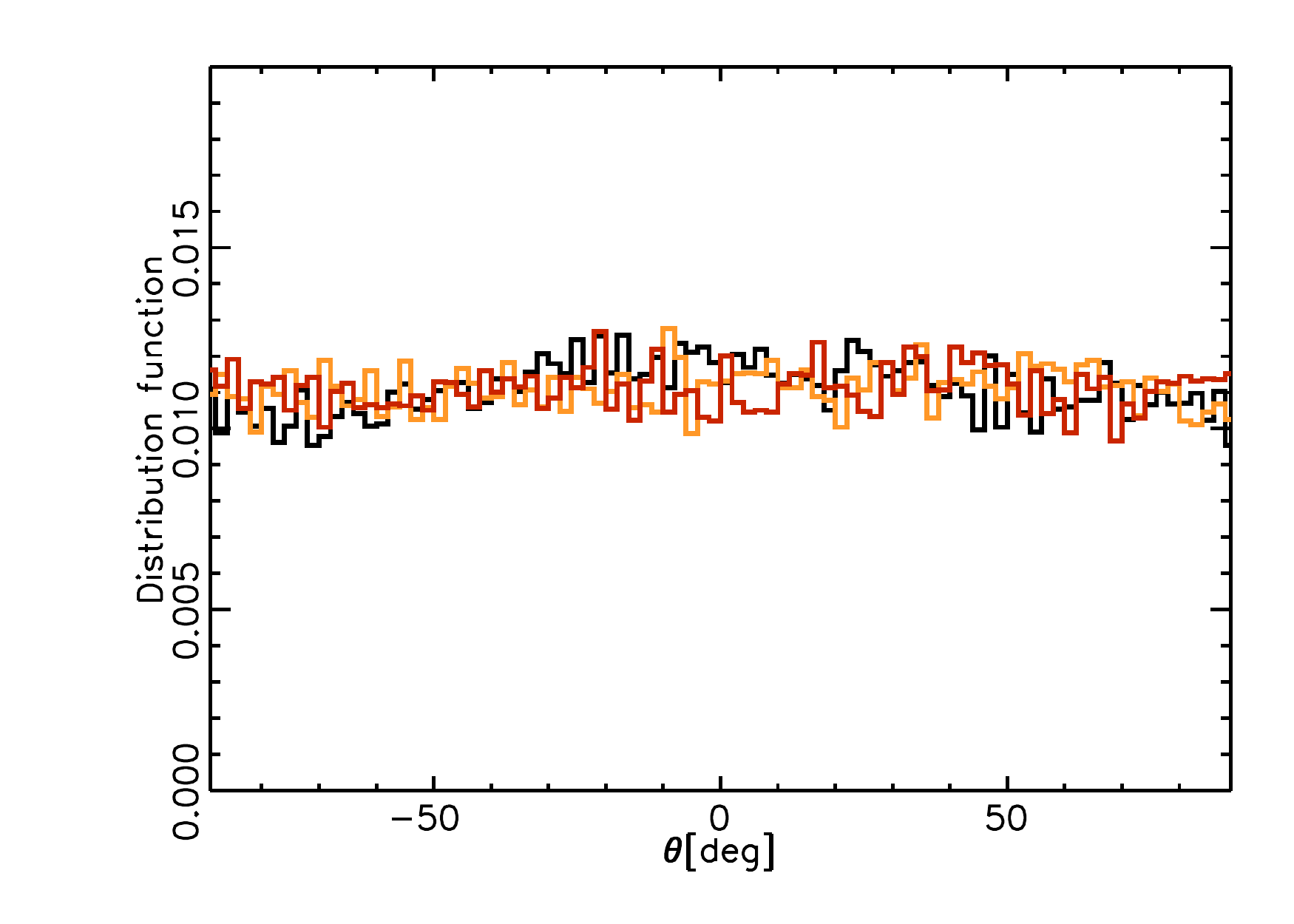}}
 \caption{Distribution functions of orientation angles for the selected ridges derived
   from the Hessian analysis on $T_{353}$. They are computed within three bins of
   Galactic latitudes with equal numbers of data points. The mean absolute latitudes 
 are $7^\circ$, $17^\circ$, and $33^\circ$ for the DFs plotted in black, orange and
 red, respectively. The image shows that the method does not introduce
any bias as a function of Galactic latitude in estimating $\theta$.}
 \label{fig:His_theta_glat}
\end{figure}

\subsection{Test on a toy model of the sky}
\label{subsec:gaussian}

We have tested the Hessian algorithm  on a toy model of the full sky $T_{353}$ built from a realization of a Gaussian map, $G_{\rm map}$,  with an angular power spectrum,
equal to a power law of index  $-2.8$, computed with the procedure {\tt SYNFAST} of \healpix\ . The spectral index used is within the range of values found for power spectra  of dust maps \citep{Miville07}. 
The mean value of $G_{\rm map}$ is 0. We form,

\begin{equation}
\label{eq:Gsky}
T_{353} = A \, (G_{\rm map}+D)/(\sin{|b|}+B \, |l| )+C,
\end{equation}
where $A$, $B$, $C$, and $D$ are four factors chosen to fit the latitude profile, and the longitude profile at Galactic latitude $b=0^\circ$, measured on the $D_{353}$
map.  The values of these factors, including their sign,  depend  on the realization of $G_{\rm map}$. 
The toy model matches the large-scale structure of the Galactic dust emission, but it does not have its filamentary structure because
it is computed from an isotropic Gaussian map. It assumes that the
amplitude of the brightness fluctuations at a given scale is
proportional to the brightness. This is in agreement with what has
been reported for the emission at 100$\,\mu$m of the diffuse ISM \citep{Miville07}.
We use the toy model to quantify the impact of the structure of the diffuse Galactic emission and data noise on the Hessian analysis.

We run the Hessian analysis on the  $T_{353}$ map with and without data noise computed as 
a Gaussian realization of the noise map on $D_{353}$.  
The DFs of orientation angles, computed for the model with added noise, are flat at all 
Galactic latitudes, as shown in Fig.~\ref{fig:His_theta_glat}. 
We find no bias on the angle introduced by
the large-scale gradient of the emission with Galactic latitude. We
use the difference between the two angle maps computed on the $T_{353}$ map, with and without noise,
to estimate the contribution of the data noise  to the uncertainty in the Hessian angle (black dots in Fig.~\ref{fig:sig_Hes}). 
This uncertainty decreases for increasing values of $T_{353}$ (i.e. increasing signal-to-noise ratio). The fact that it is higher than the
uncertainty estimated from the errors on the coefficients of the Hessian matrix, for most values of the sky brightness, 
is likely to be due to the non-Gaussian distribution of the uncertainties in $\theta $. 

\begin{figure}[h]
\centerline{\includegraphics[width=8.8cm,trim=45 0 20 0,clip=true]{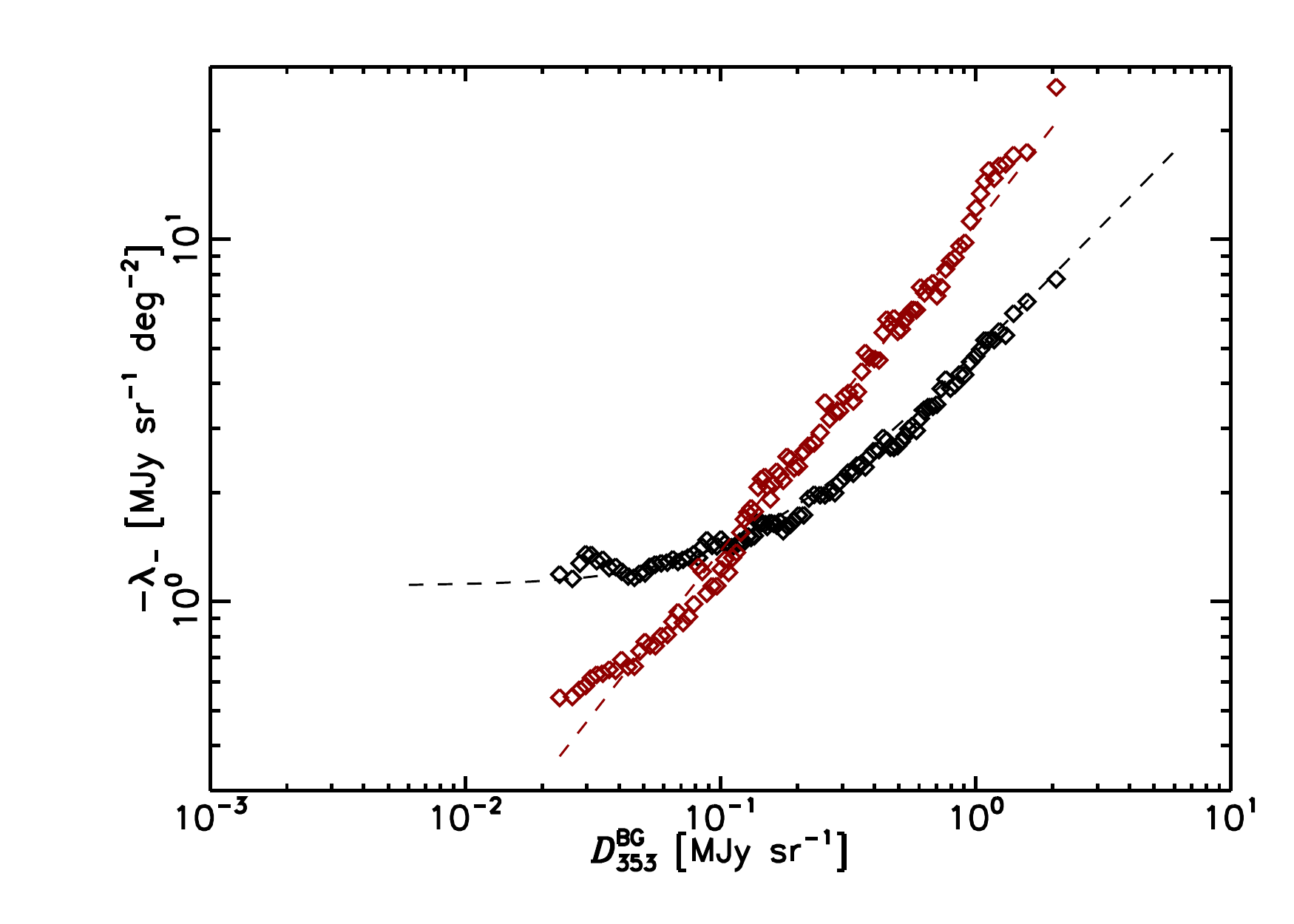}}
 \caption{Results of a Hessian analysis of the toy model of the sky
   $T_{353}$ with instrumental noise. The black (red)
 squares represent the $3\sigma$ level of the distribution of  
 $-\lambda_-(T_{353})$ ($-\lambda_-(D_{353})$) within bins of $D^{\rm
   BG}_{353}$ computed with the mask used in our data analysis. The dashed lines represent the analytical fits
 in Eq.~(\ref{eq:threshold})  (black dashed line) and Eq.~(\ref{eq:median_curvature}) (red dashed line).}
 \label{fig:curv-threshold}
\end{figure}

The toy model is also used to determine the threshold on the curvature we adopt to select ridges in the sky. 
The map of negative curvature $\lambda_-(T_{353})$ computed on  $T_{353}$  shows  filament-like ridges 
on small angular scales. We consider that the amplitude of these ridges provide an estimate of the noise on $\lambda_-$ from the structure of the
background emission.  To quantify this noise  we compute 
the histogram of the $\lambda_-(T_{353})$ map 
for 100 bins of  $T_{353}$ values, with equal numbers of data points. 
In Fig.~\ref{fig:curv-threshold}, we plot the $3\sigma$ of the
distribution of $-\lambda_-$ versus $T_{353}$, and the 
analytical fit to these values,

\begin{equation}
\label{eq:threshold}
\frac{C_{\rm T}}{{\rm MJy\,sr^{-1} \, deg^{-2}}} = 13 \, \sqrt{0.007+\left(\frac{T_{353}}{4\, {\rm MJy\,sr^{-1}}}\right)^{1.45}}.
\end{equation}
To select  brightness ridges, we use  $C_{\rm T}$ as a threshold on the negative curvature (see Sect.~\ref{sec:filaments}). The difference between the maximum and
minimum eigenvalues of the Hessian matrix, $\lambda_+$ and $\lambda_-$
respectively, is greater than $C_{\rm T}$  for $94\,\%$ of the
selected ridges. This shows that we select elongated ridges rather
than rotationally symmetric structures, for which $\lambda_+$ would be
comparable with $\lambda_-$.
Figure~\ref{fig:curv-threshold} also shows the median value of $-\lambda_-$, computed on the 
$D_{353}$ map within the  mask defined by the contrast parameter $\zeta = 1$  in Sect.~\ref{subsec:mask}, and the power law fit

\begin{equation}
\label{eq:median_curvature}
\frac{C_{\rm M}}{[{\rm MJy\,sr^{-1} \, deg^{-2}}]} = (D_{353}^{\rm BG} / [0.07\,{\rm MJy \,sr^{-1} }])^{0.9}.
\end{equation}
The comparison of $C_{\rm T}$ and $C_{\rm M}$ shows that our selection of ridges relies mainly on the brightness contrast. 
The  threshold $C_{\rm T}$ on the curvature introduces a significant selection of sky pixels only for structures at high Galactic latitudes.

\subsection{Comparison with an alternative method}
\label{subsec:inertia}

To test the robustness of our methodology in finding the orientation $\theta$ of the ridges, we use 
an independent algorithm  to compute the orientation of structures in the dust map. 
\citet{Hennebelle13} used the inertia matrix of the gas density to analyse structures in his MHD simulations.
We adapted this method, computing the inertia matrix of the dust $D_{353}$ map. 
The off-diagonal coefficients of the inertia matrix are defined as 

\begin{equation}
\label{eq:inertia}
I_{xy}=\sum_{A} \Delta x (i) \, \Delta y (i) \, K (i) \, D_{353}(i),
\end{equation}
where the sum is performed over sky pixels within an area $A$ centred on the pixels of a \healpix\ grid, and 
$\Delta x (i) $ and $\Delta y (i) $ are the 
offsets of the coordinates $x = b $ and $y=l \, \cos(b)$  of the pixel $i$ 
with respect to the barycentre of the dust emission  $x_{\rm c}$ and
$y_{\rm c}$.
The equations for the diagonal terms of the inertia matrix are obtained by substitution of the $\Delta x (i) \, \Delta y (i)$ product by the squares
of $\Delta x (i) $ and $\Delta y (i) $ in Eq.~(\ref{eq:inertia}). The coordinates of the barycentre are computed with the equations

\begin{align}
\label{eq:barycenter}
& x_{\rm c}=\sum_{A} x (i) \, K (i) \, D_{353}(i); 
& y_{\rm c}=\sum_{A} y (i) \, K (i) \, D_{353}(i).
\end{align}

In Eqs.~(\ref{eq:inertia}) and (\ref{eq:barycenter}),   $D_{353}(i)$ is the dust emission at pixel $i$ 
and $ K (i) $ is a  kernel, which depends on the distance to the central pixel of  the area $A$, defining the angular resolution of the
computation of the inertia matrix. We use a Gaussian kernel with a full width at half maximum of $15\arcmin$, which matches the angular resolution of the 
 $D_{353}$ map used for the computation of the Hessian matrix, and a circular area $A$ with a diameter of $30\arcmin$.
A good sampling of the brightness structure within the area $A$  is
needed to compute  the sum in Eq.~(\ref{eq:inertia}) with the required 
accuracy.  The orientation angle
$\theta_{\rm I} $  of the brightness structures is 
computed using Eq.~(\ref{eq:dirfil}), where the terms of the Hessian matrix are substituted by those of the inertia matrix.
When the inertia matrix is computed on a small number of pixels, the
map of $\theta_{\rm I}$
shows systematic patterns associated with the \healpix\ pixelization.
To circumvent this problem,  we use here for  $D_{353}$ the full resolution map of \citet{planck2013-p06b} at $5\arcmin$ resolution on a \healpix\ grid with $N_{\rm side} = 2048$.  

\begin{figure*}[!h]
\centering
  \begin{tabular}{r l}
     \includegraphics[width=8.8cm,trim=40 0 22 0,clip=true]{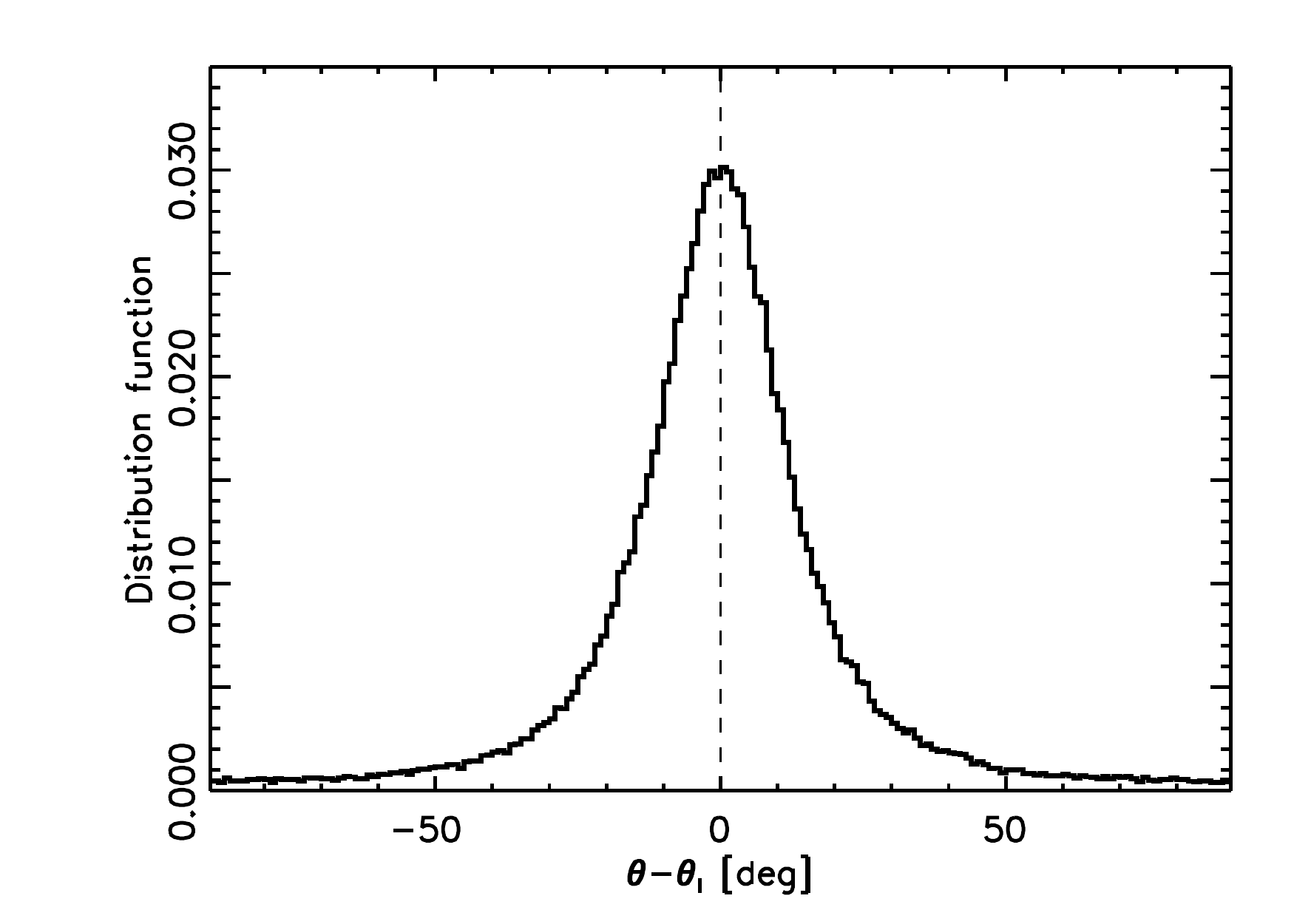}
&\includegraphics[width=8.8cm,trim=40 0 22 0,clip=true]{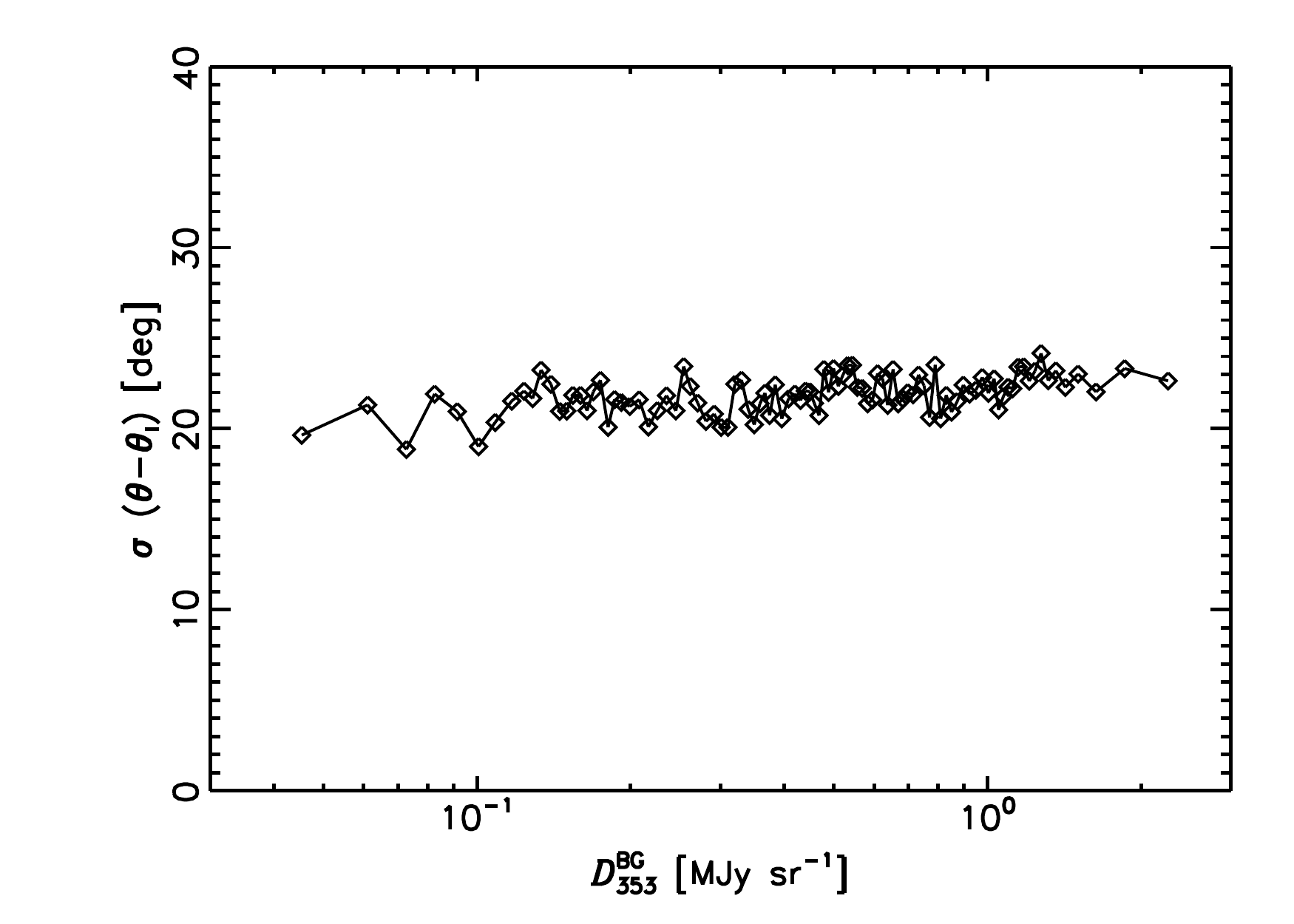}\\
    \end{tabular}
\caption{{\it Left}: distribution function of the difference between the angles of brightness structures in the dust emission map 
 derived from the Hessian ($\theta$) and inertia ($\theta_{\rm I}$) analysis. This comparison between the two
   independent algorithms provides the uncertainty of the Hessian angles that is smaller than the dispersion 
  between the Hessian and polarization angles (Sect.~\ref{sec:alignment}).
{\it Right}: the standard deviation of the difference between the
Hessian and inertia angles ($\theta -  \theta_{\rm I}$) computed within bins
of $D^{\rm BG}_{353}$.}
\label{fig:HesIN}
\end{figure*}

We computed the standard deviations of $\theta-\theta_{\rm I}$ for our selected ridges, within bins of sky brightness with equal numbers of sky pixels.
We found a fixed value of about $20^\circ$, independent of the sky
brightness (see the bottom panel in Fig.~\ref{fig:HesIN}).
The probability distribution of $\theta-\theta_{\rm I}$ computed over
our data selection is plotted in the top panel of Fig.~\ref{fig:HesIN}. It is much narrower than the distributions computed when
comparing the Hessian and polarization angles in Sect.~\ref{sec:alignment}. Finally, we also checked
that the results of our data analysis are robust against removing from the analysis the
pixels in the wings of the distribution of $\theta-\theta_{\rm I}$.    

\section{Gaussian models}
\label{appendix:gaussian}

\begin{figure}[!h]
\centering
\centerline{\includegraphics[width=8.8cm,trim=40 0 20 0,clip=true]{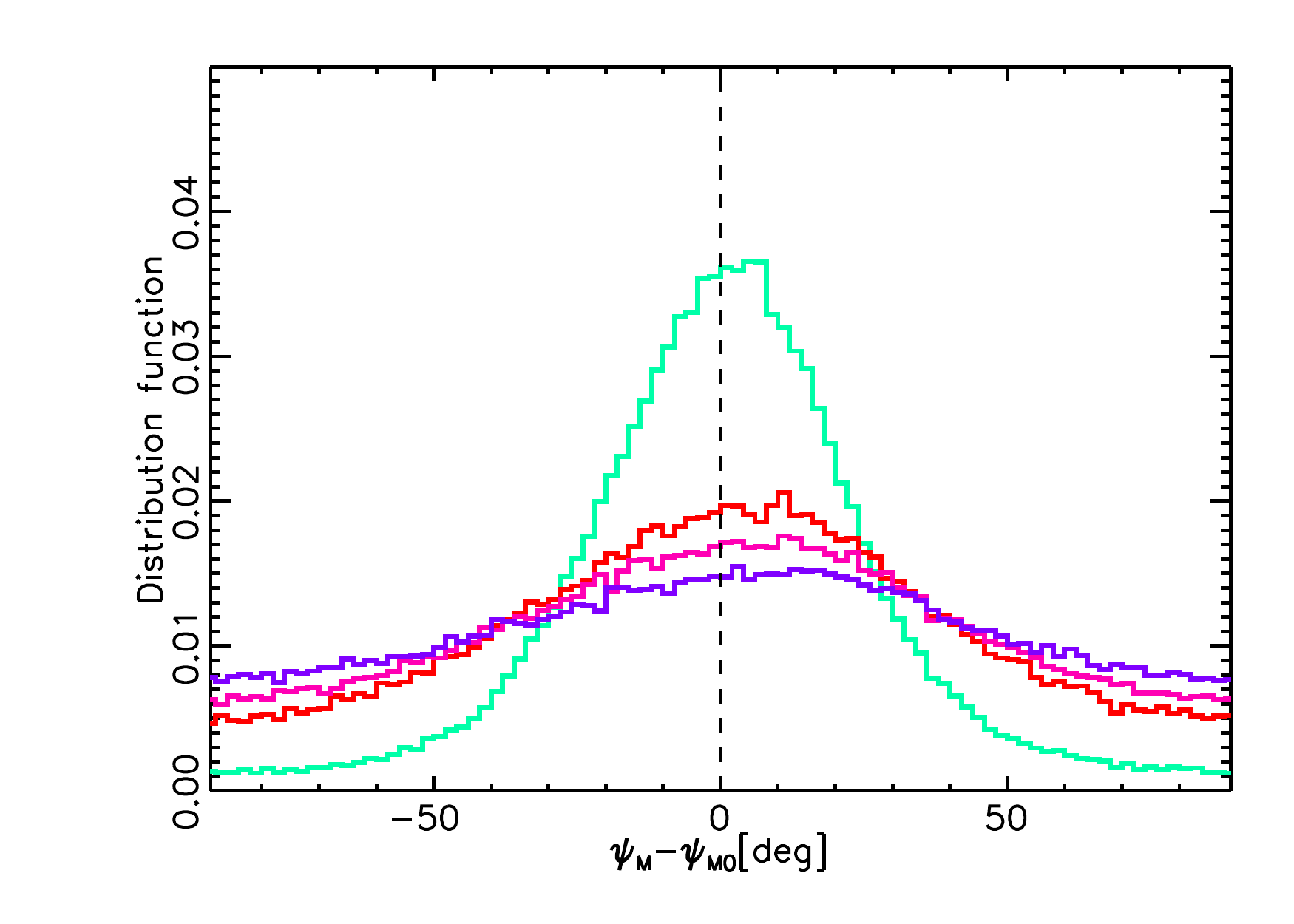}}
\caption{Distribution functions for the Gaussian models for four values of the dispersion of the $\vec{V}_{\rm M}$ direction about $\vec{V}_{\rm M0}$ . The dispersion of angles in the models is parametrized by the value of $f_{\rm M}$ ($\sigma_{\rm M}$). It is the strongest in the
  purple case, $f_{\rm M} = 1.5$ ($\sigma_{\rm M} = 38^{\circ}$). The other curves correspond to $f_{\rm M} = 1.2$ ($\sigma_{\rm M} = 33^{\circ}$) in magenta, $f_{\rm M} = 1.0$ ($\sigma_{\rm M} = 29^{\circ}$) in red, and the weakest, $f_{\rm M}=0.5$ ($\sigma_{\rm M} = 15^{\circ}$), in green. Each DF is
computed over the same set of pixels used for the data analysis. The plot shows the results for
the index of the power spectrum $-1.5$.}
\label{fig:Gauss_models}
\end{figure}

\begin{figure*}[!h]
  \centering
  \begin{tabular}{r l}
      \includegraphics[width=8.8cm,trim=40 0 20 0,clip=true]{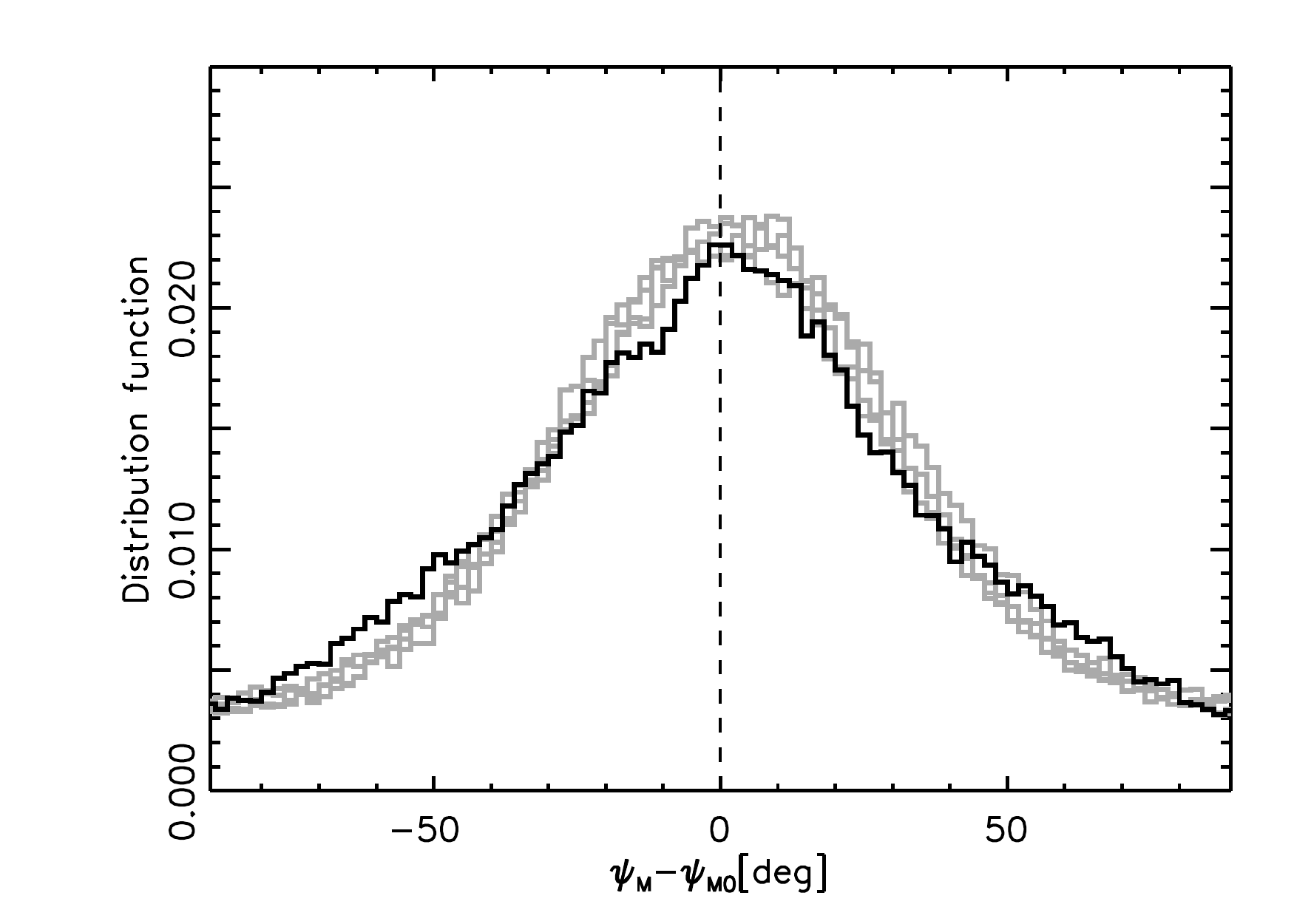}
      & \includegraphics[width=8.8cm,trim=40 0 20 0,clip=true]{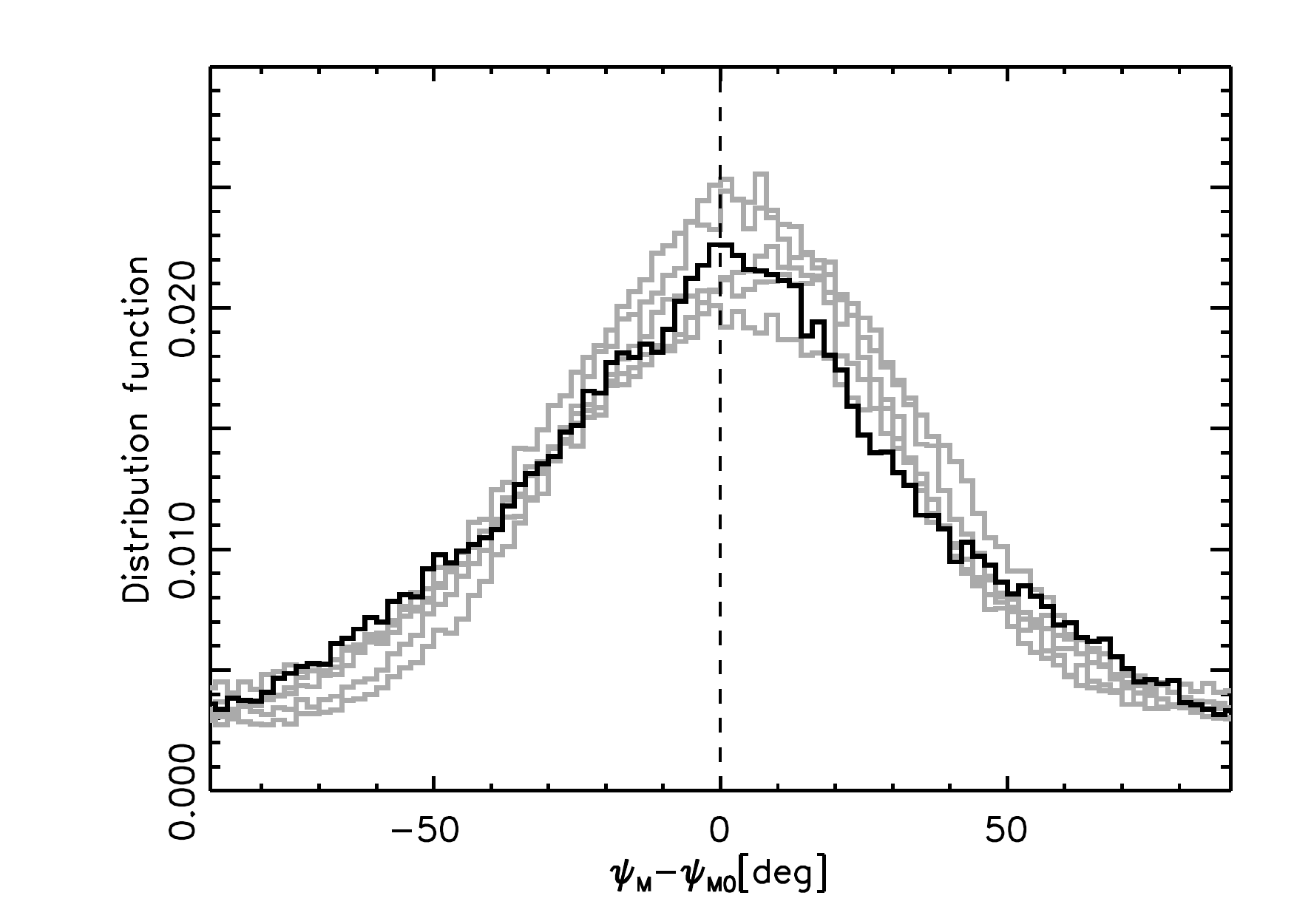}\\
    \end{tabular}
       \caption{Comparison of the distribution functions of relative orientations for the data, in black (see Fig.~\ref{fig:diff_st_bg}), and the Gaussian models with $f_{\rm M} = 0.8$, in grey, only considering the selected pixels. The two plots show five realizations of the Gaussian models with an angular power spectrum of power law index $\alpha_{\rm M}=-1.5$ ({\it left}) and $\alpha_{\rm M}=-2.0$ ({\it right}). Because of sample variance, the steeper power spectrum produces asymmetries and skewness in the distributions.}
      \label{fig:variance}
\end{figure*}

\begin{figure}[!h]
\centering
\centerline{\includegraphics[width=8.8cm,trim=40 0 20 0,clip=true]{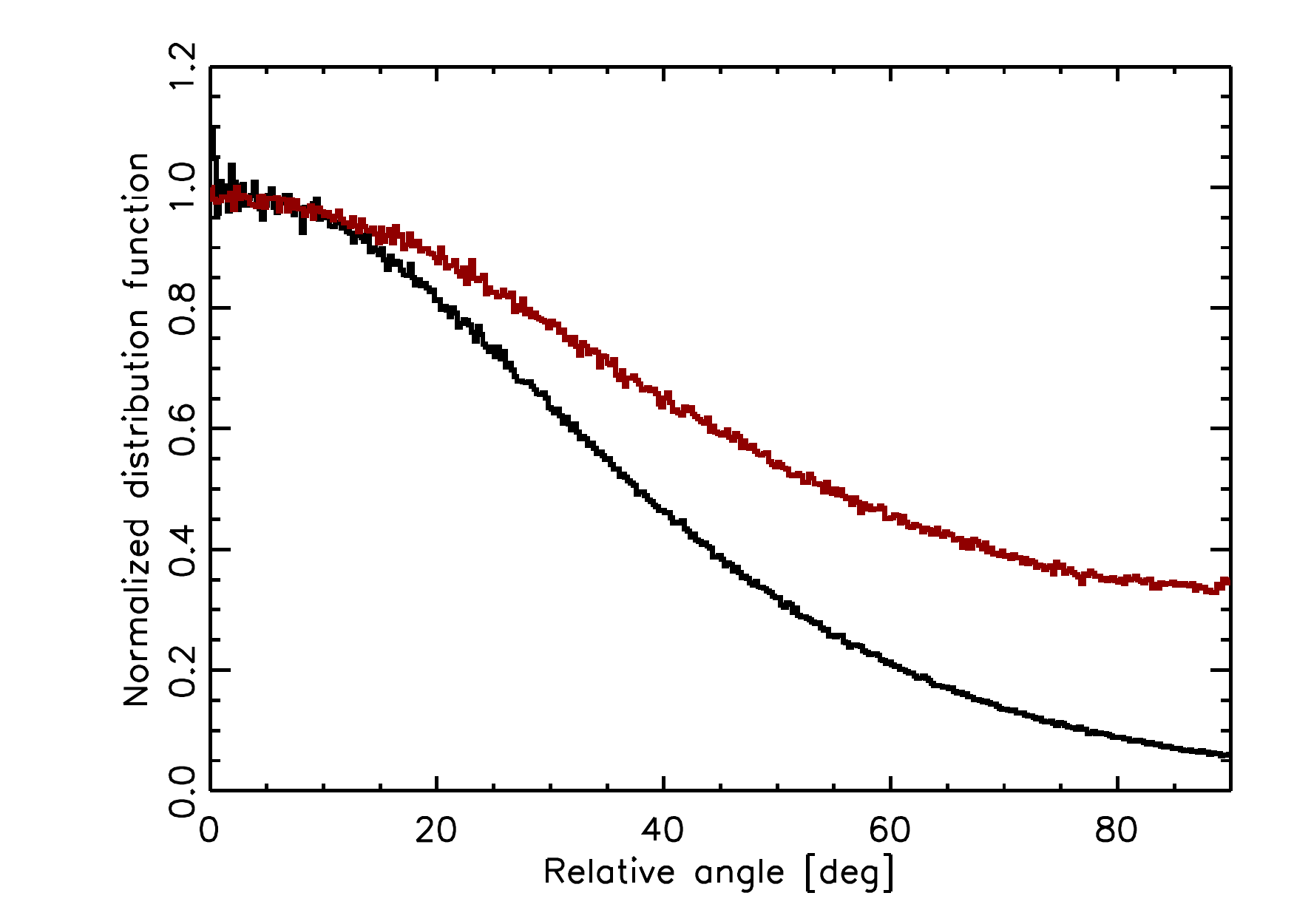}}
\caption{The distribution of angles between $\vec{V}_{\rm M} $ and $\vec{V}_{\rm M0}$ per unit solid angle for $f_{\rm M} = 1.2$ and $\alpha_{\rm M} = -1.5$, in black, is compared to the distribution of angles 
between the projections of the two vectors on the plane of the sky, in red. Because of projection effects the dispersion in 2D is larger than in 3D.}
\label{fig:histo_2vs3D}
\end{figure}

In this Appendix, we present the Gaussian models that are compared with the data in Sects.~\ref{sec:magfield}, \ref{sec:alignment} and \ref{sec:align_nh}.
These models provide a framework to quantify projection effects and interpret the DFs of the 
angle between the magnetic field and the brightness ridges on the sky. We describe the models in Sect.~\ref{sec:description} and their application in Sect.~\ref{sec:application}.

\subsection{Description of the models} 
\label{sec:description}

Each model is built from 3D vectors $\vec{V}_{\rm M} $ with a Gaussian distribution of orientations about a mean direction $\vec{V}_{\rm M0}$. 
The three components of $\vec{V}_{\rm M}$  are independent realizations of a Gaussian field on a full-sky \healpix\ grid, at a $15'$ resolution, with an angular power spectrum
having a power law of index $\alpha_{\rm M}$, to which we add the components of $\vec{V}_{\rm M0}$ \footnote{These realizations are computed with the procedure {\tt{SYNFAST}} of \healpix\ at $N_{\rm side}=512$.}.
The mean of $\vec{V}_{\rm M}$ is  $\vec{V}_{\rm M0}$.  
We computed several sets of models for a range 
of ratios $f_{\rm M}$ between the standard deviation of $|\vec{V}_{\rm M}|$ and  $|\vec{V}_{\rm M0}|$, and for different values of the spectral index $\alpha_{\rm M}$.  
The ratio $f_{\rm M}$ determines the amplitude of the scatter of $\vec{V}_{\rm M}$ with respect to $\vec{V}_{\rm M0}$, while 
the spectral index $\alpha_{\rm M}$ controls the correlation across the sky of the direction of $\vec{V}_{\rm M} $.
The distribution function of angles between $\vec{V}_{\rm M}$ and $\vec{V}_{\rm M0}$ per unit solid angle 
is close to Gaussian with a standard deviation, $\sigma_{\rm M}$, that increases from $9\pdeg7$ to  $29\pdeg5$ and $38^\circ$ for $f_{\rm M}=0.3$, $1.0$ and $1.5$.

For each model, we compute maps of the 
projections of $\vec{V}_{\rm M}$ and $\vec{V}_{\rm M0}$ onto the sky with respect to the local direction of the north Galactic pole, $\psi_{\rm M}$ and $\psi_{\rm M0}$, respectively.
We use the trigonometric 
formula in Eq.~(\ref{eq:diffang_st_bg}) to compute the difference between these angle maps with $\alpha=\psi_{\rm M}$ and $\beta=\psi_{\rm M0}$. Over the \healpix\ grid, we sample
uniformly the relative angle between the line of sight and the mean vector  $\vec{V}_{\rm M0}$.

DFs of the difference of angles between the projections of 
$\vec{V}_{\rm M}$ and $\vec{V}_{\rm M0}$ onto the plane of the sky, computed with the same mask as that used in the data analysis,
are presented, for a few models, in Fig.~\ref{fig:Gauss_models}.
We show models with increasing values of $f_{\rm M}$ for a spectral index $\alpha_{\rm M} = -1.5$. Our choice of this spectral index is arbitrary because it is not tuned to reproduce power spectra computed from observations. When we use a steeper power spectrum the results are similar, although less regular and slightly asymmetric about the origin. This asymmetry arises from sample variance because, for decreasing values of $\alpha_{\rm M}$, the distribution of the $\vec{V}_{\rm M}$ direction  is  less  sampled as a result of the correlation of $\vec{V}_{\rm M}$ on the sky. In Fig.~\ref{fig:variance} we illustrate the effect of sample variance comparing the DFs of relative orientations of the data (see Fig.~\ref{fig:diff_st_bg}) with those of multiple Gaussian realizations for both $\alpha_{\rm M} = -1.5$ and $\alpha_{\rm M} = -2.0$, with a fixed value of $f_{\rm M}=0.8$. Because of sample variance, the models for the steeper power spectrum show a larger asymmetry and skewness than those for the shallower power spectrum, and than what is observed for the data. 
Therefore, in the analysis, we make use of the Gaussian models with an angular power spectrum of power law index $\alpha_{\rm M}=-1.5$.      
In Fig.~\ref{fig:histo_2vs3D}, the DF of the angle between the projections of $\vec{V}_{\rm M}$ and $\vec{V}_{\rm M0}$ is shown to be broader than that computed in 3D.
The DFs in Fig.~\ref{fig:Gauss_models}  show a pedestal that extends to $-90^\circ$ and $90^\circ$, corresponding to positions in the sky where $\vec{V}_{\rm M}$ or $\vec{V}_{\rm M0}$
are close to the line-of-sight orientation. In this case $\vec{V}_{\rm M}$ or $\vec{V}_{\rm M0}$ are along the line of sight and
the angle between the projections of the two vectors onto the plane of the sky can have any value
independently of the 3D angle.  The impact of projection on the DF of polarization angles has also been quantified with numerical simulations \citep{Falceta08}. 

\subsection{Use of the models}
\label{sec:application}

The Gaussian models are used in different contexts, where the vectors $\vec{V}_{\rm M}$ and $\vec{V}_{\rm M0}$ have different physical interpretations. 
In Sect.~\ref{sec:magfield}, we use the Gaussian models to interpret the DF of relative orientations between the local and
mean magnetic field directions. In this case,  $\vec{V}_{\rm M}$ is the direction of the local field and  $\vec{V}_{\rm M0}$ that of the background field assumed to be fixed over the sky.
Here, $f_{\rm M}$ is the ratio between the random, referred to as turbulent, and mean components of the magnetic field. 
In Sect.~\ref{sec:alignment},  we use the Gaussian models to interpret the DF of relative orientations between the magnetic field and ridges. In this case,  
$\vec{V}_{\rm M} $ is the orientation of the field and $\vec{V}_{\rm M0}$ that of the ridge. Here, the parameter of the models is $\sigma_{\rm M}$,  which measures the degree of alignment between the field and the ridges.
In both cases, the statistics over the sky provide a homogeneous sampling of the relative orientation of the line of sight and the reference (the background magnetic field or the filaments) even when we do not  
introduce any variation of  $\vec{V}_{\rm M0}$ on the sky. 

\begin{figure*}[!h]
  \centering
  \begin{tabular}{r l}
      \includegraphics[width=8.5cm,trim=5 0 5 0,clip=true]{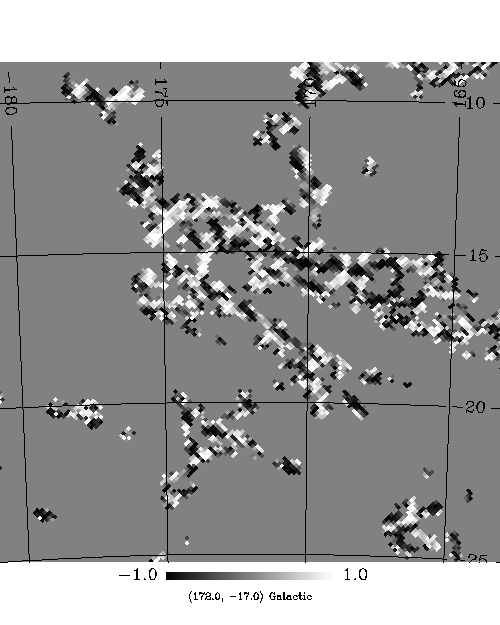}
      & \includegraphics[width=8.5cm,trim=5 0 5 0,clip=true]{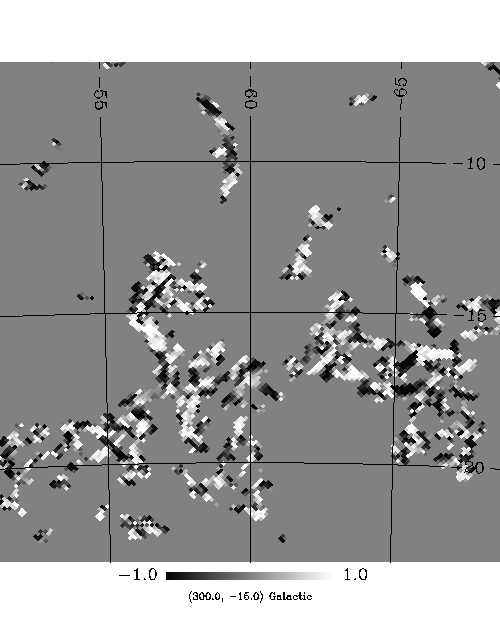}\\
    \end{tabular}
\caption{Maps of relative orientations between the orientation angle, $\theta$, of the selected structures and the projection of $\vec{V}_{\rm M}$ onto the plane of the sky, for the Gaussian model in which $\vec{V}_{\rm M0}$ is set to 0. The mask described in Sect.~\ref{subsec:mask} is applied. We show the two fields of view presented in Fig.~\ref{fig:tau_mus}: Taurus ({\it left}) and Chamaeleon ({\it right}) molecular complexes. The white (black) structures correspond to alignment (anti-alignment) between the projection of $\vec{V}_{\rm M}$ onto the plane of the sky and $\theta$. This figure shows that projection effects cannot account for the elongated and coherent structures in the relative orientation maps of Fig.~\ref{fig:tau_mus} ({\it right}).}
\label{fig:relativemodel}
\end{figure*}

To study the orientations of the filaments with respect to the magnetic field, we run models with three different configurations for $\vec{V}_{\rm M0}$. In the first case, we fix the direction of $\vec{V}_{\rm M0}$. In the second case, we use two directions of $\vec{V}_{\rm M0}$ perpendicular to each other, as explained in Sect.~\ref{subsec:bimodal}. In the third case, we run a model in which we set $\vec{V}_{\rm M0}$ to 0 to quantify the impact of the projection when 
the magnetic field and the matter structures orientations are uncorrelated. 
This third configuration allows us to test whether or not, in the highest column density regions, the correlation between the magnetic field and the distribution of interstellar matter is lost. 
In Fig.~\ref{fig:relativemodel}, we present the maps of relative orientations between $\psi_{\rm M}$ and $\theta$, in the Taurus and Chamaeleon fields shown in Fig.~\ref{fig:tau_mus}, for one realization of this model after applying the mask described in Sect.~\ref{subsec:mask}. Although these maps show black and white patterns, which resemble the data, we do not find any elongated and coherent structures such as those in Fig.~\ref{fig:tau_mus}, like the Musca filament.

\raggedright

\end{document}

%% file: polar_definitions.tex
\newcommand{\StokesI}{I}                    
\newcommand{\StokesQ}{Q}                    
\newcommand{\StokesU}{U}                    

%% file: Planck.tex
\def\setsymbol#1#2{\expandafter\def\csname #1\endcsname{#2}}
\def\getsymbol#1{\csname #1\endcsname}

\def\Planck{\textit{Planck}}

\def\all2013resultspapers{\nocite{planck2013-p01, planck2013-p02, planck2013-p02a, planck2013-p02d, planck2013-p02b, planck2013-p03, planck2013-p03c, planck2013-p03f, planck2013-p03d, planck2013-p03e, planck2013-p01a, planck2013-p06, planck2013-p03a, planck2013-pip88, planck2013-p08, planck2013-p11, planck2013-p12, planck2013-p13, planck2013-p14, planck2013-p15, planck2013-p05b, planck2013-p17, planck2013-p09, planck2013-p09a, planck2013-p20, planck2013-p19, planck2013-pipaberration, planck2013-p05, planck2013-p05a, planck2013-pip56, planck2013-p06b}}

\newbox\tablebox    \newdimen\tablewidth
\def\leaderfil{\leaders\hbox to 5pt{\hss.\hss}\hfil}
\def\tablenote#1 #2\par{\begingroup \parindent=0.8em
    \abovedisplayshortskip=0pt\belowdisplayshortskip=0pt
    \noindent
    $$\hss\vbox{\hsize\tablewidth \hangindent=\parindent \hangafter=1 \noindent
    \hbox to \parindent{$^#1$\hss}\strut#2\strut\par}\hss$$
    \endgroup}

\def\L2{\ifmmode L_2\else $L_2$\fi}

\def\DeltaT{\ifmmode \Delta T\else $\Delta T$\fi}
\def\deltat{\ifmmode \Delta t\else $\Delta t$\fi}
\def\fknee{\ifmmode f_{\rm knee}\else $f_{\rm knee}$\fi}
\def\Fmax{\ifmmode F_{\rm max}\else $F_{\rm max}$\fi}
\def\solar{\ifmmode{\rm M}_{\mathord\odot}\else${\rm M}_{\mathord\odot}$\fi}
\def\Msolar{\ifmmode{\rm M}_{\mathord\odot}\else${\rm M}_{\mathord\odot}$\fi}
\def\Lsolar{\ifmmode{\rm L}_{\mathord\odot}\else${\rm L}_{\mathord\odot}$\fi}

\def\inv{\ifmmode^{-1}\else$^{-1}$\fi}
\def\mo{\ifmmode^{-1}\else$^{-1}$\fi}
\def\sup#1{\ifmmode ^{\rm #1}\else $^{\rm #1}$\fi}
\def\expo#1{\ifmmode \times 10^{#1}\else $\times 10^{#1}$\fi}
\def\,{\thinspace}
\def\lsim{\mathrel{\raise .4ex\hbox{\rlap{$<$}\lower 1.2ex\hbox{$\sim$}}}}
\def\gsim{\mathrel{\raise .4ex\hbox{\rlap{$>$}\lower 1.2ex\hbox{$\sim$}}}}

\def\simprop{\mathrel{\raise .4ex\hbox{\rlap{$\propto$}\lower 1.2ex\hbox{$\sim$}}}}
\def\deg{\ifmmode^\circ\else$^\circ$\fi}
\def\pdeg{\ifmmode $\setbox0=\hbox{$^{\circ}$}\rlap{\hskip.11\wd0 .}$^{\circ}
          \else \setbox0=\hbox{$^{\circ}$}\rlap{\hskip.11\wd0 .}$^{\circ}$\fi}
\def\arcs{\ifmmode {^{\scriptstyle\prime\prime}}
          \else $^{\scriptstyle\prime\prime}$\fi}
\def\arcm{\ifmmode {^{\scriptstyle\prime}}
          \else $^{\scriptstyle\prime}$\fi}
\newdimen\sa  \newdimen\sb
\def\parcs{\sa=.07em \sb=.03em
     \ifmmode \hbox{\rlap{.}}^{\scriptstyle\prime\kern -\sb\prime}\hbox{\kern -\sa}
     \else \rlap{.}$^{\scriptstyle\prime\kern -\sb\prime}$\kern -\sa\fi}
\def\parcm{\sa=.08em \sb=.03em
     \ifmmode \hbox{\rlap{.}\kern\sa}^{\scriptstyle\prime}\hbox{\kern-\sb}
     \else \rlap{.}\kern\sa$^{\scriptstyle\prime}$\kern-\sb\fi}
\def\ra[#1 #2 #3.#4]{#1\sup{h}#2\sup{m}#3\sup{s}\llap.#4}
\def\dec[#1 #2 #3.#4]{#1\deg#2\arcm#3\arcs\llap.#4}
\def\deco[#1 #2 #3]{#1\deg#2\arcm#3\arcs}
\def\rra[#1 #2]{#1\sup{h}#2\sup{m}}

\def\dots{\relax\ifmmode \ldots\else $\ldots$\fi}
\def\WHzsr{\ifmmode $W\,Hz\mo\,sr\mo$\else W\,Hz\mo\,sr\mo\fi}
\def\mHz{\ifmmode $\,mHz$\else \,mHz\fi}
\def\GHz{\ifmmode $\,GHz$\else \,GHz\fi}
\def\mKs{\ifmmode $\,mK\,s$^{1/2}\else \,mK\,s$^{1/2}$\fi}
\def\muKs{\ifmmode \,\mu$K\,s$^{1/2}\else \,$\mu$K\,s$^{1/2}$\fi}
\def\muKRJs{\ifmmode \,\mu$K$_{\rm RJ}$\,s$^{1/2}\else \,$\mu$K$_{\rm RJ}$\,s$^{1/2}$\fi}
\def\muKHz{\ifmmode \,\mu$K\,Hz$^{-1/2}\else \,$\mu$K\,Hz$^{-1/2}$\fi}
\def\MJysr{\ifmmode \,$MJy\,sr\mo$\else \,MJy\,sr\mo\fi}
\def\MJysrmK{\ifmmode \,$MJy\,sr\mo$\,mK$_{\rm CMB}\mo\else \,MJy\,sr\mo\,mK$_{\rm CMB}\mo$\fi}
\def\microns{\ifmmode \,\mu$m$\else \,$\mu$m\fi}

\def\muK{\ifmmode \,\mu$K$\else \,$\mu$\hbox{K}\fi}
\def\microK{\ifmmode \,\mu$K$\else \,$\mu$\hbox{K}\fi}
\def\muW{\ifmmode \,\mu$W$\else \,$\mu$\hbox{W}\fi}
\def\kms{\ifmmode $\,km\,s$^{-1}\else \,km\,s$^{-1}$\fi}
\def\kmsMpc{\ifmmode $\,\kms\,Mpc\mo$\else \,\kms\,Mpc\mo\fi}
\setsymbol{LFI:center:frequency:70GHz:units}{70.3\,GHz}
\setsymbol{LFI:center:frequency:44GHz:units}{44.1\,GHz}
\setsymbol{LFI:center:frequency:30GHz:units}{28.5\,GHz}

\setsymbol{LFI:center:frequency:70GHz}{70.3}
\setsymbol{LFI:center:frequency:44GHz}{44.1}
\setsymbol{LFI:center:frequency:30GHz}{28.5}

\setsymbol{LFI:center:frequency:LFI18:Rad:M:units}{71.7\GHz}
\setsymbol{LFI:center:frequency:LFI19:Rad:M:units}{67.5\GHz}
\setsymbol{LFI:center:frequency:LFI20:Rad:M:units}{69.2\GHz}
\setsymbol{LFI:center:frequency:LFI21:Rad:M:units}{70.4\GHz}
\setsymbol{LFI:center:frequency:LFI22:Rad:M:units}{71.5\GHz}
\setsymbol{LFI:center:frequency:LFI23:Rad:M:units}{70.8\GHz}
\setsymbol{LFI:center:frequency:LFI24:Rad:M:units}{44.4\GHz}
\setsymbol{LFI:center:frequency:LFI25:Rad:M:units}{44.0\GHz}
\setsymbol{LFI:center:frequency:LFI26:Rad:M:units}{43.9\GHz}
\setsymbol{LFI:center:frequency:LFI27:Rad:M:units}{28.3\GHz}
\setsymbol{LFI:center:frequency:LFI28:Rad:M:units}{28.8\GHz}
\setsymbol{LFI:center:frequency:LFI18:Rad:S:units}{70.1\GHz}
\setsymbol{LFI:center:frequency:LFI19:Rad:S:units}{69.6\GHz}
\setsymbol{LFI:center:frequency:LFI20:Rad:S:units}{69.5\GHz}
\setsymbol{LFI:center:frequency:LFI21:Rad:S:units}{69.5\GHz}
\setsymbol{LFI:center:frequency:LFI22:Rad:S:units}{72.8\GHz}
\setsymbol{LFI:center:frequency:LFI23:Rad:S:units}{71.3\GHz}
\setsymbol{LFI:center:frequency:LFI24:Rad:S:units}{44.1\GHz}
\setsymbol{LFI:center:frequency:LFI25:Rad:S:units}{44.1\GHz}
\setsymbol{LFI:center:frequency:LFI26:Rad:S:units}{44.1\GHz}
\setsymbol{LFI:center:frequency:LFI27:Rad:S:units}{28.5\GHz}
\setsymbol{LFI:center:frequency:LFI28:Rad:S:units}{28.2\GHz}

\setsymbol{LFI:center:frequency:LFI18:Rad:M}{71.7}
\setsymbol{LFI:center:frequency:LFI19:Rad:M}{67.5}
\setsymbol{LFI:center:frequency:LFI20:Rad:M}{69.2}
\setsymbol{LFI:center:frequency:LFI21:Rad:M}{70.4}
\setsymbol{LFI:center:frequency:LFI22:Rad:M}{71.5}
\setsymbol{LFI:center:frequency:LFI23:Rad:M}{70.8}
\setsymbol{LFI:center:frequency:LFI24:Rad:M}{44.4}
\setsymbol{LFI:center:frequency:LFI25:Rad:M}{44.0}
\setsymbol{LFI:center:frequency:LFI26:Rad:M}{43.9}
\setsymbol{LFI:center:frequency:LFI27:Rad:M}{28.3}
\setsymbol{LFI:center:frequency:LFI28:Rad:M}{28.8}
\setsymbol{LFI:center:frequency:LFI18:Rad:S}{70.1}
\setsymbol{LFI:center:frequency:LFI19:Rad:S}{69.6}
\setsymbol{LFI:center:frequency:LFI20:Rad:S}{69.5}
\setsymbol{LFI:center:frequency:LFI21:Rad:S}{69.5}
\setsymbol{LFI:center:frequency:LFI22:Rad:S}{72.8}
\setsymbol{LFI:center:frequency:LFI23:Rad:S}{71.3}
\setsymbol{LFI:center:frequency:LFI24:Rad:S}{44.1}
\setsymbol{LFI:center:frequency:LFI25:Rad:S}{44.1}
\setsymbol{LFI:center:frequency:LFI26:Rad:S}{44.1}
\setsymbol{LFI:center:frequency:LFI27:Rad:S}{28.5}
\setsymbol{LFI:center:frequency:LFI28:Rad:S}{28.2}

\setsymbol{LFI:white:noise:sensitivity:70GHz:units}{134.7\muKs}
\setsymbol{LFI:white:noise:sensitivity:44GHz:units}{164.7\muKs}
\setsymbol{LFI:white:noise:sensitivity:30GHz:units}{143.4\muKs}

\setsymbol{LFI:white:noise:sensitivity:70GHz}{134.7}
\setsymbol{LFI:white:noise:sensitivity:44GHz}{164.7}
\setsymbol{LFI:white:noise:sensitivity:30GHz}{143.4}

\setsymbol{LFI:white:noise:sensitivity:LFI18:Rad:M:units}{512.0\muKs}
\setsymbol{LFI:white:noise:sensitivity:LFI19:Rad:M:units}{581.4\muKs}
\setsymbol{LFI:white:noise:sensitivity:LFI20:Rad:M:units}{590.8\muKs}
\setsymbol{LFI:white:noise:sensitivity:LFI21:Rad:M:units}{455.2\muKs}
\setsymbol{LFI:white:noise:sensitivity:LFI22:Rad:M:units}{492.0\muKs}
\setsymbol{LFI:white:noise:sensitivity:LFI23:Rad:M:units}{507.7\muKs}
\setsymbol{LFI:white:noise:sensitivity:LFI24:Rad:M:units}{462.2\muKs}
\setsymbol{LFI:white:noise:sensitivity:LFI25:Rad:M:units}{413.6\muKs}
\setsymbol{LFI:white:noise:sensitivity:LFI26:Rad:M:units}{478.6\muKs}
\setsymbol{LFI:white:noise:sensitivity:LFI27:Rad:M:units}{277.7\muKs}
\setsymbol{LFI:white:noise:sensitivity:LFI28:Rad:M:units}{312.3\muKs}
\setsymbol{LFI:white:noise:sensitivity:LFI18:Rad:S:units}{465.7\muKs}
\setsymbol{LFI:white:noise:sensitivity:LFI19:Rad:S:units}{555.6\muKs}
\setsymbol{LFI:white:noise:sensitivity:LFI20:Rad:S:units}{623.2\muKs}
\setsymbol{LFI:white:noise:sensitivity:LFI21:Rad:S:units}{564.1\muKs}
\setsymbol{LFI:white:noise:sensitivity:LFI22:Rad:S:units}{534.4\muKs}
\setsymbol{LFI:white:noise:sensitivity:LFI23:Rad:S:units}{542.4\muKs}
\setsymbol{LFI:white:noise:sensitivity:LFI24:Rad:S:units}{399.2\muKs}
\setsymbol{LFI:white:noise:sensitivity:LFI25:Rad:S:units}{392.6\muKs}
\setsymbol{LFI:white:noise:sensitivity:LFI26:Rad:S:units}{418.6\muKs}
\setsymbol{LFI:white:noise:sensitivity:LFI27:Rad:S:units}{302.9\muKs}
\setsymbol{LFI:white:noise:sensitivity:LFI28:Rad:S:units}{285.3\muKs}

\setsymbol{LFI:white:noise:sensitivity:LFI18:Rad:M}{512.0}
\setsymbol{LFI:white:noise:sensitivity:LFI19:Rad:M}{581.4}
\setsymbol{LFI:white:noise:sensitivity:LFI20:Rad:M}{590.8}
\setsymbol{LFI:white:noise:sensitivity:LFI21:Rad:M}{455.2}
\setsymbol{LFI:white:noise:sensitivity:LFI22:Rad:M}{492.0}
\setsymbol{LFI:white:noise:sensitivity:LFI23:Rad:M}{507.7}
\setsymbol{LFI:white:noise:sensitivity:LFI24:Rad:M}{462.2}
\setsymbol{LFI:white:noise:sensitivity:LFI25:Rad:M}{413.6}
\setsymbol{LFI:white:noise:sensitivity:LFI26:Rad:M}{478.6}
\setsymbol{LFI:white:noise:sensitivity:LFI27:Rad:M}{277.7}
\setsymbol{LFI:white:noise:sensitivity:LFI28:Rad:M}{312.3}
\setsymbol{LFI:white:noise:sensitivity:LFI18:Rad:S}{465.7}
\setsymbol{LFI:white:noise:sensitivity:LFI19:Rad:S}{555.6}
\setsymbol{LFI:white:noise:sensitivity:LFI20:Rad:S}{623.2}
\setsymbol{LFI:white:noise:sensitivity:LFI21:Rad:S}{564.1}
\setsymbol{LFI:white:noise:sensitivity:LFI22:Rad:S}{534.4}
\setsymbol{LFI:white:noise:sensitivity:LFI23:Rad:S}{542.4}
\setsymbol{LFI:white:noise:sensitivity:LFI24:Rad:S}{399.2}
\setsymbol{LFI:white:noise:sensitivity:LFI25:Rad:S}{392.6}
\setsymbol{LFI:white:noise:sensitivity:LFI26:Rad:S}{418.6}
\setsymbol{LFI:white:noise:sensitivity:LFI27:Rad:S}{302.9}
\setsymbol{LFI:white:noise:sensitivity:LFI28:Rad:S}{285.3}

\setsymbol{LFI:knee:frequency:70GHz:units}{29.5\mHz}
\setsymbol{LFI:knee:frequency:44GHz:units}{56.2\mHz}
\setsymbol{LFI:knee:frequency:30GHz:units}{113.7\mHz}

\setsymbol{LFI:knee:frequency:70GHz}{29.5}
\setsymbol{LFI:knee:frequency:44GHz}{56.2}
\setsymbol{LFI:knee:frequency:30GHz}{113.7}

\setsymbol{LFI:knee:frequency:LFI18:Rad:M:units}{16.3\mHz}
\setsymbol{LFI:knee:frequency:LFI19:Rad:M:units}{15.1\mHz}
\setsymbol{LFI:knee:frequency:LFI20:Rad:M:units}{18.7\mHz}
\setsymbol{LFI:knee:frequency:LFI21:Rad:M:units}{37.2\mHz}
\setsymbol{LFI:knee:frequency:LFI22:Rad:M:units}{12.7\mHz}
\setsymbol{LFI:knee:frequency:LFI23:Rad:M:units}{34.6\mHz}
\setsymbol{LFI:knee:frequency:LFI24:Rad:M:units}{46.2\mHz}
\setsymbol{LFI:knee:frequency:LFI25:Rad:M:units}{24.9\mHz}
\setsymbol{LFI:knee:frequency:LFI26:Rad:M:units}{67.6\mHz}
\setsymbol{LFI:knee:frequency:LFI27:Rad:M:units}{187.4\mHz}
\setsymbol{LFI:knee:frequency:LFI28:Rad:M:units}{122.2\mHz}
\setsymbol{LFI:knee:frequency:LFI18:Rad:S:units}{17.7\mHz}
\setsymbol{LFI:knee:frequency:LFI19:Rad:S:units}{22.0\mHz}
\setsymbol{LFI:knee:frequency:LFI20:Rad:S:units}{8.7\mHz}
\setsymbol{LFI:knee:frequency:LFI21:Rad:S:units}{25.9\mHz}
\setsymbol{LFI:knee:frequency:LFI22:Rad:S:units}{15.8\mHz}
\setsymbol{LFI:knee:frequency:LFI23:Rad:S:units}{129.8\mHz}
\setsymbol{LFI:knee:frequency:LFI24:Rad:S:units}{100.9\mHz}
\setsymbol{LFI:knee:frequency:LFI25:Rad:S:units}{38.9\mHz}
\setsymbol{LFI:knee:frequency:LFI26:Rad:S:units}{58.9\mHz}
\setsymbol{LFI:knee:frequency:LFI27:Rad:S:units}{104.4\mHz}
\setsymbol{LFI:knee:frequency:LFI28:Rad:S:units}{40.7\mHz}

\setsymbol{LFI:knee:frequency:LFI18:Rad:M}{16.3}
\setsymbol{LFI:knee:frequency:LFI19:Rad:M}{15.1}
\setsymbol{LFI:knee:frequency:LFI20:Rad:M}{18.7}
\setsymbol{LFI:knee:frequency:LFI21:Rad:M}{37.2}
\setsymbol{LFI:knee:frequency:LFI22:Rad:M}{12.7}
\setsymbol{LFI:knee:frequency:LFI23:Rad:M}{34.6}
\setsymbol{LFI:knee:frequency:LFI24:Rad:M}{46.2}
\setsymbol{LFI:knee:frequency:LFI25:Rad:M}{24.9}
\setsymbol{LFI:knee:frequency:LFI26:Rad:M}{67.6}
\setsymbol{LFI:knee:frequency:LFI27:Rad:M}{187.4}
\setsymbol{LFI:knee:frequency:LFI28:Rad:M}{122.2}
\setsymbol{LFI:knee:frequency:LFI18:Rad:S}{17.7}
\setsymbol{LFI:knee:frequency:LFI19:Rad:S}{22.0}
\setsymbol{LFI:knee:frequency:LFI20:Rad:S}{8.7}
\setsymbol{LFI:knee:frequency:LFI21:Rad:S}{25.9}
\setsymbol{LFI:knee:frequency:LFI22:Rad:S}{15.8}
\setsymbol{LFI:knee:frequency:LFI23:Rad:S}{129.8}
\setsymbol{LFI:knee:frequency:LFI24:Rad:S}{100.9}
\setsymbol{LFI:knee:frequency:LFI25:Rad:S}{38.9}
\setsymbol{LFI:knee:frequency:LFI26:Rad:S}{58.9}
\setsymbol{LFI:knee:frequency:LFI27:Rad:S}{104.4}
\setsymbol{LFI:knee:frequency:LFI28:Rad:S}{40.7}

\setsymbol{LFI:slope:70GHz:units}{$-1.03$\mHz}
\setsymbol{LFI:slope:44GHz:units}{$-0.89$\mHz}
\setsymbol{LFI:slope:30GHz:units}{$-0.87$\mHz}

\setsymbol{LFI:slope:70GHz}{$-1.03$}
\setsymbol{LFI:slope:44GHz}{$-0.89$}
\setsymbol{LFI:slope:30GHz}{$-0.87$}

\setsymbol{LFI:slope:LFI18:Rad:M:units}{$-1.04$\mHz}
\setsymbol{LFI:slope:LFI19:Rad:M:units}{$-1.09$\mHz}
\setsymbol{LFI:slope:LFI20:Rad:M:units}{$-0.69$\mHz}
\setsymbol{LFI:slope:LFI21:Rad:M:units}{$-1.56$\mHz}
\setsymbol{LFI:slope:LFI22:Rad:M:units}{$-1.01$\mHz}
\setsymbol{LFI:slope:LFI23:Rad:M:units}{$-0.96$\mHz}
\setsymbol{LFI:slope:LFI24:Rad:M:units}{$-0.83$\mHz}
\setsymbol{LFI:slope:LFI25:Rad:M:units}{$-0.91$\mHz}
\setsymbol{LFI:slope:LFI26:Rad:M:units}{$-0.95$\mHz}
\setsymbol{LFI:slope:LFI27:Rad:M:units}{$-0.87$\mHz}
\setsymbol{LFI:slope:LFI28:Rad:M:units}{$-0.88$\mHz}
\setsymbol{LFI:slope:LFI18:Rad:S:units}{$-1.15$\mHz}
\setsymbol{LFI:slope:LFI19:Rad:S:units}{$-1.00$\mHz}
\setsymbol{LFI:slope:LFI20:Rad:S:units}{$-0.95$\mHz}
\setsymbol{LFI:slope:LFI21:Rad:S:units}{$-0.92$\mHz}
\setsymbol{LFI:slope:LFI22:Rad:S:units}{$-1.01$\mHz}
\setsymbol{LFI:slope:LFI23:Rad:S:units}{$-0.95$\mHz}
\setsymbol{LFI:slope:LFI24:Rad:S:units}{$-0.73$\mHz}
\setsymbol{LFI:slope:LFI25:Rad:S:units}{$-1.16$\mHz}
\setsymbol{LFI:slope:LFI26:Rad:S:units}{$-0.79$\mHz}
\setsymbol{LFI:slope:LFI27:Rad:S:units}{$-0.82$\mHz}
\setsymbol{LFI:slope:LFI28:Rad:S:units}{$-0.91$\mHz}

\setsymbol{LFI:slope:LFI18:Rad:M}{$-1.04$}
\setsymbol{LFI:slope:LFI19:Rad:M}{$-1.09$}
\setsymbol{LFI:slope:LFI20:Rad:M}{$-0.69$}
\setsymbol{LFI:slope:LFI21:Rad:M}{$-1.56$}
\setsymbol{LFI:slope:LFI22:Rad:M}{$-1.01$}
\setsymbol{LFI:slope:LFI23:Rad:M}{$-0.96$}
\setsymbol{LFI:slope:LFI24:Rad:M}{$-0.83$}
\setsymbol{LFI:slope:LFI25:Rad:M}{$-0.91$}
\setsymbol{LFI:slope:LFI26:Rad:M}{$-0.95$}
\setsymbol{LFI:slope:LFI27:Rad:M}{$-0.87$}
\setsymbol{LFI:slope:LFI28:Rad:M}{$-0.88$}
\setsymbol{LFI:slope:LFI18:Rad:S}{$-1.15$}
\setsymbol{LFI:slope:LFI19:Rad:S}{$-1.00$}
\setsymbol{LFI:slope:LFI20:Rad:S}{$-0.95$}
\setsymbol{LFI:slope:LFI21:Rad:S}{$-0.92$}
\setsymbol{LFI:slope:LFI22:Rad:S}{$-1.01$}
\setsymbol{LFI:slope:LFI23:Rad:S}{$-0.95$}
\setsymbol{LFI:slope:LFI24:Rad:S}{$-0.73$}
\setsymbol{LFI:slope:LFI25:Rad:S}{$-1.16$}
\setsymbol{LFI:slope:LFI26:Rad:S}{$-0.79$}
\setsymbol{LFI:slope:LFI27:Rad:S}{$-0.82$}
\setsymbol{LFI:slope:LFI28:Rad:S}{$-0.91$}

\setsymbol{LFI:FWHM:70GHz:units}{13\parcm01}
\setsymbol{LFI:FWHM:44GHz:units}{27\parcm92}
\setsymbol{LFI:FWHM:30GHz:units}{32\parcm65}

\setsymbol{LFI:FWHM:70GHz}{13.01}
\setsymbol{LFI:FWHM:44GHz}{27.92}
\setsymbol{LFI:FWHM:30GHz}{32.65}

\setsymbol{LFI:FWHM:LFI18:units}{13\parcm39}
\setsymbol{LFI:FWHM:LFI19:units}{13\parcm01}
\setsymbol{LFI:FWHM:LFI20:units}{12\parcm75}
\setsymbol{LFI:FWHM:LFI21:units}{12\parcm74}
\setsymbol{LFI:FWHM:LFI22:units}{12\parcm87}
\setsymbol{LFI:FWHM:LFI23:units}{13\parcm27}
\setsymbol{LFI:FWHM:LFI24:units}{22\parcm98}
\setsymbol{LFI:FWHM:LFI25:units}{30\parcm46}
\setsymbol{LFI:FWHM:LFI26:units}{30\parcm31}
\setsymbol{LFI:FWHM:LFI27:units}{32\parcm65}
\setsymbol{LFI:FWHM:LFI28:units}{32\parcm66}

\setsymbol{LFI:FWHM:LFI18}{13.39}
\setsymbol{LFI:FWHM:LFI19}{13.01}
\setsymbol{LFI:FWHM:LFI20}{12.75}
\setsymbol{LFI:FWHM:LFI21}{12.74}
\setsymbol{LFI:FWHM:LFI22}{12.87}
\setsymbol{LFI:FWHM:LFI23}{13.27}
\setsymbol{LFI:FWHM:LFI24}{22.98}
\setsymbol{LFI:FWHM:LFI25}{30.46}
\setsymbol{LFI:FWHM:LFI26}{30.31}
\setsymbol{LFI:FWHM:LFI27}{32.65}
\setsymbol{LFI:FWHM:LFI28}{32.66}

\setsymbol{LFI:FWHM:uncertainty:LFI18:units}{0.170\arcm}
\setsymbol{LFI:FWHM:uncertainty:LFI19:units}{0.174\arcm}
\setsymbol{LFI:FWHM:uncertainty:LFI20:units}{0.170\arcm}
\setsymbol{LFI:FWHM:uncertainty:LFI21:units}{0.156\arcm}
\setsymbol{LFI:FWHM:uncertainty:LFI22:units}{0.164\arcm}
\setsymbol{LFI:FWHM:uncertainty:LFI23:units}{0.171\arcm}
\setsymbol{LFI:FWHM:uncertainty:LFI24:units}{0.652\arcm}
\setsymbol{LFI:FWHM:uncertainty:LFI25:units}{1.075\arcm}
\setsymbol{LFI:FWHM:uncertainty:LFI26:units}{1.131\arcm}
\setsymbol{LFI:FWHM:uncertainty:LFI27:units}{1.266\arcm}
\setsymbol{LFI:FWHM:uncertainty:LFI28:units}{1.287\arcm}

\setsymbol{LFI:FWHM:uncertainty:LFI18}{0.170}
\setsymbol{LFI:FWHM:uncertainty:LFI19}{0.174}
\setsymbol{LFI:FWHM:uncertainty:LFI20}{0.170}
\setsymbol{LFI:FWHM:uncertainty:LFI21}{0.156}
\setsymbol{LFI:FWHM:uncertainty:LFI22}{0.164}
\setsymbol{LFI:FWHM:uncertainty:LFI23}{0.171}
\setsymbol{LFI:FWHM:uncertainty:LFI24}{0.652}
\setsymbol{LFI:FWHM:uncertainty:LFI25}{1.075}
\setsymbol{LFI:FWHM:uncertainty:LFI26}{1.131}
\setsymbol{LFI:FWHM:uncertainty:LFI27}{1.266}
\setsymbol{LFI:FWHM:uncertainty:LFI28}{1.287}

\setsymbol{HFI:center:frequency:100GHz:units}{100\,GHz}
\setsymbol{HFI:center:frequency:143GHz:units}{143\,GHz}
\setsymbol{HFI:center:frequency:217GHz:units}{217\,GHz}
\setsymbol{HFI:center:frequency:353GHz:units}{353\,GHz}
\setsymbol{HFI:center:frequency:545GHz:units}{545\,GHz}
\setsymbol{HFI:center:frequency:857GHz:units}{857\,GHz}

\setsymbol{HFI:center:frequency:100GHz}{100}
\setsymbol{HFI:center:frequency:143GHz}{143}
\setsymbol{HFI:center:frequency:217GHz}{217}
\setsymbol{HFI:center:frequency:353GHz}{353}
\setsymbol{HFI:center:frequency:545GHz}{545}
\setsymbol{HFI:center:frequency:857GHz}{857}

\setsymbol{HFI:Ndetectors:100GHz}{8}
\setsymbol{HFI:Ndetectors:143GHz}{11}
\setsymbol{HFI:Ndetectors:217GHz}{12}
\setsymbol{HFI:Ndetectors:353GHz}{12}
\setsymbol{HFI:Ndetectors:545GHz}{3}
\setsymbol{HFI:Ndetectors:857GHz}{4}

\setsymbol{HFI:FWHM:Maps:100GHz:units}{9\parcm88}
\setsymbol{HFI:FWHM:Maps:143GHz:units}{7\parcm18}
\setsymbol{HFI:FWHM:Maps:217GHz:units}{4\parcm87}
\setsymbol{HFI:FWHM:Maps:353GHz:units}{4\parcm65}
\setsymbol{HFI:FWHM:Maps:545GHz:units}{4\parcm72}
\setsymbol{HFI:FWHM:Maps:857GHz:units}{4\parcm39}
\setsymbol{HFI:FWHM:Maps:100GHz}{9.88}
\setsymbol{HFI:FWHM:Maps:143GHz}{7.18}
\setsymbol{HFI:FWHM:Maps:217GHz}{4.87}
\setsymbol{HFI:FWHM:Maps:353GHz}{4.65}
\setsymbol{HFI:FWHM:Maps:545GHz}{4.72}
\setsymbol{HFI:FWHM:Maps:857GHz}{4.39}

\setsymbol{HFI:beam:ellipticity:Maps:100GHz}{1.15}
\setsymbol{HFI:beam:ellipticity:Maps:143GHz}{1.01}
\setsymbol{HFI:beam:ellipticity:Maps:217GHz}{1.06}
\setsymbol{HFI:beam:ellipticity:Maps:353GHz}{1.05}
\setsymbol{HFI:beam:ellipticity:Maps:545GHz}{1.14}
\setsymbol{HFI:beam:ellipticity:Maps:857GHz}{1.19}

\setsymbol{HFI:FWHM:Mars:100GHz:units}{9\parcm37}
\setsymbol{HFI:FWHM:Mars:143GHz:units}{7\parcm04}
\setsymbol{HFI:FWHM:Mars:217GHz:units}{4\parcm68}
\setsymbol{HFI:FWHM:Mars:353GHz:units}{4\parcm43}
\setsymbol{HFI:FWHM:Mars:545GHz:units}{3\parcm80}
\setsymbol{HFI:FWHM:Mars:857GHz:units}{3\parcm67}

\setsymbol{HFI:FWHM:Mars:100GHz}{9.37}
\setsymbol{HFI:FWHM:Mars:143GHz}{7.04}
\setsymbol{HFI:FWHM:Mars:217GHz}{4.68}
\setsymbol{HFI:FWHM:Mars:353GHz}{4.43}
\setsymbol{HFI:FWHM:Mars:545GHz}{3.80}
\setsymbol{HFI:FWHM:Mars:857GHz}{3.67}

\setsymbol{HFI:beam:ellipticity:Mars:100GHz}{1.18}
\setsymbol{HFI:beam:ellipticity:Mars:143GHz}{1.03}
\setsymbol{HFI:beam:ellipticity:Mars:217GHz}{1.14}
\setsymbol{HFI:beam:ellipticity:Mars:353GHz}{1.09}
\setsymbol{HFI:beam:ellipticity:Mars:545GHz}{1.25}
\setsymbol{HFI:beam:ellipticity:Mars:857GHz}{1.03}

\setsymbol{HFI:CMB:relative:calibration:100GHz}{$\lsim 1\%$}
\setsymbol{HFI:CMB:relative:calibration:143GHz}{$\lsim 1\%$}
\setsymbol{HFI:CMB:relative:calibration:217GHz}{$\lsim 1\%$}
\setsymbol{HFI:CMB:relative:calibration:353GHz}{$\lsim 1\%$}
\setsymbol{HFI:CMB:relative:calibration:545GHz}{}
\setsymbol{HFI:CMB:relative:calibration:857GHz}{}

\setsymbol{HFI:CMB:absolute:calibration:100GHz}{$\lsim 2\%$}
\setsymbol{HFI:CMB:absolute:calibration:143GHz}{$\lsim 2\%$}
\setsymbol{HFI:CMB:absolute:calibration:217GHz}{$\lsim 2\%$}
\setsymbol{HFI:CMB:absolute:calibration:353GHz}{$\lsim 2\%$}
\setsymbol{HFI:CMB:absolute:calibration:545GHz}{}
\setsymbol{HFI:CMB:absolute:calibration:857GHz}{}

\setsymbol{HFI:FIRAS:gain:calibration:accuracy:statistical:100GHz}{}
\setsymbol{HFI:FIRAS:gain:calibration:accuracy:statistical:143GHz}{}
\setsymbol{HFI:FIRAS:gain:calibration:accuracy:statistical:217GHz}{}
\setsymbol{HFI:FIRAS:gain:calibration:accuracy:statistical:353GHz}{2.5\%}
\setsymbol{HFI:FIRAS:gain:calibration:accuracy:statistical:545GHz}{1\%}
\setsymbol{HFI:FIRAS:gain:calibration:accuracy:statistical:857GHz}{0.5\%}

\setsymbol{HFI:FIRAS:gain:calibration:accuracy:systematic:100GHz}{}
\setsymbol{HFI:FIRAS:gain:calibration:accuracy:systematic:143GHz}{}
\setsymbol{HFI:FIRAS:gain:calibration:accuracy:systematic:217GHz}{}
\setsymbol{HFI:FIRAS:gain:calibration:accuracy:systematic:353GHz}{}
\setsymbol{HFI:FIRAS:gain:calibration:accuracy:systematic:545GHz}{7\%}
\setsymbol{HFI:FIRAS:gain:calibration:accuracy:systematic:857GHz}{7\%}

\setsymbol{HFI:FIRAS:zero:point:accuracy:100GHz:units}{0.8\MJysr}
\setsymbol{HFI:FIRAS:zero:point:accuracy:143GHz:units}{}
\setsymbol{HFI:FIRAS:zero:point:accuracy:217GHz:units}{}
\setsymbol{HFI:FIRAS:zero:point:accuracy:353GHz:units}{1.4\MJysr}
\setsymbol{HFI:FIRAS:zero:point:accuracy:545GHz:units}{2.2\MJysr}
\setsymbol{HFI:FIRAS:zero:point:accuracy:857GHz:units}{1.7\MJysr}

\setsymbol{HFI:FIRAS:zero:point:accuracy:100GHz}{0.8}
\setsymbol{HFI:FIRAS:zero:point:accuracy:143GHz}{}
\setsymbol{HFI:FIRAS:zero:point:accuracy:217GHz}{}
\setsymbol{HFI:FIRAS:zero:point:accuracy:353GHz}{1.4}
\setsymbol{HFI:FIRAS:zero:point:accuracy:545GHz}{2.2}
\setsymbol{HFI:FIRAS:zero:point:accuracy:857GHz}{1.7}

\setsymbol{HFI:unit:conversion:100GHz:units}{0.2415\MJysrmK}
\setsymbol{HFI:unit:conversion:143GHz:units}{0.3694\MJysrmK}
\setsymbol{HFI:unit:conversion:217GHz:units}{0.4811\MJysrmK}
\setsymbol{HFI:unit:conversion:353GHz:units}{0.2883\MJysrmK}
\setsymbol{HFI:unit:conversion:545GHz:units}{0.05826\MJysrmK}
\setsymbol{HFI:unit:conversion:857GHz:units}{0.002238\MJysrmK}

\setsymbol{HFI:unit:conversion:100GHz}{0.2415}
\setsymbol{HFI:unit:conversion:143GHz}{0.3694}
\setsymbol{HFI:unit:conversion:217GHz}{0.4811}
\setsymbol{HFI:unit:conversion:353GHz}{0.2883}
\setsymbol{HFI:unit:conversion:545GHz}{0.05826}
\setsymbol{HFI:unit:conversion:857GHz}{0.002238}

\setsymbol{HFI:colour:correction:alpha=-2:V1.01:100GHz}{0.9893}
\setsymbol{HFI:colour:correction:alpha=-2:V1.01:143GHz}{0.9759}
\setsymbol{HFI:colour:correction:alpha=-2:V1.01:217GHz}{1.0007}
\setsymbol{HFI:colour:correction:alpha=-2:V1.01:353GHz}{1.0028}
\setsymbol{HFI:colour:correction:alpha=-2:V1.01:545GHz}{1.0019}
\setsymbol{HFI:colour:correction:alpha=-2:V1.01:857GHz}{0.9889}

\setsymbol{HFI:colour:correction:alpha=0:V1.01:100GHz}{1.0008}
\setsymbol{HFI:colour:correction:alpha=0:V1.01:143GHz}{1.0148}
\setsymbol{HFI:colour:correction:alpha=0:V1.01:217GHz}{0.9909}
\setsymbol{HFI:colour:correction:alpha=0:V1.01:353GHz}{0.9888}
\setsymbol{HFI:colour:correction:alpha=0:V1.01:545GHz}{0.9878}
\setsymbol{HFI:colour:correction:alpha=0:V1.01:857GHz}{1.0014}

\providecommand{\sorthelp}[1]{}

%% file: PIP_94_Boulanger_authors_and_institutes.tex
\author{\small
Planck Collaboration:
R.~Adam\inst{68}
\and
P.~A.~R.~Ade\inst{79}
\and
N.~Aghanim\inst{54}
\and
M.~I.~R.~Alves\inst{54}
\and
M.~Arnaud\inst{66}
\and
D.~Arzoumanian\inst{54}
\and
M.~Ashdown\inst{63, 6}
\and
J.~Aumont\inst{54}
\and
C.~Baccigalupi\inst{78}
\and
A.~J.~Banday\inst{85, 10}
% ...
}
\institute{\small
APC, AstroParticule et Cosmologie, Universit\'{e} Paris Diderot, CNRS/IN2P3, CEA/lrfu, Observatoire de Paris, Sorbonne Paris Cit\'{e}, 10, rue Alice Domon et L\'{e}onie Duquet, 75205 Paris Cedex 13, France\goodbreak
\and
African Institute for Mathematical Sciences, 6-8 Melrose Road, Muizenberg, Cape Town, South Africa\goodbreak
\and
Agenzia Spaziale Italiana Science Data Center, Via del Politecnico snc, 00133, Roma, Italy\goodbreak
\and
Agenzia Spaziale Italiana, Viale Liegi 26, Roma, Italy\goodbreak
\and
Aix Marseille Universit\'{e}, CNRS, LAM (Laboratoire d'Astrophysique de Marseille) UMR 7326, 13388, Marseille, France\goodbreak
\and
Astrophysics Group, Cavendish Laboratory, University of Cambridge, J J Thomson Avenue, Cambridge CB3 0HE, U.K.\goodbreak
\and
Astrophysics \& Cosmology Research Unit, School of Mathematics, Statistics \& Computer Science, University of KwaZulu-Natal, Westville Campus, Private Bag X54001, Durban 4000, South Africa\goodbreak
\and
Atacama Large Millimeter/submillimeter Array, ALMA Santiago Central Offices, Alonso de Cordova 3107, Vitacura, Casilla 763 0355, Santiago, Chile\goodbreak
\and
CITA, University of Toronto, 60 St. George St., Toronto, ON M5S 3H8, Canada\goodbreak
\and
CNRS, IRAP, 9 Av. colonel Roche, BP 44346, F-31028 Toulouse cedex 4, France\goodbreak
\and
California Institute of Technology, Pasadena, California, U.S.A.\goodbreak
\and
Centro de Estudios de F\'{i}sica del Cosmos de Arag\'{o}n (CEFCA), Plaza San Juan, 1, planta 2, E-44001, Teruel, Spain\goodbreak
\and
Computational Cosmology Center, Lawrence Berkeley National Laboratory, Berkeley, California, U.S.A.\goodbreak
\and
DSM/Irfu/SPP, CEA-Saclay, F-91191 Gif-sur-Yvette Cedex, France\goodbreak
\and
DTU Space, National Space Institute, Technical University of Denmark, Elektrovej 327, DK-2800 Kgs. Lyngby, Denmark\goodbreak
\and
D\'{e}partement de Physique Th\'{e}orique, Universit\'{e} de Gen\`{e}ve, 24, Quai E. Ansermet,1211 Gen\`{e}ve 4, Switzerland\goodbreak
\and
Departamento de F\'{\i}sica, Universidad de Oviedo, Avda. Calvo Sotelo s/n, Oviedo, Spain\goodbreak
\and
Department of Astrophysics/IMAPP, Radboud University Nijmegen, P.O. Box 9010, 6500 GL Nijmegen, The Netherlands\goodbreak
\and
Department of Physics \& Astronomy, University of British Columbia, 6224 Agricultural Road, Vancouver, British Columbia, Canada\goodbreak
\and
Department of Physics and Astronomy, Dana and David Dornsife College of Letter, Arts and Sciences, University of Southern California, Los Angeles, CA 90089, U.S.A.\goodbreak
\and
Department of Physics and Astronomy, University College London, London WC1E 6BT, U.K.\goodbreak
\and
Department of Physics, Florida State University, Keen Physics Building, 77 Chieftan Way, Tallahassee, Florida, U.S.A.\goodbreak
\and
Department of Physics, Gustaf H\"{a}llstr\"{o}min katu 2a, University of Helsinki, Helsinki, Finland\goodbreak
\and
Department of Physics, Princeton University, Princeton, New Jersey, U.S.A.\goodbreak
\and
Department of Physics, University of California, Santa Barbara, California, U.S.A.\goodbreak
\and
Department of Physics, University of Illinois at Urbana-Champaign, 1110 West Green Street, Urbana, Illinois, U.S.A.\goodbreak
\and
Dipartimento di Fisica e Astronomia G. Galilei, Universit\`{a} degli Studi di Padova, via Marzolo 8, 35131 Padova, Italy\goodbreak
\and
Dipartimento di Fisica e Scienze della Terra, Universit\`{a} di Ferrara, Via Saragat 1, 44122 Ferrara, Italy\goodbreak
\and
Dipartimento di Fisica, Universit\`{a} La Sapienza, P. le A. Moro 2, Roma, Italy\goodbreak
\and
Dipartimento di Fisica, Universit\`{a} degli Studi di Milano, Via Celoria, 16, Milano, Italy\goodbreak
\and
Dipartimento di Fisica, Universit\`{a} degli Studi di Trieste, via A. Valerio 2, Trieste, Italy\goodbreak
\and
Dipartimento di Fisica, Universit\`{a} di Roma Tor Vergata, Via della Ricerca Scientifica, 1, Roma, Italy\goodbreak
\and
Discovery Center, Niels Bohr Institute, Blegdamsvej 17, Copenhagen, Denmark\goodbreak
\and
Dpto. Astrof\'{i}sica, Universidad de La Laguna (ULL), E-38206 La Laguna, Tenerife, Spain\goodbreak
\and
European Southern Observatory, ESO Vitacura, Alonso de Cordova 3107, Vitacura, Casilla 19001, Santiago, Chile\goodbreak
\and
European Space Agency, ESAC, Planck Science Office, Camino bajo del Castillo, s/n, Urbanizaci\'{o}n Villafranca del Castillo, Villanueva de la Ca\~{n}ada, Madrid, Spain\goodbreak
\and
European Space Agency, ESTEC, Keplerlaan 1, 2201 AZ Noordwijk, The Netherlands\goodbreak
\and
Facolt\`{a} di Ingegneria, Universit\`{a} degli Studi e-Campus, Via Isimbardi 10, Novedrate (CO), 22060, Italy\goodbreak
\and
HGSFP and University of Heidelberg, Theoretical Physics Department, Philosophenweg 16, 69120, Heidelberg, Germany\goodbreak
\and
Helsinki Institute of Physics, Gustaf H\"{a}llstr\"{o}min katu 2, University of Helsinki, Helsinki, Finland\goodbreak
\and
INAF - Osservatorio Astrofisico di Catania, Via S. Sofia 78, Catania, Italy\goodbreak
\and
INAF - Osservatorio Astronomico di Padova, Vicolo dell'Osservatorio 5, Padova, Italy\goodbreak
\and
INAF - Osservatorio Astronomico di Roma, via di Frascati 33, Monte Porzio Catone, Italy\goodbreak
\and
INAF - Osservatorio Astronomico di Trieste, Via G.B. Tiepolo 11, Trieste, Italy\goodbreak
\and
INAF/IASF Bologna, Via Gobetti 101, Bologna, Italy\goodbreak
\and
INAF/IASF Milano, Via E. Bassini 15, Milano, Italy\goodbreak
\and
INFN, Sezione di Bologna, Via Irnerio 46, I-40126, Bologna, Italy\goodbreak
\and
INFN, Sezione di Roma 1, Universit\`{a} di Roma Sapienza, Piazzale Aldo Moro 2, 00185, Roma, Italy\goodbreak
\and
INFN/National Institute for Nuclear Physics, Via Valerio 2, I-34127 Trieste, Italy\goodbreak
\and
IPAG: Institut de Plan\'{e}tologie et d'Astrophysique de Grenoble, Universit\'{e} Grenoble Alpes, IPAG, F-38000 Grenoble, France, CNRS, IPAG, F-38000 Grenoble, France\goodbreak
\and
Imperial College London, Astrophysics group, Blackett Laboratory, Prince Consort Road, London, SW7 2AZ, U.K.\goodbreak
\and
Infrared Processing and Analysis Center, California Institute of Technology, Pasadena, CA 91125, U.S.A.\goodbreak
\and
Institut Universitaire de France, 103, bd Saint-Michel, 75005, Paris, France\goodbreak
\and
Institut d'Astrophysique Spatiale, CNRS (UMR8617) Universit\'{e} Paris-Sud 11, B\^{a}timent 121, Orsay, France\goodbreak
\and
Institut d'Astrophysique de Paris, CNRS (UMR7095), 98 bis Boulevard Arago, F-75014, Paris, France\goodbreak
\and
Institute for Space Sciences, Bucharest-Magurale, Romania\goodbreak
\and
Institute of Astronomy, University of Cambridge, Madingley Road, Cambridge CB3 0HA, U.K.\goodbreak
\and
Institute of Theoretical Astrophysics, University of Oslo, Blindern, Oslo, Norway\goodbreak
\and
Instituto de Astrof\'{\i}sica de Canarias, C/V\'{\i}a L\'{a}ctea s/n, La Laguna, Tenerife, Spain\goodbreak
\and
Instituto de F\'{\i}sica de Cantabria (CSIC-Universidad de Cantabria), Avda. de los Castros s/n, Santander, Spain\goodbreak
\and
Jet Propulsion Laboratory, California Institute of Technology, 4800 Oak Grove Drive, Pasadena, California, U.S.A.\goodbreak
\and
Jodrell Bank Centre for Astrophysics, Alan Turing Building, School of Physics and Astronomy, The University of Manchester, Oxford Road, Manchester, M13 9PL, U.K.\goodbreak
\and
Kavli Institute for Cosmology Cambridge, Madingley Road, Cambridge, CB3 0HA, U.K.\goodbreak
\and
LAL, Universit\'{e} Paris-Sud, CNRS/IN2P3, Orsay, France\goodbreak
\and
LERMA, CNRS, Observatoire de Paris, 61 Avenue de l'Observatoire, Paris, France\goodbreak
\and
Laboratoire AIM, IRFU/Service d'Astrophysique - CEA/DSM - CNRS - Universit\'{e} Paris Diderot, B\^{a}t. 709, CEA-Saclay, F-91191 Gif-sur-Yvette Cedex, France\goodbreak
\and
Laboratoire Traitement et Communication de l'Information, CNRS (UMR 5141) and T\'{e}l\'{e}com ParisTech, 46 rue Barrault F-75634 Paris Cedex 13, France\goodbreak
\and
Laboratoire de Physique Subatomique et de Cosmologie, Universit\'{e} Joseph Fourier Grenoble I, CNRS/IN2P3, Institut National Polytechnique de Grenoble, 53 rue des Martyrs, 38026 Grenoble cedex, France\goodbreak
\and
Laboratoire de Physique Th\'{e}orique, Universit\'{e} Paris-Sud 11 \& CNRS, B\^{a}timent 210, 91405 Orsay, France\goodbreak
\and
Lawrence Berkeley National Laboratory, Berkeley, California, U.S.A.\goodbreak
\and
Max-Planck-Institut f\"{u}r Astrophysik, Karl-Schwarzschild-Str. 1, 85741 Garching, Germany\goodbreak
\and
Max-Planck-Institut f\"{u}r Radioastronomie, Auf dem H\"{u}gel 69, 53121 Bonn, Germany\goodbreak
\and
McGill Physics, Ernest Rutherford Physics Building, McGill University, 3600 rue University, Montr\'{e}al, QC, H3A 2T8, Canada\goodbreak
\and
National University of Ireland, Department of Experimental Physics, Maynooth, Co. Kildare, Ireland\goodbreak
\and
Niels Bohr Institute, Blegdamsvej 17, Copenhagen, Denmark\goodbreak
\and
Observational Cosmology, Mail Stop 367-17, California Institute of Technology, Pasadena, CA, 91125, U.S.A.\goodbreak
\and
Optical Science Laboratory, University College London, Gower Street, London, U.K.\goodbreak
\and
SISSA, Astrophysics Sector, via Bonomea 265, 34136, Trieste, Italy\goodbreak
\and
School of Physics and Astronomy, Cardiff University, Queens Buildings, The Parade, Cardiff, CF24 3AA, U.K.\goodbreak
\and
Space Research Institute (IKI), Russian Academy of Sciences, Profsoyuznaya Str, 84/32, Moscow, 117997, Russia\goodbreak
\and
Space Sciences Laboratory, University of California, Berkeley, California, U.S.A.\goodbreak
\and
Special Astrophysical Observatory, Russian Academy of Sciences, Nizhnij Arkhyz, Zelenchukskiy region, Karachai-Cherkessian Republic, 369167, Russia\goodbreak
\and
Sub-Department of Astrophysics, University of Oxford, Keble Road, Oxford OX1 3RH, U.K.\goodbreak
\and
UPMC Univ Paris 06, UMR7095, 98 bis Boulevard Arago, F-75014, Paris, France\goodbreak
\and
Universit\'{e} de Toulouse, UPS-OMP, IRAP, F-31028 Toulouse cedex 4, France\goodbreak
% ...
}